\documentclass[aps,prx,reprint,superscriptaddress,longbibliography,floatfix,nofootinbib]{revtex4-2}
\usepackage[T1]{fontenc}
\usepackage{amsmath,amssymb,mathtools,graphicx,bm,microtype,xcolor,booktabs}
\usepackage{amsthm}
\usepackage[colorlinks=true,linkcolor=blue!55!black,citecolor=blue!55!black,urlcolor=blue!55!black]{hyperref}
\newcommand{\dd}{\mathrm d}
\newcommand{\Var}{\operatorname{Var}}
\newcommand{\Cov}{\operatorname{Cov}}
\newcommand{\R}{\mathbb R}
\newcommand{\Z}{\mathbb Z}
\newcommand{\one}{\mathbf 1}
\newcommand{\Tr}{\operatorname{Tr}}
\newcommand{\cC}{\mathcal C}

\newcommand{\infosep}{\,{:}\,}
\newcommand{\given}{\,{\mid}\,}
\newcommand{\relentsep}{\,{\Vert}\,}
\hypersetup{pdftitle={Local-channel inequivalence within analytic symmetry-broken phases},pdfauthor={Guanxi Xun, Rintaro Masaoka, Weiguang Cao, and Haruki Watanabe}}
\newtheorem{theorem}{Theorem}
\newtheorem{proposition}[theorem]{Proposition}
\newtheorem{lemma}[theorem]{Lemma}

\newcommand{\smcontentsentry}[2]{%
  \noindent\hyperref[#1]{\makebox[2.7em][l]{\ref*{#1}.}#2}\dotfill\pageref{#1}\par}

\begin{document}
\title{Local-channel inequivalence within analytic symmetry-broken phases}
\author{Guanxi Xun}
\affiliation{Department of Physics, Hong Kong University of Science and Technology, Clear Water Bay, Hong Kong, China}
\author{Rintaro Masaoka}
\affiliation{Department of Physics, Hong Kong University of Science and Technology, Clear Water Bay, Hong Kong, China}
\author{Weiguang Cao}
\affiliation{Department of Physics, Hong Kong University of Science and Technology, Clear Water Bay, Hong Kong, China}
\affiliation{Center for Theoretical Condensed Matter Physics, Hong Kong University of Science and Technology, Clear Water Bay, Hong Kong, China}
\author{Haruki Watanabe}
\email{hwatanabe@ust.hk}
\affiliation{Department of Physics, Hong Kong University of Science and Technology, Clear Water Bay, Hong Kong, China}
\affiliation{Center for Theoretical Condensed Matter Physics, Hong Kong University of Science and Technology, Clear Water Bay, Hong Kong, China}
\affiliation{Institute for Advanced Study, Hong Kong University of Science and Technology, Clear Water Bay, Hong Kong, China}
\date{November 13, 2026}
\begin{abstract}
Gibbs states at different temperatures can be inequivalent under two-way
quasi-local channels even within a single phase with spontaneously broken
continuous symmetry and analytic free energy. We prove this for the symmetry-invariant Gibbs states of a canonical
ideal Bose gas at fixed density and a classical vortex-free Villain model
on hypercubic lattices in $d\ge3$ dimensions.
At lower temperatures, two distant observers can estimate the shared
direction of order more accurately. The maximum mutual information they
can obtain by separate measurements grows logarithmically with the radius
of their regions. The leading term is the same throughout each ordered
phase, while the constant term
$\cC(T)=\tfrac12\ln[\Upsilon(T)/T]+c_{\mathrm{geom}}$
is strictly larger at lower temperatures. Here $\Upsilon(T)$ is the phase
stiffness and $c_{\mathrm{geom}}$ depends only on the region geometry.
A quasi-local channel cannot increase this constant, so it is invariant
under two-way conversion. Cooling would increase it and is therefore
impossible between any two distinct temperatures in the ordered phase.
On finite lattices, cooling with vanishing error requires a circuit whose
causal radius grows at least linearly with the lattice side length.
We also obtain the invariant for quantum and multicomponent Villain models
and conjecture a stiffness formula for more general patterns of continuous
symmetry breaking.
\end{abstract}
\maketitle
\section{Introduction}
\label{sec:intro}

Changing the temperature of an ordered system can leave its free energy
analytic and its broken symmetry unchanged. Can quasi-local operations
then convert its equilibrium states into one another? We show that even
cooling by an arbitrarily small fixed temperature difference can be
impossible. In the condensed phase of a canonical ideal Bose gas and in
the ordered regime of a vortex-free Villain magnet, colder states share
more information about the order direction between sufficiently large,
distant regions.
This difference separates every pair of distinct temperatures under
two-way quasi-local channels.

For gapped ground states, quasiadiabatic continuation links the
Hamiltonian and state descriptions of phases. A smooth,
symmetry-preserving local Hamiltonian path that keeps the excitation
gap open transports the ground-state subspace by a quasi-local
unitary~\cite{HastingsWen2005,Bachmann2012}. This correspondence
underlies the classification of gapped phases through local unitary
transformations~\cite{ChenGuWen2010}.

At finite temperature, two natural definitions invite a similar
comparison. The thermodynamic definition connects Gibbs states by a
path of local Hamiltonians and temperatures along which the bulk
free-energy density is analytic. The operational definition requires a
physical transformation in each direction. We require these transformations
to be quasi-local. For any two sufficiently distant regions, independent
operations on their respective input neighborhoods must reproduce the joint
output state to high accuracy. The two directions may use different
channels~\cite{Coser2019}. Related approaches use local
recovery~\cite{SangHsieh2025}, mixed-state symmetry
breaking~\cite{Lessa2025}, the response to
decoherence~\cite{LeeJianXu2023}, and local reversibility~\cite{Sang2026}.

Stable clustering provides a bridge between these definitions.
Ma, Khemani, and Sang construct two-way quasi-local channels along
Gibbs-state paths where exponential correlation decay survives small
allowed perturbations~\cite{MKS2026,Capel2025}. Thermodynamic stability
and two-way channel connectivity have also been established in
low-temperature domains of classical $\Z_2$ lattice gauge
models~\cite{SengokuCaoWatanabe2026,XunCaoWatanabe2026}. Continuous symmetry breaking
suggests a different possibility: long-wavelength fluctuations
persist throughout the ordered phase. Our examples turn this
possibility into an explicit separation between thermodynamic and
channel equivalence.\footnote{Here analyticity concerns temperature at
zero ordering field. Ref.~\cite{MKS2026} treats finite-dimensional local
spaces and bounded interactions. Our models have unbounded local spaces,
with a fixed total particle number in the Bose gas and a local flatness
constraint in the Villain model.
Their Gibbs states violate the clustering condition even for bounded
symmetric observables, as shown in Supplemental Material (SM),
Secs.~\ref{sm:bose-bounded-clustering}
and~\ref{sm:classical-bond-clustering}.}

Consider two observers measuring cubes of radius $R$ at infinite
separation. In our symmetric Gibbs states every order direction is
equally likely. The two cubes share the same random direction: conditional
on that direction their readings are independent, but knowing one reading
helps predict the other when the direction is unknown.
In the classical Villain magnet, each bond carries an integer and a
local constraint forbids vortices. The integers let each observer
unwrap the spin angles and take a weighted average. Subtracting the
two readings cancels the common angle modulo $2\pi$, leaving the
difference of two independent thermal errors. At lower temperatures
this difference is more tightly concentrated around zero, so one
reading predicts the other more accurately.

In the ideal Bose gas, noninteracting particles hop between neighboring
sites. Each observer uses heterodyne detection to read out complex
amplitudes and takes the phase of their weighted average. In either
model each observer can
report just the angular interval containing their estimate. The
classical mutual information of these two labels measures how much
knowing one reduces uncertainty about the other.

Suppose a quasi-local channel could cool the system. Each observer
could then reproduce their cold measurement by acting on a slightly
larger cube of the warm input. At large separation these two operations
act independently, so they would reproduce the joint law of the cold
labels with a vanishing error in mutual information. To bound what any
warm measurements can achieve, we give each observer all the spin and
bond data in their cube, or, in the Bose gas, the complex amplitudes
used to prepare its quantum state. These data suffice to simulate
every local measurement. For any fixed pair of ordered temperatures,
we find that the cold angle labels from cubes of radius $R$ share more
information than even these warm data on cubes of radius $R+o(R)$.
The gap stays positive as $R$ grows, ruling out the cooling channel.

To quantify this gap, we write $M_T(R)$ for the supremum of the mutual
information over the two observers' separate finite-outcome measurements
with the outcome budget specified in Sec.~\ref{sec:local-readers}.
Optimized local information is an established
correlation measure~\cite{DiVincenzo2004}, and optimized distant
correlations have yielded channel invariants in gauge and topological
models~\cite{XunCaoWatanabe2026,Watanabe2026}. Here the key information
is the constant term in the large-block limit.

\begin{samepage}
For the canonical Bose gas and one-component Villain models on a
hypercubic lattice in dimension $d\ge3$, we find
\begin{equation}
 M_T(R)=\frac{d-2}{2}\ln R+
 \cC(T)+o(1),
 \label{eq:intro-result}
\end{equation}
\begin{equation}
 \cC(T)
 =\frac12\ln\frac{\Upsilon(T)}{T}
 +\frac12\ln\frac{\pi\kappa_d}{e}.
 \label{eq:intro-invariant}
\end{equation}
\end{samepage}
Here $\Upsilon(T)$ is the phase stiffness, defined by the free-energy
cost of a boundary twist, and $\kappa_d$ is the electrostatic capacity
of the unit cube $[-1,1]^d$. The remainder $o(1)$ vanishes
as $R\to\infty$. The leading growth is the same throughout the
ordered phase, while the constant retains the temperature dependence.

Enlarging the warm cubes changes the leading logarithm by $o(1)$,
so the gap between the cold readings and the warm upper bound tends to
$\cC(T_{\rm cold})-\cC(T_{\rm warm})$. This difference is positive
because $\Upsilon(T)/T$ decreases strictly with temperature in both
models. Section~\ref{sec:local-readers} specifies the locality condition
and relates it to the disjoint input regions that influence distant
outputs of a local circuit. Section~\ref{sec:reader-monotonicity}
proves that $\cC$ cannot increase under any such channel and is
therefore invariant under two-way conversion. For circuits whose
range grows with system size, the finite-size bounds of
Secs.~\ref{sec:optimal-cooling} and~\ref{sec:spin-reader} establish
the cooling obstruction using growing blocks.

The two models realize this separation in complementary ways. We
vary only the temperature of each fixed Hamiltonian, keeping the
density fixed for the canonical Bose gas. Its condensed phase
$0<T<T_c$ has analytic free energy and stiffness
$\Upsilon(T)=2J\varrho_0(T)$, where $J$ is the hopping amplitude and
$\varrho_0(T)$ is the condensate density, which vanishes at
$T_c$~\cite{PathriaBeale2011,Barber1977}. The classical vortex-free
Villain model instead has $\Upsilon(T)=J$, with $J$ its interaction
strength, and remains ordered with analytic free energy at every
positive temperature. Spin measurements alone attain its invariant.
Both models thus contain a continuum of channel classes within one
analytic ordered regime.

This distinction also constrains attempts to reconcile the two phase
definitions. Local reversibility was proposed to protect information
that ordinary two-way channels can erase~\cite{Sang2026}. Adding that
requirement cannot connect our already inequivalent states. At the
other extreme, permitting depth $O(L)$ on a system of linear size
$L$ can erase thermodynamic phase distinctions: our Bose preparation
connects condensed and noncondensed states.\footnote{The preparation
allows unbounded on-site occupation and general local gates. Its
spatial-depth bound and resources are specified in
Sec.~\ref{sec:optimal-cooling} and SM
Sec.~\ref{sm:bose-linear-preparation}.}
For cooling within the ordered regimes, linear depth is optimal:
fixed-range circuits of depth $o(L)$ leave a nonvanishing error.

\begin{figure}[t]
\centering
\includegraphics[width=\columnwidth]{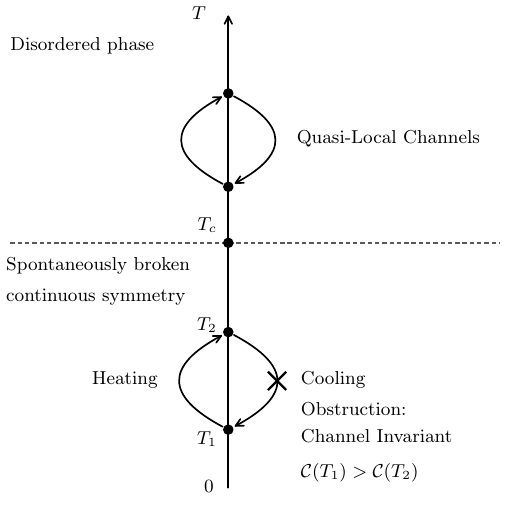}
\caption{Proposed channel classification in $d>2$. In both models the
crossed arrow is rigorously excluded because cooling from $T_2$ to
$T_1$ would increase the mutual information accessible to two distant
observers. Heating for general temperature pairs and
two-way conversion throughout the disordered phase are conjectural in
this diagram. The vortex-free Villain model remains ordered at every $T>0$.}
\label{fig:phase-qlc}
\end{figure}

The exact quantum and coupled multicomponent Villain extensions show
that this mechanism survives quantum fluctuations and multiple shared
angles. We then conjecture its extension to a general pattern of
spontaneous symmetry breaking $G\to H$. Here $G$ is the internal
symmetry group and $H$ is the unbroken subgroup that preserves a chosen
order direction; the distinct ordered directions form the space $G/H$.
When there are several independent tilts of the order direction, the
scalar phase stiffness becomes a matrix that measures the free-energy
cost of slowly varying these directions in space.

For ordered intervals satisfying the conditions of
Sec.~\ref{sec:general}, we formulate two independent conjectures:
the free-energy density remains analytic, and the information constant
is governed by the determinant of the thermal stiffness matrix,
obtained by dividing the stiffness matrix by $T$.
If both hold, temperatures with different values of
this determinant belong to different channel classes even within
one analytic thermodynamic phase. Figure~\ref{fig:phase-qlc} sketches
this proposed extension of our exact examples. Establishing these
conjectures for finite-spin magnets remains an open problem.

Table~\ref{tab:model-summary} summarizes the results and conjectures.

\begin{table*}[t]
\caption{Summary of the ordered regimes for $d\ge3$. The first four rows are rigorously derived; the last gives the conjectures under the assumptions of Sec.~\ref{sec:general}.}
\label{tab:model-summary}
\begingroup
\small
\setlength{\tabcolsep}{3pt}
\renewcommand{\arraystretch}{1.35}
\begin{tabular}{@{}p{0.14\textwidth}p{0.115\textwidth}p{0.22\textwidth}p{0.285\textwidth}p{0.18\textwidth}@{}}
\toprule
Model & Breaking\newline pattern & Free energy $f(T)$ & Connected correlations\newline $r\to\infty$ & Channel invariant\newline $\cC(T)$ \\
\midrule
Ideal Bose gas\newline Canonical ensemble\newline Sec.~\ref{sec:bose}\newline $0<T<T_c$\newline Fixed density $\varrho$
& $U(1)\!\to\!\{1\}$\newline $\mathcal M=S^1$\newline $n=1$
& $\displaystyle T\int_k\ln\!\bigl(1-e^{-\epsilon(k)/T}\bigr)$\newline Eq.~\eqref{eq:bose-free}
& $\displaystyle C_\parallel=C_\perp\sim\frac{T A_d}{2Jr^{d-2}}$\newline SM Sec.~\ref{sm:bose-correlations}
& $\displaystyle\frac12\ln\frac{\Upsilon}T+c_{\rm geom}$\newline $c_{\rm geom}=\tfrac12\ln\frac{\pi\kappa_d}{e}$\newline $\Upsilon=2J\varrho_0$\newline Eqs.~\eqref{eq:helicity-condensate}, \eqref{eq:invariant-stiffness} \\
\addlinespace
Classical vortex-free Villain\newline Secs.~\ref{sec:model}--\ref{sec:info}\newline $T>0$
& $O(2)\!\to\!O(1)$\newline $\mathcal M=S^1$\newline $n=1$
& $\displaystyle\frac T2\!\Bigl(\ln\frac{2\pi J}T+\ell_d\Bigr)$\newline Eq.~\eqref{eq:free-energy}
& $\displaystyle\begin{aligned}C_\perp&\sim\frac{m^2T A_d}{Jr^{d-2}},\\ C_\parallel&\sim\frac{m^2T^2 A_d^2}{2J^2r^{2(d-2)}}.\end{aligned}$\newline SM Sec.~\ref{sm:classical-correlations}
& $\displaystyle\frac12\ln\frac{\Upsilon}T+c_{\rm geom}$\newline $c_{\rm geom}=\tfrac12\ln\frac{\pi\kappa_d}{e}$\newline $\Upsilon=J$\newline Eqs.~\eqref{eq:stiffness}, \eqref{eq:value} \\
\addlinespace
Quantum vortex-free Villain\newline Sec.~\ref{sec:quantum}\newline $T>0$, $\chi>0$
& $O(2)\!\to\!O(1)$\newline $\mathcal M=S^1$\newline $n=1$
& $\displaystyle T\int_k\ln\!\Bigl(2\sinh\frac{\omega_k}{2T}\Bigr)$\newline Eq.~\eqref{eq:quantum-free-energy}
& $\displaystyle\begin{aligned}C_\perp&\sim\frac{m_{\rm q}^2T A_d}{Jr^{d-2}},\\ C_\parallel&\sim\frac{m_{\rm q}^2T^2 A_d^2}{2J^2r^{2(d-2)}}.\end{aligned}$\newline SM Sec.~\ref{sm:quantum-details}
& $\displaystyle\frac12\ln\frac{\Upsilon}T+c_{\rm geom}$\newline $c_{\rm geom}=\tfrac12\ln\frac{\pi\kappa_d}{e}$\newline $\Upsilon=J$\newline Eq.~\eqref{eq:quantum-information} \\
\addlinespace
Multicomponent classical vortex-free Villain\newline Sec.~\ref{sec:multicomponent}\newline $T>0$, $n\ge1$
& $U(1)^n\!\to\!\{1\}$\newline $\mathcal M=\mathbb T^n$
& \mbox{$\displaystyle\frac T2\biggl[\ln\det\!\Bigl(\frac{2\pi K}T\Bigr)+n\ell_d\biggr]$}\newline Eq.~\eqref{eq:multi-free}
& $\displaystyle\begin{aligned}C_\perp^{aa}&\sim\frac{m_a^2T(K^{-1})_{aa}A_d}{r^{d-2}},\\ C_\parallel^{aa}&\sim\frac{m_a^2}{2}\!\biggl[\frac{T(K^{-1})_{aa}A_d}{r^{d-2}}\biggr]^2.\end{aligned}$\newline SM Sec.~\ref{sm:multicomponent-details}
& $\displaystyle\tfrac12\ln\det\!\Bigl(\frac{\boldsymbol\Upsilon}T\Bigr)+c_{\rm geom}$\newline $c_{\rm geom}=\tfrac n2\ln\frac{\pi\kappa_d}{e}$\newline $\boldsymbol\Upsilon=K$\newline \mbox{Eqs.~\eqref{eq:multi-stiffness},\,\eqref{eq:multi-invariant},\,\eqref{eq:multi-geometry}} \\
\addlinespace
General ordered phase\newline Sec.~\ref{sec:general}\newline Conjecture
& $G\to H$\newline $\mathcal M=G/H$\newline $n=\dim\mathcal M$
& Model dependent\newline Analyticity is a separate\newline conjecture.
& Tangent fields: $r^{-(d-2)}$\newline (Gaussian prediction).\newline Longitudinal: model dependent.\newline Sec.~\ref{sec:general-information}
& $\displaystyle\tfrac12\ln\det_{g_0}\!\Bigl(\frac{\boldsymbol\Upsilon}T\Bigr)+c_{\rm geom}$\newline $\displaystyle c_{\rm geom}=\tfrac n2\ln\frac{\kappa_d}{4\pi e}$\newline\hspace*{\fill}$\displaystyle{}+\ln\operatorname{Vol}(\mathcal M,g_0)$\newline Eqs.~\eqref{eq:general-conjecture}, \eqref{eq:coset-geometry} \\
\bottomrule
\end{tabular}
\par\smallskip
\begin{minipage}{\textwidth}\footnotesize\raggedright
$f$ is per site; $\int_k$, $\epsilon(k)$, and $\ell_d$ are defined in Eqs.~\eqref{eq:bose-integral}, \eqref{eq:bose-dispersion}, and~\eqref{eq:green}. For $k\ne0$ the oscillator frequency is $\omega_k=\sqrt{\epsilon(k)/\chi}$, with inertia $\chi$. Classical free energies use angular measure $\dd\theta/(2\pi)$ per component; quantum free energies include zero-point energy.

$\mathcal M$ is the space of order directions and $n=\dim\mathcal M$.
$\cC$ is the information constant after subtracting $n(d-2)\ln R/2$.
$\varrho_0$ is the condensate density; $m,m_{\rm q},m_a$ are the classical, quantum, and component-$a$ magnetization magnitudes; $\Upsilon$ (or $\boldsymbol\Upsilon$) is the stiffness. $K$ is a positive-definite symmetric coupling matrix. $A_d\coloneqq\Gamma(d/2-1)/(4\pi^{d/2})$, and $\kappa_d$ is the capacity of $[-1,1]^d$ [Eq.~\eqref{eq:bose-capacity}]. $g_0$ is a temperature-independent, $G$-invariant reference metric on $\mathcal M$; $\det_{g_0}$ and $\operatorname{Vol}$ denote the determinant relative to this metric and its total volume. Here $O(1)\simeq\Z_2$; the multicomponent row lists only continuous symmetries.

$C_\parallel,C_\perp$ are longitudinal and transverse connected correlations in a selected ordered state; $aa$ specifies component $a=1,\ldots,n$, without summation; the invariant refers to the symmetric Gibbs state. The limit is $r=|x|\to\infty$ at fixed $T>0$, after the thermodynamic limit; $\sim$ denotes a ratio tending to one. Temperature arguments are suppressed; in the last row, $\boldsymbol\Upsilon$ denotes $\boldsymbol\Upsilon(T,q)$ at any order direction $q$, and the relative determinant is independent of $q$. Reader and channel assumptions are given in Sec.~\ref{sec:bose-information}.
\end{minipage}
\endgroup
\end{table*}

Section~\ref{sec:bose} treats the Bose gas, Sec.~\ref{sec:villain} the
Villain models, and Sec.~\ref{sec:general} the conjectures for $G\to H$,
followed by conclusions in Sec.~\ref{sec:end}. The main text derives
the free energies and channel invariants of the canonical Bose gas
and one-component classical Villain model. The Supplemental Material
contains the detailed estimates, extension proofs, and circuit
constructions.
\section{Ideal Bose gas}
\label{sec:bose}

We derive thermodynamics and the information invariant for a
fixed-density canonical family, introducing the reader and channel
definitions used throughout the paper. A grand-canonical generating
function gives its free energy and phase stiffness.

\subsection{Model}
\label{sec:bose-model}

Consider a periodic $d$-dimensional hypercubic lattice with $L^d$ sites,
where the spatial dimension $d$ is any fixed integer at least three.
At each site put a bosonic Fock space with occupations $0,1,2,\ldots$,
annihilation operator $b_x$, and $[b_x,b_y^\dagger]=\delta_{xy}$. Fix a common nearest-neighbor hopping strength $J>0$ and
the particle density $\varrho>0$. The Hamiltonian and canonical state are
\begin{align}
 H_L&\coloneqq J\sum_{x,i}
 (b_{x+\hat e_i}^\dagger-b_x^\dagger)(b_{x+\hat e_i}-b_x),
 \label{eq:bose-H}\\
 \rho_L(T)&\coloneqq
 \frac{P_Ne^{-H_L/T}}{Z_{L,N}(T)},\qquad
 N\coloneqq\lfloor\varrho L^d\rfloor,
 \label{eq:bose-state}
\end{align}
where $P_N$ projects onto total particle number $N$ and
$Z_{L,N}(T)\coloneqq\Tr(P_Ne^{-H_L/T})$. The floor $\lfloor x\rfloor$ is the greatest integer not exceeding $x$, and we set Boltzmann's constant and the lattice spacing to one. The sum runs over sites $x$ and coordinate directions $i=1,\ldots,d$;
$\hat e_i$ is a unit step in direction $i$.
We use $\varrho$ for particle density and $\rho_L(T)$ for the density operator. The Hamiltonian is
nearest neighbor, translation invariant, and invariant under the
global phase rotation $b_x\mapsto e^{i\alpha}b_x$.
The particles are noninteracting, with fixed total number $N$.
Channels and local measurements act on the full tensor product of Fock
spaces, so a local measurement may compare the field phase with a fixed reference.

For any integrable function $u(k)$ on the Brillouin zone, we write
\begin{equation}
 \int_k u(k)\coloneqq\int_{[-\pi,\pi]^d}\frac{\dd^d k}{(2\pi)^d}u(k).
 \label{eq:bose-integral}
\end{equation}
The dispersion, saturation density of excited particles, and
condensate density are
\begin{align}
 \epsilon(k)&\coloneqq2J\sum_{i=1}^d(1-\cos k_i),
 \label{eq:bose-dispersion}\\
 \varrho_{\rm ex}(T)&\coloneqq\int_k\frac1{e^{\epsilon(k)/T}-1},
 \label{eq:bose-excited-density}\\
 \varrho_0(T)&\coloneqq\max\{\varrho-\varrho_{\rm ex}(T),0\}.
 \label{eq:bose-density}
\end{align}
The density integral is finite because $\epsilon(k)=J|k|^2+O(|k|^4)$ near zero and $d>2$. Here $O(|k|^4)$ is bounded in magnitude by a constant times $|k|^4$ as $k\to0$. The density $\varrho_{\rm ex}(T)$ is strictly
increasing from zero to infinity.
Thus $\varrho_{\rm ex}(T_c)=\varrho$ defines a unique $0<T_c<\infty$.
The condensate density $\varrho_0(T)$ is positive for $0<T<T_c$
and vanishes for $T\ge T_c$. In a component with a selected phase,
the field expectation has magnitude $m(T)=\sqrt{\varrho_0(T)}$
[Sec.~\ref{sec:bose-local}]; it tends to zero at $T_c$ and
to $\sqrt\varrho$ as $T\downarrow0$.

\subsection{Thermodynamics}
\label{sec:bose-thermo}

We derive the canonical free energy and its analyticity, then the phase
stiffness from a boundary twist.

\subsubsection{Analyticity of the free energy}
\label{sec:bose-free-energy}

The grand-canonical calculation uses independent momentum
occupations~\cite{PathriaBeale2011}.

Let $\hat N\coloneqq\sum_x b_x^\dagger b_x$ be the total-number
operator and choose a chemical potential $\mu<0$, below the lowest
one-particle energy $\epsilon(0)=0$. At finite $L$, momenta lie in
$(2\pi/L)\Z^d$ modulo $2\pi$ in each coordinate. Summing the occupations
$0,1,2,\ldots$ of each momentum mode gives
\begin{equation}
 \begin{aligned}
 \Xi_L(T,\mu)&\coloneqq\Tr e^{-(H_L-\mu\hat N)/T}\\
 &=\prod_k\bigl(1-e^{-(\epsilon(k)-\mu)/T}\bigr)^{-1}.
 \end{aligned}
 \label{eq:bose-grand-partition}
\end{equation}
The relation to the canonical ensemble is already exact at finite size:
\begin{equation}
 \Xi_L(T,\mu)=\sum_{n=0}^{\infty}e^{\mu n/T}Z_{L,n}(T).
 \label{eq:bose-ensemble-partitions}
\end{equation}
Here $Z_{L,n}$ is the partition function restricted to total particle
number $n$. Equivalently, conditioning the grand-canonical state on
the outcome $N$ of a total-number measurement gives
\begin{equation}
 \frac{P_Ne^{-(H_L-\mu\hat N)/T}}
 {\Tr\!\bigl(P_Ne^{-(H_L-\mu\hat N)/T}\bigr)}=\rho_L(T).
 \label{eq:bose-ensemble-conditioning}
\end{equation}
The factor $e^{\mu N/T}$ cancels in this ratio.
Taking the logarithm of the product in Eq.~\eqref{eq:bose-grand-partition}
turns it into a sum. The grand potential
per site is therefore
\begin{equation}
 \begin{aligned}
 g(T,\mu)&\coloneqq-T\lim_{L\to\infty}L^{-d}\ln\Xi_L(T,\mu)\\
 &=T\int_k\ln\bigl(1-e^{-(\epsilon(k)-\mu)/T}\bigr).
 \end{aligned}
 \label{eq:bose-grand-potential}
\end{equation}
Here $g(T,0)$ means the limit $\mu\uparrow0$ after $L\to\infty$;
at fixed $\mu<0$, a single mode contributes zero to this density.

To implement the prescribed density at finite $L$, choose the unique
$\mu_L(T)<0$ for which the mean occupations obey
\begin{equation}
 n_{L,k}\coloneqq\frac1{e^{[\epsilon(k)-\mu_L(T)]/T}-1},
 \qquad \frac1{L^d}\sum_kn_{L,k}=\varrho.
 \label{eq:helicity-gc-density}
\end{equation}
These definitions specify an auxiliary grand-canonical ensemble at any $T>0$.
The occupation sum increases continuously from zero to infinity as
the chemical potential increases from $-\infty$ to zero, so this choice
is unique.

For $0<T<T_c$, the density constraint forces $\mu_L(T)\to0$:
a negative limiting chemical potential would carry less than
$\varrho_{\rm ex}(T)<\varrho$ particles per site.
For $k\ne0$, the occupations are bounded by
$n_T(k)\coloneqq(e^{\epsilon(k)/T}-1)^{-1}\le C|k|^{-2}$,
where $C$ is finite and independent of $k,L$ at fixed $T,J,d$.
Their Riemann sum therefore converges to $\varrho_{\rm ex}(T)$ for
$d>2$, leaving
\begin{equation}
 \frac{n_{L,0}}{L^d}\longrightarrow\varrho_0(T).
 \label{eq:helicity-gc-condensation}
\end{equation}
Equivalently, $\mu_L(T)=-T/[\varrho_0(T)L^d]+o(L^{-d})$.

At the thermodynamic level, the corresponding fixed-density free energy
is obtained by the Legendre transform
$f(T)=\sup_{\mu\le0}\{g(T,\mu)+\mu\varrho\}$.
The density carried by modes at chemical potential $\mu$ is
$-\partial_{\mu}g(T,\mu)=\int_k(e^{(\epsilon(k)-\mu)/T}-1)^{-1}$.
It increases to $\varrho_{\rm ex}(T)$ as $\mu\uparrow0$.
For $T<T_c$, this is less than the prescribed density $\varrho$,
so $g(T,\mu)+\mu\varrho$ increases all the way to its maximum
at $\mu=0$. The excess density $\varrho_0(T)$ occupies the zero-energy
mode. Consequently the Helmholtz free energy per site is
\begin{equation}
 \boxed{\begin{aligned}
 f(T)&\coloneqq
 -T\lim_{L\to\infty}L^{-d}\ln Z_{L,N}(T)\\
 &=T\int_k\ln(1-e^{-\epsilon(k)/T}).
 \end{aligned}}
 \label{eq:bose-free}
\end{equation}
SM Sec.~\ref{sm:bose-canonical-occupations} also verifies
Eq.~\eqref{eq:bose-free} directly at fixed $N$ and supplies the
occupation estimates used below.

\begin{figure}[tbp]
\centering
\includegraphics[width=\columnwidth]{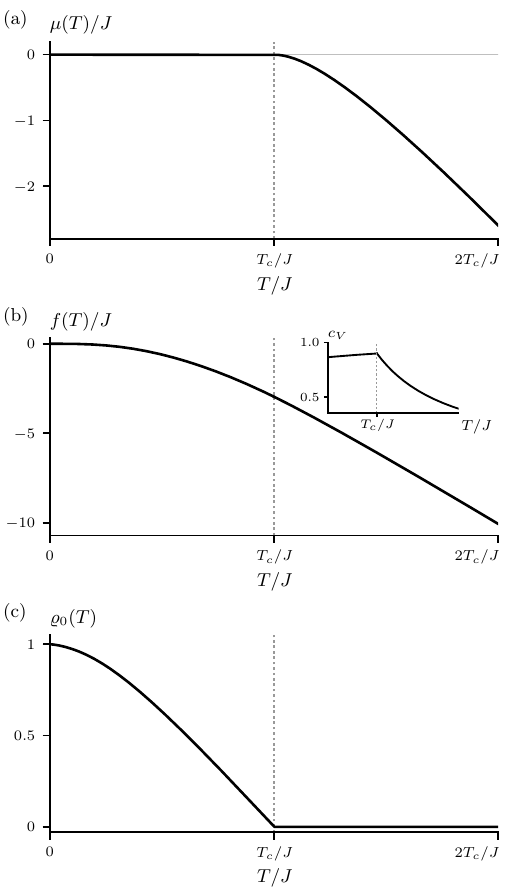}
\caption{Three-dimensional ideal Bose gas at density $\varrho=1$,
plotted against $T/J$.
(a) Chemical potential $\mu(T)/J$, zero for $T\le T_c$ and determined by
Eq.~\eqref{eq:bose-normal-density} above $T_c$.
(b) Free energy $f(T)/J$ [exact Eq.~\eqref{eq:bose-free-both}].
The inset shows the specific heat $c_V=-Tf''(T)$.
(c) Condensate density $\varrho_0(T)$ [Eq.~\eqref{eq:bose-density}].
Dotted lines mark $T_c/J\simeq5.59097$.}
\label{fig:bose-temperature}
\end{figure}

The free energy in Eq.~\eqref{eq:bose-free} is real analytic at every $0<T<T_c$. Indeed, split the logarithm
into $\ln\epsilon(k)-\ln T+\ln((1-e^{-z})/z)$, with
$z=\epsilon(k)/T$. The last logarithm has a removable singularity
at $z=0$. For complex $T$ near any fixed positive real temperature,
it has a common holomorphic branch over the bounded lattice band. The only
momentum singularity is the integrable, temperature-independent
$\ln\epsilon(k)$. This proves analyticity along every temperature
path strictly inside the condensed interval.
To continue the free energy above $T_c$ at the same density, the chemical
potential must change with temperature. Its thermodynamic value is
$\mu(T)\coloneqq\lim_{L\to\infty}\mu_L(T)$, which is zero for $T\le T_c$;
for $T>T_c$, the unique $\mu(T)<0$ is determined by
\begin{equation}
 \varrho=\int_k\frac{1}{e^{(\epsilon(k)-\mu(T))/T}-1}.
 \label{eq:bose-normal-density}
\end{equation}
Figure~\ref{fig:bose-temperature}(a) shows $\mu(T)$, which decreases
continuously from zero above $T_c$.
The fixed-density free energy on both sides of the transition is
\begin{equation}
 f(T)=\mu(T)\varrho +T\int_k\ln(1-e^{-(\epsilon(k)-\mu(T))/T}).
 \label{eq:bose-free-both}
\end{equation}
The term $\mu(T)\varrho$ converts the grand-canonical expression to the
canonical free energy for this fixed Hamiltonian and density.
For $\mu<0$ the density integral is analytic with a strictly positive,
finite derivative with respect to $\mu$. The implicit-function theorem
therefore also gives analyticity at every $T>T_c$.

The transition singularity depends on dimension: $f$ is $C^2$
at $T_c$ for $d=3,4$, and $C^1$ with a jump of its second derivative
for $d\ge5$ (SM Sec.~\ref{sm:bose-critical-thermo}).

Panels (b) and (c) of Fig.~\ref{fig:bose-temperature} show the free energy and condensate
density in three dimensions, respectively. The inset in panel (b) shows the continuous cusp in the fixed-density heat capacity
per site, $c_V(T)\coloneqq-T\,\dd^2f(T)/\dd T^2$
\cite{RakhimovAskerzade2015}.

\subsubsection{Phase stiffness}
\label{sec:stiffness}

We next identify the thermodynamic response that will enter the
information constant. The phase stiffness, also called the helicity
modulus, measures the free-energy cost of twisting the bosonic phase
across the sample~\cite{Fisher1973}.\footnote{This static equilibrium
response does not imply a positive Landau critical velocity for the ideal gas.}
Its relation to condensate density is known for the continuum ideal
gas~\cite{Barber1977}. We calculate the lattice response using the
grand partition function; SM Sec.~\ref{sm:bose-twist} proves that
it agrees with the canonical stiffness in the thermodynamic limit.

Insert a total angle $\vartheta$ around the first periodic direction.
In each bond difference along that direction, replace
$b_{x+\hat e_1}$ by $e^{i\vartheta/L}b_{x+\hat e_1}$ and make the
conjugate replacement in its adjoint. This uniformly distributes
the boundary twist and preserves nearest-neighbor locality.
With $b_k\coloneqq L^{-d/2}\sum_xe^{-ik\cdot x}b_x$, the twisted Hamiltonian is
\begin{equation}
 H_L(\vartheta)\coloneqq
 \sum_k\epsilon(k+\vartheta\hat e_1/L)b_k^\dagger b_k.
 \label{eq:twisted-H}
\end{equation}
Here $k$ remains the periodic Fourier label in $(2\pi/L)\Z^d$ modulo
$2\pi$, and $N$ remains fixed as $\vartheta$ changes. Define
the total free energy and the finite-size stiffness by
\begin{align}
 F_{L,N}(T,\vartheta)&\coloneqq
 -T\ln\Tr\!\bigl(P_Ne^{-H_L(\vartheta)/T}\bigr),
 \label{eq:twist-free}\\
 \Upsilon_L(T)&\coloneqq
 L^{2-d}\left.\frac{\partial^2F_{L,N}(T,\vartheta)}{\partial\vartheta^2}
 \right|_{\vartheta=0},
 \label{eq:helicity-finite-definition}\\
 \Upsilon(T)&\coloneqq\lim_{L\to\infty}\Upsilon_L(T).
 \label{eq:helicity-definition}
\end{align}
Thus the quadratic twist cost at finite $L$ is
$\frac12\Upsilon_L(T)L^{d-2}\vartheta^2$.
The derivative precedes the thermodynamic limit to retain this
subextensive cost.

The grand partition function has the same product form as
Eq.~\eqref{eq:bose-grand-partition}, with twisted one-particle energies:
\begin{equation}
 \Xi_L(T,\mu,\vartheta)\coloneqq
 \prod_k\bigl(1-e^{-[\epsilon(k+\vartheta\hat e_1/L)-\mu]/T}\bigr)^{-1}.
 \label{eq:helicity-gc-partition}
\end{equation}
Differentiate this partition function with $\mu_L(T)$ fixed by the
zero-twist density constraint in Eq.~\eqref{eq:helicity-gc-density},
then set $\vartheta=0$.
With $v_k\coloneqq\partial_{k_1}\epsilon(k)=2J\sin k_1$ and
$w_k\coloneqq\partial_{k_1}^2\epsilon(k)=2J\cos k_1$, the auxiliary
grand-canonical response is
\begin{equation}
 \begin{aligned}
 \Upsilon_L^{\rm gc}(T)&\coloneqq
 -TL^{2-d}\left.\partial_\vartheta^2
 \ln\Xi_L(T,\mu_L(T),\vartheta)\right|_{\vartheta=0}\\
 &=\frac1{L^d}\sum_k\biggl[w_kn_{L,k}
       -\frac{v_k^2}{T}n_{L,k}(1+n_{L,k})\biggr].
 \end{aligned}
 \label{eq:helicity-gc-finite}
\end{equation}
The second term is the equilibrium current fluctuation.

For $0<T<T_c$, the zero-mode limit was established in
Eq.~\eqref{eq:helicity-gc-condensation}. Since $v_0=0$ and $w_0=2J$,
separating that mode in Eq.~\eqref{eq:helicity-gc-finite} gives
\begin{equation}
 \begin{aligned}
 \lim_{L\to\infty}\Upsilon_L^{\rm gc}(T)
 &=2J\varrho_0(T)+\int_k w_kn_T(k)\\
 &\quad-\frac1T\int_k v_k^2n_T(k)(1+n_T(k)).
 \end{aligned}
 \label{eq:helicity-integrals}
\end{equation}
Both integrands are bounded near zero by $C|k|^{-2}$.
The thermal terms cancel by periodic integration by parts:
\begin{equation}
 \partial_{k_1}\bigl(v_kn_T(k)\bigr)
 =w_kn_T(k)-\frac{v_k^2}{T}n_T(k)(1+n_T(k)).
 \label{eq:helicity-cancellation}
\end{equation}
Removing a ball of radius $r$ around $k=0$ leaves a boundary term
$O(r^{d-2})$, which vanishes as $r\downarrow0$.
SM Sec.~\ref{sm:bose-twist} shows that
$\Upsilon_L(T)-\Upsilon_L^{\rm gc}(T)\to0$ by controlling canonical
conditioning of the excited occupations.
Thus the canonical stiffness defined above is exactly
\begin{equation}
 \Upsilon(T)=2J\varrho_0(T),\qquad 0<T<T_c.
 \label{eq:helicity-condensate}
\end{equation}
In particular, $\Upsilon(T)\to2J\varrho$ as $T\downarrow0$ and
$\Upsilon(T)\to0$ as $T\uparrow T_c$.
The following subsection connects this independently defined response to the optimized information constant.
\subsection{Channel invariant}
\label{sec:bose-information}

We now evaluate the information shared by distant regions in the
local limit of the canonical states in Eq.~\eqref{eq:bose-state}.
At each temperature below $T_c$ the condensate has a definite
magnitude, and its random phase is shared across the sample.
We compare a concrete measurement on the cold gas with the most
information any pair of warm measurements could supply. The resulting
gap rules out cooling and identifies the temperature-dependent constant.

\subsubsection{Readers and the locality condition}
\label{sec:local-readers}

\begin{figure}[!tbp]
\centering
\includegraphics[width=\columnwidth]{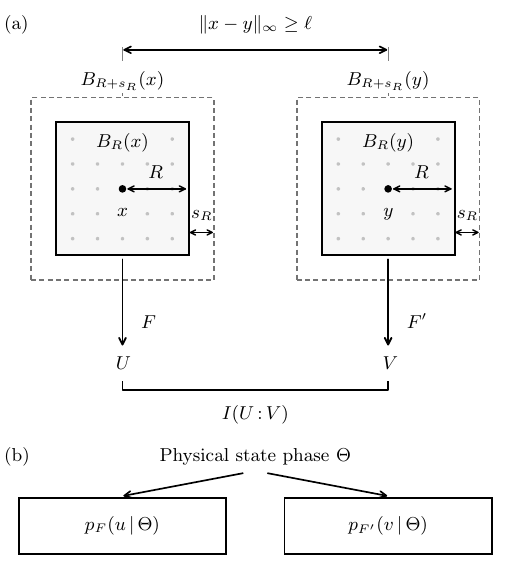}
\caption{Block measurements and shared phase.
(a) Planar sections of cubic blocks $B_R(x),B_R(y)$ of radius $R$;
their center separation is at least $\ell$ in maximum-coordinate distance.
Dashed boundaries mark input blocks enlarged by width $s_R$
[Eq.~\eqref{eq:channel-law}]. Separate readers $F,F'$ return $U,V$.
(b) In the condensed canonical Bose gas and the ordered classical
vortex-free Villain model, the two blocks share a uniformly distributed phase $\Theta$.
After the thermodynamic limit, sending their separation to infinity makes
the two readings independent conditional on $\Theta$, with output
probabilities $p_F$ and $p_{F'}$ [Eq.~\eqref{eq:shared-phase-reader-law}].}
\label{fig:reader-geometry}
\end{figure}

Let $\rho(T)$ denote the consistent family of local reduced
states obtained from $\rho_L(T)$ as $L\to\infty$; the
existence and form of this limit are proved in Sec.~\ref{sec:bose-local}.
For a positive integer radius $R$, define the cubic block
\begin{equation}
 B_R(x)\coloneqq x+([-R,R]^d\cap\Z^d),\qquad B_R\coloneqq B_R(0).
 \label{eq:reader-block}
\end{equation}
Figure~\ref{fig:reader-geometry}(a) shows the two measurement blocks.
A local reader in the Bose gas is any quantum measurement within one block with at most
$N_R\coloneqq\lceil R^\gamma\rceil$ outcomes, where $\gamma>(d-2)/2$ is fixed
once and for all. The ceiling denotes the smallest integer not less
than its argument. By the Born rule, on a block in state $\sigma_B$ the probability
of outcome $u$ is $\Tr(\sigma_B F_u)$. The operators $F_u\ge0$,
$u=1,\ldots,N_R$, sum to the block identity, $\sum_{u=1}^{N_R}F_u=I$,
so the probabilities sum to one. This collection is a
positive-operator-valued measure (POVM)~\cite{NielsenChuang2010}.
Unused labels have $F_u=0$.
Each reader may act jointly on all sites of its block, and may
use private randomness. No particle-number superselection rule~\cite{Bartlett2007},
occupation cutoff, or energy bound is imposed.

For two readers $F=\{F_u\}_{u=1}^{N_R}$ and
$F'=\{F'_v\}_{v=1}^{N_R}$, let $U,V$ be their outcome labels and
$p_{uv}$ their joint probabilities. Their mutual information, in
natural logarithms, is
\begin{equation}
 I(U\infosep V)\coloneqq\sum_{u=1}^{N_R}\sum_{v=1}^{N_R}p_{uv}\ln\frac{p_{uv}}{p_up_v},
 \label{eq:reader-mutual-information}
\end{equation}
where $p_u,p_v$ are the probabilities for each reader separately and
a zero-probability summand is zero. Independent readings give
$p_{uv}=p_up_v$ and hence zero mutual information.
For continuous variables with a joint density, the same definition
uses an integral instead of the sum.
The measurements $F,F'$ are supported on $B_R(x),B_R(y)$, respectively.
Write $\|x-y\|_\infty$ for the largest absolute coordinate separation.
At fixed $R$ the most mutual information two distant readers can obtain is
\begin{equation}
 M_T(R)\coloneqq
 \lim_{\ell\to\infty}\ \sup_{\|x-y\|_\infty\ge\ell}
 \ \sup_{F,F'} I(U\infosep V).
 \label{eq:accessible-info}
\end{equation}
The limit exists because increasing the minimum separation restricts the
supremum. We take $L\to\infty$ first, then $\ell\to\infty$ at fixed $R$,
and only then let $R$ grow. For a general input state $\sigma$ we write
$M_\sigma(R)$.

For a classical state, a reader is instead a stochastic map from the
variables in its block to a finite label. The same definitions apply
with these stochastic readers. For models with site and bond variables,
a cell anchored at a site contains that site's variable and the
positively oriented outgoing bond variables; blocks are sets of such
cells. This convention is used for the Villain model below.

The locality conditions below express the causal structure of standard
local channel circuits. Write $\sigma$ for a general input state.
A region subscript denotes
the reduced state on that region: in finite volume,
$\sigma_A\coloneqq\Tr_{A^c}\sigma$, where $A^c$ is the region outside
$A$ within that volume. For an infinite-lattice state, $\sigma_A$
denotes the corresponding local reduced state; classically, it is
the marginal distribution on $A$. A union of two blocks specifies
their joint state, including correlations between them.
For each large integer $R$, let $s_R$ be a nonnegative integer. Assume that the joint reduced output of a channel $\mathcal E$
on any two sufficiently distant blocks can be approximated by
independent completely positive trace-preserving (CPTP) maps on the
enlarged input blocks, indicated by the dashed boundaries in
Fig.~\ref{fig:reader-geometry}(a). The map $\mathcal E_{R,x}$ takes
the input on $B_{R+s_R}(x)$ to an output on $B_R(x)$, and similarly
for $y$. We require
\begin{equation}
 \begin{aligned}[b]
 &\tfrac12\bigl\| (\mathcal E(\sigma))_{B_R(x)\cup B_R(y)}\\
 &\quad - (\mathcal E_{R,x}\otimes\mathcal E_{R,y})
 (\sigma_{B_{R+s_R}(x)\cup B_{R+s_R}(y)})\bigr\|_1
 \le\delta_R .
 \end{aligned}
 \label{eq:channel-law}
\end{equation}
Here $\|X\|_1\coloneqq\Tr\sqrt{X^\dagger X}$ is the trace norm,
$\tfrac12\|\sigma-\sigma'\|_1$ is trace distance, and
\begin{equation}
 s_R=o(R),\qquad
 \delta_R=o(1/\ln R)\qquad(R\to\infty).
 \label{eq:channel}
\end{equation}
The added width is negligible compared with the block radius, and
the approximation error decays faster than $1/\ln R$.
For a classical channel, replace trace distance in
Eq.~\eqref{eq:channel-law} by total variation---the largest absolute
difference in the probability of any event---and the CPTP maps by
stochastic maps. All the conditions and the proof below are unchanged.
The error bound is uniform in the positions and their sufficiently
large separation. Each approximating map has its own independent
auxiliary systems (ancillas). In particular, a shared random phase or shared entangled
ancillas are not free resources. The input and output site spaces
may differ; only the spatial supports and the resulting finite
measurement labels are compared. No translation symmetry, phase
symmetry, or number conservation is assumed for the channel.

These conditions follow directly from circuit locality. For a circuit
whose gates in each layer have disjoint supports, let $r$ bound the
backward causal radius: the depth times the largest gate diameter.
For sufficiently distant output blocks, the gates that can influence
one block are disjoint from those influencing the other. Composing
the gates in each region and tracing out unused outputs gives the two
CPTP maps in Eq.~\eqref{eq:channel-law}, exactly, with $s_R=r$ and
$\delta_R=0$. Thus fixed-depth, fixed-range circuits automatically
satisfy Eq.~\eqref{eq:channel}. More generally, a joint truncation bound
$\delta_R\le cR^p e^{-s_R/\xi}$, with fixed constants $c,p,\xi>0$,
suffices: an enlargement $s_R=O(\ln R)$ with a sufficiently large
coefficient gives $\delta_R=O(R^{-2})$. This is the usual mechanism
for evolution over a fixed time generated by uniformly bounded,
finite-range Hamiltonian or Lindblad terms~\cite{BarthelKliesch2012}.
The truncations use
independent local auxiliary systems, as in the circuit case.

The circuits constructed in Ref.~\cite{MKS2026} use precisely
this product-of-local-gates structure (Sec.~II, Eq.~(4) there). Their
causal-radius bound $r_L$ is at most a fixed power of $\ln L$ when the target-state error is chosen
inverse-polynomially in $L$. At each finite $L$, Eq.~\eqref{eq:channel-law}
therefore holds exactly with $s_R=r_L$ and $\delta_R=0$. Choosing
growing blocks with $r_L=o(R)$ and $R=o(L)$ makes the enlargement
negligible. This joint finite-size limit is handled directly by the
growing-block bounds in Secs.~\ref{sec:optimal-cooling}
and~\ref{sec:spin-reader}; it does not require an infinite-volume limit
of the circuit family. In particular, the locality conditions do not
require clustering of the input or local reversibility of the channel.

How accurately does a simulation reproduce the mutual information of two readings?
If two laws $p,q$ on pairs of at most $N_R$ labels have total variation
distance $\delta\le1/2$, then
\begin{equation}
 |I_p(U\infosep V)-I_q(U\infosep V)|
 \le4\delta\ln N_R+3S(\delta),
 \label{eq:reader-continuity}
\end{equation}
where total variation is one half the sum of the absolute probability
differences, and
$S(\delta)\coloneqq-\delta\ln\delta-(1-\delta)\ln(1-\delta)$
is the Shannon entropy of the two-outcome distribution
$(\delta,1-\delta)$, with $0\ln0=0$.
To see this bound, write mutual information as the two marginal
Shannon entropies minus the joint entropy. Each marginal contributes
at most $\delta\ln N_R+S(\delta)$, and the joint alphabet contributes
at most $2\delta\ln N_R+S(\delta)$.

Applying the approximating local channel and then measuring its output defines a
single measurement on the enlarged input block.
The adjoint maps positive operators to positive operators and preserves
the identity, so this gives a valid input POVM with the same number of
labels. For large $R$, these labels are allowed by the larger input
budget $N_{R+s_R}$. Taking the separation limit and optimizing gives
\begin{equation}
 M_{\mathcal E(\sigma)}(R) \le M_\sigma(R+s_R) +4\delta_R\ln N_R+3S(\delta_R).
 \label{eq:reader-monotone-size}
\end{equation}
The error term vanishes because $\ln N_R=O(\ln R)$ and
$\delta_R=o(1/\ln R)$. Thus even one specified pair of measurements
on the output must fit within the information budget of the enlarged
input blocks, up to a vanishing error.

\subsubsection{The local canonical limit}
\label{sec:bose-local}

Below $T_c$, each observer sees the same condensate phase through
local thermal noise. We first identify the distribution of that signal
in a fixed block; we then compare the information in its preparation
amplitudes with what a measurement of the phase can recover.
The local canonical limit is a mixture of coherent product states
with positive probabilities.
A coherent state $|z_x\rangle$ of one site satisfies
$b_x|z_x\rangle=z_x|z_x\rangle$. Its complex amplitude $z_x$ gives
the mean field, and $|z_x|^2$ is its mean particle number.
Write $|z_B\rangle\coloneqq\bigotimes_{x\in B}|z_x\rangle$ and define the thermal covariance
\begin{equation}
 C_T(x-y)\coloneqq\int_k
 \frac{e^{ik\cdot(x-y)}}{e^{\epsilon(k)/T}-1}.
 \label{eq:bose-C}
\end{equation}
For a fixed block $B$ containing $|B|$ sites, set
\mbox{$(C_{T,B})_{xy}\coloneqq C_T(x-y)$} for $x,y\in B$.
Here $\dd^{2|B|}\xi$ means integration over the real and imaginary
parts of every component. The fixed-block limit holds in trace norm:
\begin{equation}
 \begin{aligned}[b]
 &\lim_{L\to\infty}(\rho_L(T))_B\\
 &=\int_0^{2\pi}\frac{\dd\Theta}{2\pi}
   \int_{\mathbb C^{|B|}}\!\dd^{2|B|}\xi\,
   \frac{e^{-\xi^\dagger C_{T,B}^{-1}\xi}}{\pi^{|B|}\det C_{T,B}}\\
 &\quad\times|\sqrt{\varrho_0}e^{i\Theta}\one_B+\xi\rangle
 \langle\sqrt{\varrho_0}e^{i\Theta}\one_B+\xi|.
 \end{aligned}
 \label{eq:bose-mixture}
\end{equation}
Here $\one_B$ is the vector with every entry equal to one on $B$.
The inner integral averages the Gaussian noise, while the outer integral
averages the common phase $\Theta$. We write $\mathcal{N}(m,s^2)$ for
the real Gaussian law with mean $m$ and variance $s^2$. A complex
noise $Z=X+iY$ has the circular Gaussian law $\mathcal{CN}(0,v)$
when $X$ and $Y$ are independent with law $\mathcal{N}(0,v/2)$.
Its density on the complex plane is $e^{-|z|^2/v}/(\pi v)$.

The fixed condensate magnitude follows from concentration of the
canonical condensate density at $\varrho_0(T)$~\cite{PuleZagrebnov2004}.
The finite-volume approximation in SM Lemma~\ref{sm:bose-block}
bounds the trace distance directly. Restricting it to the fixed
block gives Eq.~\eqref{eq:bose-mixture}
(SM Sec.~\ref{sm:bose-local-details}).

Fixing $\Theta=\alpha$ gives the field expectation
$\langle b_x\rangle_\alpha=\sqrt{\varrho_0}e^{i\alpha}$, where brackets denote
expectation and the subscript specifies the selected phase. The covariance
$C_T(x-y)$ tends to zero at large separation. Two fixed distant blocks
therefore become independent conditional on the same uniform phase
[Fig.~\ref{fig:reader-geometry}(b)].
In this separated limit, the joint reader probabilities are
\begin{equation}
 p_{uv}=\int_0^{2\pi}\frac{\dd\Theta}{2\pi}\,
 p_F(u\given \Theta)\,p_{F'}(v\given \Theta),
 \label{eq:shared-phase-reader-law}
\end{equation}
where $p_F(u\given \Theta)$ is the probability that reader $F$ returns
$u$ in the state with phase $\Theta$, and similarly for $F'$.
Convergence of the two-block states holds in trace norm: their Gaussian covariance
matrices are strictly positive, their cross terms tend to zero, and
preparing coherent states contracts total variation into trace distance.
Thus the only information surviving between the distant blocks is their information about the common condensate phase.

At fixed $R$, a trace-distance error $\delta$ changes the joint law of
any reader pair by at most $\delta$ in total variation. Since $N_R$ is
finite, Eq.~\eqref{eq:reader-continuity} makes information convergence
uniform over all readers. We may therefore optimize in the separated
state, even when the measurements depend on separation.

\subsubsection{Common phase and angular precision}
\label{sec:bose-precision-setup}

For the upper bound, imagine telling each observer the amplitudes
$z_0=\sqrt{\varrho_0}e^{i\Theta}\one+\xi$ used to prepare that
observer's block in Eq.~\eqref{eq:bose-mixture}, where $\one$ is the
all-ones vector on $B_R$. To obtain the lower
bound, we give each observer a measurement that estimates those amplitudes.

Heterodyne detection estimates the real and imaginary parts of a
single-site amplitude together~\cite{Leonhardt1997}. Its POVM is
$|z\rangle\langle z|\,\dd^2z/\pi$. For an input $|w\rangle$,
the outcome density is $|\langle z|w\rangle|^2/\pi=e^{-|z-w|^2}/\pi$.
Thus measuring every site returns $z_1=z_0+\zeta$, where the site
errors $\zeta_x$ have law $\mathcal{CN}(0,1)$ and are independent of
one another and of $z_0$. We use
$v=0$ for the preparation amplitudes and $v=1$ for the measured
amplitudes. In either case the noise at fixed phase has covariance
$K_v\coloneqq C_{T,B_R}+vI$, where $I$ is the identity on the block.

Every site amplitude $z_{v,x}$ has the same mean
$\sqrt{\varrho_0}e^{i\Theta}$ at fixed $\Theta$. To estimate this common
amplitude, take a weighted average with real weights that sum to one:
\begin{equation}
 Y_v\coloneqq\sum_{x\in B_R}w_x z_{v,x},
 \qquad \sum_{x\in B_R}w_x=1.
 \label{eq:bose-statistic}
\end{equation}
The normalization preserves the mean, so we choose the weights to
minimize the complex noise variance $w^{\mathsf T}K_vw$. At its minimum,
transferring a small weight from one site to another leaves this
variance unchanged to first order, so all entries of $K_vw$ are equal.
Normalizing the weights then gives $w=K_v^{-1}\one/Q_v$, where
$Q_v\coloneqq\sum_{x,y\in B_R}(K_v^{-1})_{xy}$ is the sum of all
entries of the inverse covariance. In particular, $w_x$ is the sum of
row $x$ divided by $Q_v$. Substitution gives a minimum variance of
$1/Q_v$, so at fixed $\Theta$ the noise $Y_v-\sqrt{\varrho_0}e^{i\Theta}$
has distribution $\mathcal{CN}(0,1/Q_v)$.

The direction of $Y_v$ estimates $\Theta$. To first order the angular
error is the noise perpendicular to the
mean amplitude divided by its length $\sqrt{\varrho_0}$.
Circular symmetry puts half the complex variance in this perpendicular
component, giving an angular variance of $1/(2\varrho_0 Q_v)$ to
linear order. Its inverse defines the angular precision
\begin{equation}
 P_{T,v}(R)\coloneqq 2\varrho_0(T)Q_v
 =2\varrho_0(T)\sum_{x,y\in B_R}(K_v^{-1})_{xy}.
 \label{eq:bose-precision}
\end{equation}
At fixed $\Theta$, the linearized error
$\operatorname{Im}(e^{-i\Theta}Y_v)/\sqrt{\varrho_0}$ has law
$\mathcal{N}(0,P_{T,v}(R)^{-1})$. For the measured phase $\arg Y_1$,
$P_{T,1}(R)^{-1}$ gives the leading variance of the angular error as $R\to\infty$
(Sec.~\ref{sec:bose-lower-bound}). We write $P_T(R)\coloneqq P_{T,1}(R)$.

The leading angular precision is governed by a geometric capacity.
Define the temperature-independent lattice Green function
\begin{equation}
 G(x)\coloneqq\int_k\frac{J e^{ik\cdot x}}{\epsilon(k)}.
 \label{eq:bose-correlation-G}
\end{equation}
For a finite block $B$, let $(G_B)_{xy}=G(x-y)$ for $x,y\in B$.
Its lattice capacity is
\begin{equation}
 \operatorname{cap}(B)\coloneqq\sum_{x,y\in B}(G_B^{-1})_{xy},
 \label{eq:capacity}
\end{equation}
the sum of all entries of the inverse Green matrix. Since charges produce
potential $G_B$ times the charge vector, this is the total charge
needed to hold $B$ at unit potential, with zero potential at infinity.
The continuum cube capacity is
\begin{equation}
 \kappa_d\coloneqq
 \inf_{\varphi=1\ {\rm near}\ [-1,1]^d}
 \int_{\R^d}|\nabla\varphi(x)|^2\,\dd^dx,
 \label{eq:bose-capacity}
\end{equation}
where $\varphi(x)$ is a smooth, real-valued trial potential that equals
one near the cube and vanishes outside a bounded region.
It fixes the large-block limit of the same lattice capacity:
\begin{equation}
 \lim_{R\to\infty}\frac{\operatorname{cap}(B_R)}{R^{d-2}}=\kappa_d.
 \label{eq:capacity-limit}
\end{equation}
A unit-potential profile on scale $R$ has gradients of order $R^{-1}$
over volume $R^d$, explaining the power $R^{d-2}$.
For the Bose observations,
\begin{equation}
 P_{T,v}(R)=\frac{\Upsilon(T)}T\operatorname{cap}(B_R)(1+o(1)),\qquad v=0,1.
 \label{eq:bose-precision-limit}
\end{equation}

This asymptotic relation uses $\Upsilon(T)=2J\varrho_0(T)$ and the small-momentum behavior
$n_T(k)=T/\epsilon(k)+O(1)$: the thermal covariance and $(T/J)G$
have the same long-wavelength singularity. Adding unit heterodyne
noise leaves that singularity unchanged. SM
Sec.~\ref{sm:capacity-details} proves both limits by matching
potential and charge bounds~\cite{LiZhuang2024}.

As the block grows, the mean signal at fixed phase retains its
magnitude $\sqrt{\varrho_0}$ while the root-mean-square noise
falls as $R^{-(d-2)/2}$.
In units of that noise the amplitude is $\sqrt{P_{T,v}/2}$.
Since $P_{T,0}/P_{T,1}\to1$, the two observations have the same
leading angular precision.

\subsubsection{Upper bound for arbitrary measurements}
\label{sec:bose-upper-bound}

The mutual information between the preparation amplitudes bounds what any
pair of quantum readers can extract. Indeed, for any POVM $\{F_u\}$, the
Born rule gives
\begin{equation}
 \Pr(U=u\given z)=\langle z|F_u|z\rangle.
 \label{eq:bose-born-reader}
\end{equation}
An observer given $z$ can therefore simulate this measurement by
drawing a label with the displayed probabilities. The two observers
do this independently.

Only the weighted average $Y_0$ carries shared information about the
phase. Subtracting $\one Y_0$ from $z_0$ cancels the condensate signal.
The minimum-variance weights in Eq.~\eqref{eq:bose-statistic} make the residual
uncorrelated with the Gaussian error in $Y_0$ and hence independent of
it. At infinite separation the two residuals are therefore private
noise, independent of both averages and of the phase. Processing each
average with private randomness cannot increase their mutual information
(the data-processing inequality~\cite{NielsenChuang2010}). Thus
\begin{equation}
 I(U\infosep V)\le I(Y_0\infosep Y'_0),
 \label{eq:bose-reader-upper}
\end{equation}
where $Y_0$ and $Y'_0$ are the statistics of the blocks read by
$F$ and $F'$, respectively. The bound holds for every reader pair and hence for the optimum
$M_T(R)$.

The scalar upper bound includes the observed radius as well as the
angle. Rescale $Y_0$ to $Z_s\coloneqq se^{i\Theta}+\xi$, where
$s^2=P_{T,0}(R)/2$ and $\xi\sim\mathcal{CN}(0,1)$.
Write $Z'_s$ for the second observation, with independent noise and the
same uniform phase. Their sum and difference divided by $\sqrt2$ are independent,
with the law of $Z_{\sqrt2s}$ and the law $\mathcal{CN}(0,1)$,
respectively. This change of
variables has unit Jacobian, so at every $s$ the mutual information is exactly
$2S(Z_s)-S(Z_{\sqrt2s})-\ln(\pi e)$, where
$S(X)=-\int p_X\ln p_X$ is differential entropy, with $p_X$ the density on the real coordinates.

For large $s$, the angle is uniform and the radius fluctuates
about $s$ with limiting variance $1/2$. The polar-coordinate
entropy, including its Jacobian, is
$S(Z_s)=\ln(2\pi s)+\tfrac12\ln(\pi e)+o(1)$
(SM Sec.~\ref{sm:bose-entropy-details}). Substituting gives
\begin{equation}
 M_T(R)\le\frac12\ln\frac{\pi P_{T,0}(R)}e+o(1).
 \label{eq:bose-scalar-information}
\end{equation}

\subsubsection{A phase measurement that rules out cooling}
\label{sec:bose-lower-bound}

For a measurement attaining this upper bound, bin the phase of
the measured weighted average $Y_1$ into $N_R$ equal intervals.
Write $\operatorname{Arg}z\in[-\pi,\pi)$ for the principal angle.
At fixed phase, the signal in units of the complex noise is
$s=\sqrt{P_T(R)/2}$. To leading order its angular error is the
transverse noise divided by $s$, so
\begin{equation}
 \Var\!\bigl(\operatorname{Arg}(e^{-i\Theta}Y_1)\bigr)
 =P_T(R)^{-1}(1+o(1)).
 \label{eq:bose-angle-variance}
\end{equation}
The expansion on $|\xi|<s/2$ controls the variance, since the
principal angle is bounded and the complementary Gaussian probability is
exponentially small. Circular symmetry and independence of the noises
make the two errors independent of each other and of $\Theta$.
For any such identically distributed errors of variance $\sigma^2$,
the two bin labels $U,V$ with width $\Delta_R\coloneqq 2\pi/N_R$ satisfy
\begin{equation}
 I(U\infosep V)\ge\ln(2\pi)
 -\frac12\ln\!\bigl(2\pi e(2\sigma^2+\Delta_R^2/3)\bigr).
 \label{eq:bin-information}
\end{equation}
Here $U$ is uniform and uncertainty about $V$ is bounded by the
entropy of the binned angle difference (SM Sec.~\ref{sm:bose-bin-details}).
Since $\gamma>(d-2)/2$, $\Delta_R^2=o(P_T(R)^{-1})$.
Substituting the variance from Eq.~\eqref{eq:bose-angle-variance} in
Eq.~\eqref{eq:bin-information} gives the lower bound
$\tfrac12\ln(\pi P_T(R)/e)+o(1)$.
The upper bound~\eqref{eq:bose-scalar-information} has the same
constant because $P_{T,0}/P_T\to1$. Hence
\begin{equation}
 M_T(R)
 =\frac12\ln\frac{\pi P_T(R)}e+o(1).
 \label{eq:bose-information}
\end{equation}
Fix two temperatures $0<T_-<T_+<T_c$ at the same density. Let
$I^{\rm het}_{T_-}(R)$ be the mutual information of the two cold
heterodyne angle labels just constructed, at infinite separation.
If a channel cooled the gas, Eq.~\eqref{eq:reader-monotone-size} would
require $I^{\rm het}_{T_-}(R)\le M_{T_+}(R+s_R)+o(1)$. Instead,
compare the cold lower bound with the upper bound for every warm
measurement in Eq.~\eqref{eq:bose-scalar-information}.
Since $P_T(R)=(\Upsilon(T)/T)\kappa_d R^{d-2}(1+o(1))$
and $\ln[(R+s_R)/R]\to0$, the common growth with block size cancels:
\begin{equation}
 \begin{aligned}
 &\liminf_{R\to\infty}\bigl[I^{\rm het}_{T_-}(R)-M_{T_+}(R+s_R)\bigr]\\
 &\qquad\ge \frac12\ln\frac{\Upsilon(T_-)\,T_+}{\Upsilon(T_+)\,T_-}>0.
 \end{aligned}
 \label{eq:direct-cooling-gap}
\end{equation}
The inequality is strict on the right because the condensate density,
and hence $\Upsilon(T)=2J\varrho_0(T)$, decreases as the gas warms.
A sufficiently large block therefore gives a contradiction. This already
rules out cooling and hence two-way conversion within the condensed phase.

\subsubsection{The information constant and its channel invariance}
\label{sec:reader-monotonicity}

The same comparison assigns a number to each state. Remove the common
logarithmic growth from the information available in a block:
\begin{equation}
 \cC(\sigma)\coloneqq
 \limsup_{R\to\infty}\bigl(M_\sigma(R)-\tfrac{d-2}{2}\ln R\bigr).
 \label{eq:invariant}
\end{equation}
For the thermal state write $\cC(T)\coloneqq\cC(\rho(T))$. In
Eq.~\eqref{eq:bose-information}, subtracting
$\tfrac{d-2}{2}\ln R$ and using the microscopic twist result
$\Upsilon(T)=2J\varrho_0(T)$ gives
\begin{equation}
 \boxed{\begin{aligned}
 \cC(T)
 &=\frac12\ln\frac{2\pi J\varrho_0(T)\kappa_d}{eT}\\
 &=\frac12\ln\frac{\Upsilon(T)}{T}+\frac12\ln\frac{\pi\kappa_d}{e}.
 \end{aligned}}
 \label{eq:invariant-stiffness}
\end{equation}
The logarithmically subtracted upper limit is therefore an ordinary limit.
For example, $\kappa_3\simeq16.60465$ in three dimensions~\cite{HelsingPerfekt2013}.

\begin{figure}[tbp]
\centering
\includegraphics[width=\columnwidth]{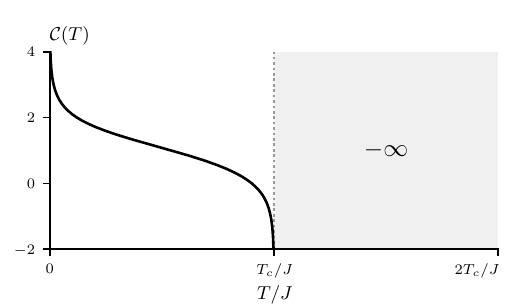}
\caption{Canonical three-dimensional ideal Bose gas at density $\varrho=1$.
The invariant follows from Eq.~\eqref{eq:invariant-stiffness} with
$\Upsilon(T)=2J\varrho_0(T)$ and cube capacity
$\kappa_3\simeq8\pi\times0.6606781541$
[Eq.~\eqref{eq:bose-capacity}]~\cite{HelsingPerfekt2013}.
Shading denotes $-\infty$ for $T>T_c$
[Eq.~\eqref{eq:bose-normal-invariant}]; the dotted line marks
$T_c/J\simeq5.59097$.}
\label{fig:bose-channel}
\end{figure}

For any channel satisfying Eqs.~\eqref{eq:channel-law} and~\eqref{eq:channel},
subtract the same logarithm on the two sides of
Eq.~\eqref{eq:reader-monotone-size}. The input radius $R+s_R$ changes
the subtraction by $o(1)$, and the limsup along these radii cannot exceed the limsup
over all radii. With the measurement error also vanishing, this gives
\begin{equation}
 \cC(\mathcal E(\sigma))\le\cC(\sigma).
 \label{eq:monotone}
\end{equation}
If one such channel takes $\rho(T_1)$ to $\rho(T_2)$, then
$\cC(T_2)\le\cC(T_1)$. A channel in the reverse direction gives the
opposite inequality. Thus two-way conversion requires
$\cC(T_1)=\cC(T_2)$ and, for the ideal gas, equal values of $\Upsilon/T$.
For this fixed Hamiltonian and density,
\begin{equation}
 \frac{\dd\cC}{\dd T}
 =-\frac12\left(\frac{\varrho_{\rm ex}'(T)}{\varrho_0(T)}+\frac1T\right)<0.
 \label{eq:bose-monotonic}
\end{equation}
The invariant thus distinguishes every pair along this analytic temperature
path. It ranges over all of $\R$: it tends to $+\infty$ as $T\downarrow0$ and to $-\infty$
as $T\uparrow T_c$.
Figure~\ref{fig:bose-channel} shows this strict decrease for the
same Hamiltonian and density as Fig.~\ref{fig:bose-temperature}.
The divergence is logarithmic: $\varrho_0(T)$ vanishes linearly as
$T\uparrow T_c$, giving
$\cC(T)=\tfrac12\ln(1-T/T_c)+O(1)$.

Above the transition, the local canonical limit is a centered
Gaussian state with no shared condensate variable. Its covariance is
obtained from Eq.~\eqref{eq:bose-C} by replacing $\epsilon(k)$ with
$\epsilon(k)-\mu(T)$, using Eq.~\eqref{eq:bose-normal-density}.
The cross covariance vanishes at infinite separation, so the Gaussian
preparation amplitudes have vanishing mutual information. The Born-rule bound gives
this conclusion uniformly over all readers, and hence
\begin{equation}
 M_T(R)=0,\qquad \cC(T)=-\infty,\qquad T>T_c.
 \label{eq:bose-normal-invariant}
\end{equation}
The first equality holds for every fixed block radius; subtracting
$\tfrac{d-2}{2}\ln R$ gives the second.
SM Sec.~\ref{sm:bose-normal-details} gives the local-limit argument
for $T>T_c$.

\subsubsection{Optimal finite-size cooling depth}
\label{sec:optimal-cooling}

Fix two condensed temperatures $T_-<T_+$ at the same density. On
the side-$L$ torus let $\varepsilon_L$ be the trace distance between
the circuit output on $\rho_L(T_+)$ and the target $\rho_L(T_-)$.
For disjoint gates in each layer, independent local ancillas and local
feedforward, the backward causal radius $r_L$ is at most the depth
$D_L$ times the maximum gate diameter. We obtain
\begin{equation}
 r_L=o(L)\quad\Longrightarrow\quad
 \liminf_{L\to\infty}\varepsilon_L>0.
 \label{eq:cooling-depth-lower}
\end{equation}
This bound uses a separate estimate on growing blocks of the finite
torus: even after enlarging the input blocks by $r_L$, readers that
reproduce the cold state's uniform angle-label marginals fall short of
its probability of reporting the same label
(SM Sec.~\ref{sec:bose-circuits}). Conversely, a nearest-neighbor
circuit prepares the cold state exactly, with unbounded on-site
occupation and general occupation-dependent gates. Thus for fixed
gate diameter and any prescribed error tending to zero,
\begin{equation}
 \min D_L=\Theta(L).
 \label{eq:cooling-depth-optimal}
\end{equation}
The construction and its resource requirements are given in
SM Sec.~\ref{sm:bose-linear-preparation}.\footnote{For locality-based
lower bounds on preparing topological order, see Ref.~\cite{KoenigPastawski2014}.}

\subsubsection{Field correlations and the selected phase}
\label{sec:bose-correlations}

The symmetric state has $\langle b_x\rangle=0$, while selecting a
phase $\alpha$ gives $\langle b_x\rangle_\alpha=\sqrt{\varrho_0}e^{i\alpha}$.
Both states have $\langle b_0^\dagger b_x\rangle=\varrho_0+C_T(x)$:
the long-distance plateau survives in the connected correlation only
in the symmetric state. Table~\ref{tab:model-summary} lists the
longitudinal and transverse tails in a selected state.
Even the bounded, $U(1)$-invariant empty-site projector
$|0\rangle_x\langle0|$ has a connected correlation decaying as
$|x|^{-(d-2)}$. This violates the exponential clustering bound of
Ref.~\cite{MKS2026}, including its finite-size version
(SM Secs.~\ref{sm:bose-correlations} and~\ref{sm:bose-bounded-clustering}).
\section{Vortex-free Villain model}
\label{sec:villain}

The second example is a classical magnet with local constraints whose
free energy is analytic at every positive temperature, yet no two
distinct temperatures admit two-way conversion by quasi-local channels. Exact Gaussian
integration gives both the free energy and the channel invariant.
Order persists throughout $0<T<\infty$, with stiffness $J$ and the
same geometric information constant as for the Bose gas. We then
extend the result to quantum and coupled multicomponent models.

\subsection{Model}
\label{sec:model}

\begin{figure}[!t]
\centering
\includegraphics[width=150pt]{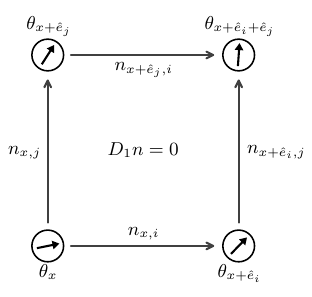}
\caption{A plaquette of the vortex-free Villain model.
Vertex arrows show spin angles measured counterclockwise from the right;
edge arrows orient the bond integers. The constraint sets their signed
plaquette sum $(D_1n)_{x,ij}$ to zero [Eq.~\eqref{eq:edges}].}
\label{fig:phase}
\end{figure}

We again consider a $d$-dimensional hypercubic lattice, $\Z^d$,
with $d\ge3$. Figure~\ref{fig:phase} shows one
plaquette (elementary square). A planar spin sits at each vertex, and
each oriented nearest-neighbor bond carries an integer. The signed sum
of the four integers around a plaquette defines its vortex charge,
which we set to zero.

We write the spin at site $x$ as the unit complex number
$s_x\coloneqq e^{i\theta_x}$, with angle $\theta_x\in[0,2\pi)$. Give each bond
a reference orientation from $x$ to its neighbor $x+\hat e_i$, one
lattice spacing along the positive $i$th coordinate axis. Its integer
$n_{x,i}\in\Z$ specifies how many full turns to add to the difference
of the two site angles. The real angle increment and the lattice curl
of the integers are
\begin{align}
 a_{x,i}&\coloneqq \theta_{x+\hat e_i}-\theta_x+2\pi n_{x,i}, \label{eq:bond-increment}\\
 (D_1n)_{x,ij}&\coloneqq n_{x,i}+n_{x+\hat e_i,j}
                  -n_{x+\hat e_j,i}-n_{x,j}. \label{eq:edges}
\end{align}
For the bond $e=(x,x+\hat e_i)$ we also write $n_e$ and $a_e$ for
$n_{x,i}$ and $a_{x,i}$. Traversing $e$ in reverse gives increment
$-a_e$. For the plaquette $p$ based at $x$ in directions $i<j$,
we write $(D_1n)_p=(D_1n)_{x,ij}$.
In a finite box of side $L$ the sites are
$x\in\{0,\ldots,L-1\}^d$. Write $E_L$ for the set of bonds, each counted
once, and $P_L$ for the set of plaquettes. With free
boundaries we include only bonds and plaquettes in the box. With
periodic boundaries we take the coordinates modulo $L$. Allowed
configurations satisfy the flatness condition
\begin{equation}
 (D_1n)_p=0\qquad(p\in P_L).
 \label{eq:flat-space}
\end{equation}
With interaction strength $J>0$, the Hamiltonian is
\begin{equation}
 H_L\coloneqq \frac J2\sum_{e\in E_L}a_e^2.
 \label{eq:H}
\end{equation}
The flatness restriction is implemented by
\begin{equation}
 P_{\mathrm{flat},L}(n)\coloneqq
 \prod_{p\in P_L}\delta_{(D_1n)_p,0}.
 \label{eq:flat-projector}
\end{equation}
Since $P_{\mathrm{flat},L}^2=P_{\mathrm{flat},L}$, this indicator projects
onto flat configurations. In contrast to the
global particle-number projection $P_N$ of Sec.~\ref{sec:bose-model},
it is a product of commuting local constraints on the four bond integers
of each plaquette. No particle number or global winding sector is fixed.

With Boltzmann's constant and the lattice spacing set to one,
the partition function at temperature $T$ is
\begin{equation}
 Z_L(T)\coloneqq \int\prod_x\frac{\dd\theta_x}{2\pi}
 \sum_{\{n_e\}} P_{\mathrm{flat},L}(n)e^{-H_L/T}.
 \label{eq:constrained-Z}
\end{equation}
The sum includes all integer bond assignments; $P_{\mathrm{flat},L}$
selects the allowed configurations from the product of uniform angle
and integer counting measures.
On the torus, the winding vector $w\in\Z^d$ has components
$w_i\coloneqq\sum_{t=0}^{L-1}n_{x+t\hat e_i,i}$ along the straight
coordinate cycles. Flatness makes this sum independent of the chosen
cycle: the difference between parallel cycles is the sum of plaquette
curls on the intervening strip. The site angles telescope, so the
lifted circulation is $2\pi w_i$. All winding sectors are included.

The bond integer implements the \emph{flat gauging} of the $2\pi\Z$
shift symmetry of a real field $\varphi_x\in\R$~\cite{Sulejmanpasic2019,Gorantla2021,Cherman2024}.
A lift of a spin $s_x$ is a real angle $\varphi_x$ with
$e^{i\varphi_x}=s_x$. Any two such angles differ by a multiple of
$2\pi$. Changing representatives by $2\pi k_x$, with $k_x\in\Z$,
requires a compensating change on each bond,
\begin{equation}
 \varphi_x\mapsto\varphi_x+2\pi k_x,\qquad
 n_{x,i}\mapsto n_{x,i}+k_x-k_{x+\hat e_i}.
 \label{eq:integer-gauging}
\end{equation}
Both $s_x=e^{i\varphi_x}$ and the lifted increment $a_{x,i}$ are
invariant. Choosing the principal representative
$\varphi_x=\theta_x\in[0,2\pi)$ makes the bond integer recoverable from
the endpoint spins and $a_{x,i}$.

The internal $O(2)$ symmetry consists of a common rotation,
$s_x\mapsto e^{i\alpha}s_x$ with each $a_e$ fixed, and reflection,
$s_x\mapsto\bar s_x$, $a_e\mapsto-a_e$.

Flatness constrains the circulation of the lifted increments,
$\sum_{e\in\partial p}a_e=2\pi(D_1n)_p$, rather than the winding
inferred by taking the shortest turn between spins. Spin configurations
with nonzero XY winding can therefore remain allowed; an explicit
example is given in SM Sec.~\ref{sm:classical-vortices}.

\subsection{Thermodynamics}
\label{sec:free}

Lifting the angles to a real Gaussian field separates the local
fluctuations from the common rotation. This gives the bulk free
energy and spontaneous magnetization. A twist of the boundary
conditions determines the spin stiffness.

\begin{figure}[tbp]
\centering
\includegraphics[width=\columnwidth]{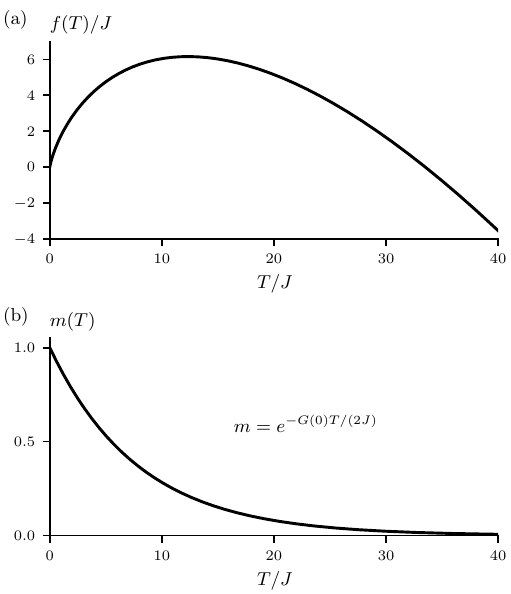}
\caption{Classical vortex-free Villain model in $d=3$, versus $T/J$.
(a) Free energy per site $f(T)/J$ [exact Eq.~\eqref{eq:free-energy}],
using angular measure $\dd\theta/(2\pi)$.
(b) Spontaneous magnetization $m(T)$ [exact Eq.~\eqref{eq:m}].
The constants $\ell_3$ and $G(0)$ are defined by the integrals in
Eq.~\eqref{eq:green}~\cite{Watson1939}.}
\label{fig:free-energy}
\end{figure}

\subsubsection{Exact Gaussian integration}
We first evaluate the zero-winding sector; the sum over all windings
will then restore the full periodic partition function.
In a free box, or a zero-winding torus, flatness gives
$n_{x,i}=m_{x+\hat e_i}-m_x$ for site integers $m_x$.
The lifted heights $h_x\coloneqq\theta_x+2\pi m_x$ are real-valued
representatives of the compact angles, chosen consistently so that
$e^{ih_x}=s_x$ and $h_{x+\hat e_i}-h_x=a_{x,i}$.
Summing over the integers unfolds the angle integrals into real-height
integrals, with a common shift by $2\pi$ identified.
The energy is diagonal in the orthonormal lattice-Laplacian modes, with
\begin{equation}
 \lambda(k)\coloneqq 2\sum_{i=1}^{d}(1-\cos k_i).
 \label{eq:lattice-dispersion}
\end{equation}
Using the Brillouin-zone integral of Eq.~\eqref{eq:bose-integral}, define
\begin{equation}
 G(x)=\int_k\frac{e^{ik\cdot x}}{\lambda(k)},\qquad
 \ell_d\coloneqq \int_k\ln\lambda(k).
 \label{eq:green}
\end{equation}
This $G$ agrees with Eq.~\eqref{eq:bose-correlation-G} since
$\epsilon(k)=J\lambda(k)$.
The orthogonal change of variables preserves $\prod_x\dd h_x/(2\pi)$.
The constant mode has period $2\pi L^{d/2}$ and contributes $L^{d/2}$;
each other real mode contributes
\begin{equation}
 \int_{\mathbb R}\frac{\dd u}{2\pi}
 e^{-J\lambda(k)u^2/(2T)}
 =\left(\frac{T}{2\pi J\lambda(k)}\right)^{1/2}.
 \label{eq:villain-mode-integral}
\end{equation}
Their product is
$L^{d/2}\prod_{k\ne0}[T/(2\pi J\lambda(k))]^{1/2}$.
On the torus, winding $w$ adds a uniform slope $2\pi w/L$ orthogonal
to periodic gradients, multiplying this product by
$\sum_{w\in\Z^d} e^{-2\pi^2JL^{d-2}|w|^2/T}=1+O(e^{-2\pi^2JL^{d-2}/T})$.
The free-box and periodic momenta are respectively $k_i=\pi r_i/L$
and $2\pi r_i/L$, with $r_i=0,\ldots,L-1$.
Both give $L^{-d}\sum_{k\ne0}\ln\lambda(k)\to\ell_d$:
$\lambda$ is even and its logarithmic singularity
$\ln\lambda(k)\sim2\ln|k|$ is integrable.
The constant-mode and winding terms vanish in $L^{-d}\ln Z_L$, so
\begin{equation}
 \boxed{f(T)\coloneqq -T\lim_{L\to\infty}\frac{\ln Z_L(T)}{L^d}
     =\tfrac T2\bigl(\ln\tfrac{2\pi J}{T}+\ell_d\bigr).}
 \label{eq:free-energy}
\end{equation}
The zero-field free energy is real analytic throughout $0<T<\infty$
[Fig.~\ref{fig:free-energy}(a)].

\subsubsection{Ordered Gibbs states}
\label{sec:ordered}
The finite-torus Gibbs density, relative to uniform angle and integer
counting measures, is
\begin{equation}
 \rho_L(T;\theta,n)\coloneqq
 \frac{P_{\mathrm{flat},L}(n)e^{-H_L/T}}{Z_L(T)}.
 \label{eq:states}
\end{equation}
Write $\langle\cdot\rangle_L$ for its expectation.
As $L\to\infty$, nonzero windings disappear, the constant mode gives a
uniform angle $\Theta$, and the other modes give an independent centered
real Gaussian field $\eta$ with covariance $(T/J)G$.
For $d\ge3$, the mode covariance converges because $1/\lambda(k)$
is integrable. Thus the limiting joint law is
\begin{equation}
 \theta_x=(\Theta+\eta_x)\bmod 2\pi,\qquad
 a_{x,i}=\eta_{x+\hat e_i}-\eta_x.
 \label{eq:selected-state}
\end{equation}
On a finite set $B$, the Gaussian field has covariance $(T/J)G_B$,
with $(G_B)_{xy}=G(x-y)$.
Since $\eta_0-\eta_x$ has variance $2(T/J)[G(0)-G(x)]$, its Gaussian
characteristic function gives
\begin{equation}
 \lim_{L\to\infty}\langle s_0\bar s_x\rangle_L
 =e^{-(T/J)\bigl(G(0)-G(x)\bigr)}.
 \label{eq:spin-correlation}
\end{equation}
As $G(x)\to0$, the spontaneous magnetization is
\begin{equation}
 m(T)\coloneqq\sqrt{\lim_{|x|\to\infty}\lim_{L\to\infty}\langle s_0\bar s_x\rangle_L}
 =e^{-TG(0)/(2J)}.
 \label{eq:m}
\end{equation}
Fixing $\Theta$ gives a Gibbs state: conditioning the lifted Gaussian
field on exterior heights gives the quadratic Boltzmann weight
\cite{Garet1999}, and folding back to spins and integers gives the
original conditional law, independent of a common $2\pi$ shift.
Because $G(x)\to0$, distant bounded local observables decorrelate in
this state, and its block-averaged spin converges to $e^{i\Theta}m(T)$. Averaging over $\Theta$ restores the
symmetry while distant blocks retain the same direction.
Since $m(T)>0$ at every finite temperature
[Fig.~\ref{fig:free-energy}(b)], $O(2)$ is broken to reflection about
the selected direction throughout $0<T<\infty$: there is no
finite-temperature ordering transition.

\subsubsection{Spin stiffness from a twist}
Impose a total twist $\vartheta$ in the first direction by replacing
$a_{x,1}$ with $a_{x,1}+\vartheta/L$. Its uniform slope is orthogonal
to periodic gradients, so
$Z_L(T,\vartheta)\propto\sum_{w_1\in\Z}
 e^{-JL^{d-2}(2\pi w_1+\vartheta)^2/(2T)}$.
Near $\vartheta=0$, the zero-winding term gives the free-energy
curvature $JL^{d-2}$, while the other terms give exponentially small
relative corrections at fixed $T>0$, $d>2$.
Thus, with $F_L=-T\ln Z_L$, the spin stiffness
(helicity modulus)~\cite{Fisher1973} is
\begin{equation}
 \Upsilon(T)\coloneqq \lim_{L\to\infty}L^{2-d}
 \left.\frac{\partial^2 F_L(T,\vartheta)}{\partial\vartheta^2}
 \right|_{\vartheta=0}=J.
 \label{eq:stiffness}
\end{equation}
The ratio $\Upsilon(T)/T=J/T$ sets the precision with which distant
blocks can estimate their common direction.

\subsubsection{Correlations in the ordered phase}
\label{sec:classical-correlations}
\label{sec:bond-clustering}
In the symmetric state, the connected spin correlation retains the
plateau $m(T)^2$ because the mean spin vanishes. In a state with a
selected direction, subtracting the mean-spin product removes the
plateau. Table~\ref{tab:model-summary} gives the longitudinal and
transverse tails, derived in SM Sec.~\ref{sm:classical-correlations}.
Even the bounded, rotation-invariant bond alignment $\cos a_e$ has
algebraically decaying connected correlations. On finite tori these
violate the exponential bound of Ref.~\cite{MKS2026}, including its
polynomial size prefactor (SM Sec.~\ref{sm:classical-bond-clustering}).

\subsection{Channel invariant}
\label{sec:info}

In the symmetric Gibbs state, spins in distant blocks fluctuate about
the same unknown direction $\Theta$. The accessible information of
Sec.~\ref{sec:local-readers} is the most one block's reading can
reveal about the other's. We bound this quantity for all readers of
spins and bonds. Reporting the angular bin of a weighted spin average
attains the bound through its constant term.

\begin{figure}[tbp]
\centering
\includegraphics[width=\columnwidth]{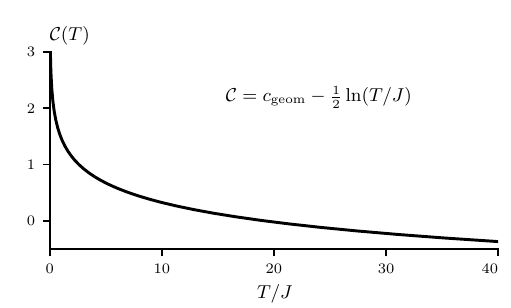}
\caption{Channel invariant $\cC$ (nats) versus $T/J$ for cubic
blocks in the three-dimensional vortex-free Villain model.
The classical and quantum curves coincide
[Eqs.~\eqref{eq:value} and~\eqref{eq:quantum-information}].
Here $c_{\rm geom}=\tfrac12\ln(\pi\kappa_3/e)\simeq1.47721$,
with cube capacity $\kappa_3$ defined in
Eq.~\eqref{eq:bose-capacity}~\cite{HelsingPerfekt2013}.}
\label{fig:blocks}
\end{figure}

\subsubsection{The local data and spin-only readers}
\label{sec:villain-readers}
A cell at $x$ consists of the spin angle $\theta_x$ and the $d$ outgoing
bond integers $n_{x,i}$. A reader on $B_R$ has access to these anchored
cells; their contents lie within the sites and internal bonds of
$B_{R+1}$. It returns at most $N_R=\lceil R^\gamma\rceil$ labels,
with the same fixed $\gamma>(d-2)/2$ as in
Sec.~\ref{sec:local-readers}. Its output probabilities may be
arbitrary measurable functions of the block configuration, with
independent private randomness for the two readers.
The quantities $M_T(R)$ and $\cC(T)$ are
therefore exactly those in Eqs.~\eqref{eq:accessible-info}
and~\eqref{eq:invariant}, interpreted for classical stochastic readers.
The order of limits is unchanged: first $L\to\infty$ on fixed blocks,
then infinite separation, and finally $R\to\infty$.

Write $M_T^{\mathrm{spin}}(R)$ for the same supremum when each reader
has access only to the spins in its block.

A classical channel is a stochastic map on the full product of cell
spaces; it need not preserve flatness at intermediate steps. We impose
the two-block approximation of Eqs.~\eqref{eq:channel-law}
and~\eqref{eq:channel} in total variation. Input and output cell spaces
may differ: a map can read full warm cells when its target is only the
cold spin law. Composing readers with the local stochastic maps gives
the finite-block bound in Eq.~\eqref{eq:reader-monotone-size} for these
classical readers as well. The channels may break translation or internal symmetry.

\subsubsection{Gaussian phase statistic and upper bound}
\label{sec:full-information}
The internal integer bonds unwrap the spins in a block $B$ into real
heights $h_B=\Theta\one+\eta_B$, up to a common multiple of $2\pi$.
Use the same lattice capacity $\operatorname{cap}(B)$ defined in
Eq.~\eqref{eq:capacity} for the Bose calculation.
The normalized equilibrium charges have components
\begin{equation}
 q_x^B\coloneqq\frac{\sum_{y\in B}(G_B^{-1})_{xy}}
 {\operatorname{cap}(B)},\qquad x\in B.
 \label{eq:equilibrium-charge-components}
\end{equation}
They sum to one and give the weighted angle
\begin{align}
 U_B&\coloneqq\sum_{x\in B}q_x^B h_x
       =\Theta+\xi_B\pmod{2\pi}, \label{eq:statistic}\\
 \Var\xi_B&=\frac{T}{J\operatorname{cap}(B)}. \label{eq:statistic-variance}
\end{align}
Here $\xi_B=\sum_{x\in B}q_x^B\eta_x$ is the Gaussian angular error.
For $B=B_R$, define the linear angular precision in the same convention
as for the Bose reader:
\begin{equation}
 P_T(R)\coloneqq\frac{J}{T}\operatorname{cap}(B_R)
 =\frac{\Upsilon(T)}T\operatorname{cap}(B_R).
 \label{eq:villain-phase-precision}
\end{equation}
At every finite $R$, the unwrapped error has the exact law
$\xi_B\sim\mathcal{N}(0,P_T(R)^{-1})$.\footnote{After
reduction modulo $2\pi$, the principal error has variance
$P_T(R)^{-1}(1+o(1))$ as $R\to\infty$.}
The angle $U_B$ is a sufficient scalar observation, not just an optimal linear
average. Indeed, $G_Bq^B=\one/\operatorname{cap}(B)$ implies that the
Gaussian residual $h_B-\one(q^B)^Th_B=\eta_B-\one\xi_B$ has zero
covariance with $\xi_B$ and is independent of $(\Theta,U_B)$.
Together with $U_B$ it reconstructs the internal data, but supplies no
additional information about $\Theta$.

Since $G(x)\to0$, infinitely separated blocks are independent
conditional on $\Theta$.\footnote{For fixed blocks, convergence of the
positive definite Gaussian covariances gives convergence in total
variation. The finite-label continuity bound of
Sec.~\ref{sec:local-readers} makes the thermodynamic and separation
limits uniform over readers.}
Their reader probabilities therefore have the shared-phase form
of Eq.~\eqref{eq:shared-phase-reader-law}.
For two translates $B,B'$ of $B_R$, the error difference
$Y\coloneqq\xi_{B'}-\xi_B$ has variance
$2/P_T(R)-2\Cov(\xi_B,\xi_{B'})$. In the infinite-separation
limit the covariance vanishes and the variance is $2/P_T(R)$.
The angle $U_B$ is uniform and independent of $Y$, giving
$I(U_B\infosep U_{B'})=\ln(2\pi)-S(Y\bmod2\pi)$.
Here $S$ denotes differential entropy with respect to Lebesgue measure.
As $R\to\infty$, wrapping changes the Gaussian entropy
$\tfrac12\ln(4\pi e/P_T(R))$ by $o(1)$
(SM Sec.~\ref{sm:wrapped-entropy}); hence
\begin{equation}
 I(U_B\infosep U_{B'})=\frac12\ln\frac{\pi P_T(R)}e+o(1).
 \label{eq:wrapped}
\end{equation}
The two residuals are independent of each other and the angle pair,
so data processing bounds every pair of internal-data readers by this
mutual information. For the anchored-cell readers defining $M_T(R)$,
apply this upper bound to the internal data of $B_{R+1}$, which contain
all cells anchored in $B_R$.

\subsubsection{Matching lower bound and the invariant}
\label{sec:villain-lower-bound}
A finite-bin reader attains the upper bound through its constant term.
Bin $U_B$ into $N_R$ equal intervals. Since
$\Delta_R=2\pi/N_R=o(P_T(R)^{-1/2})$, the finite-bin bound
\eqref{eq:bin-information} attains Eq.~\eqref{eq:wrapped} through its
constant term. The internal data of $B_R$ are available from its
anchored cells, and the capacity ratio of $B_{R+1}$ and $B_R$ tends to
one. The upper and lower bounds therefore give
$M_T(R)=\tfrac12\ln(\pi P_T(R)/e)+o(1)$,
the same asymptotic formula as Eq.~\eqref{eq:bose-information} for the
Bose gas. The capacity limit in Eq.~\eqref{eq:capacity-limit} gives
$P_T(R)=(J/T)\kappa_dR^{d-2}(1+o(1))$.

Fix $0<T_-<T_+$ and use the equal-bin readers of $U_B$ on the two cold
blocks. At infinite separation their mutual information exceeds the warm
optimum $M_{T_+}(R+s_R)$ by $\tfrac12\ln(T_+/T_-)+o(1)$, since
$s_R=o(R)$ changes the logarithm of the block capacity by $o(1)$.
If a cooling channel existed, composing its two local approximating
maps with the fixed cold readers would give readers on the enlarged
warm blocks. By Eq.~\eqref{eq:reader-continuity} their mutual
information would agree with the cold value up to $o(1)$.
The positive gap rules this out, as in Eq.~\eqref{eq:direct-cooling-gap}.

Substituting the asymptotic precision into $M_T(R)$ gives
$M_T(R)=\tfrac{d-2}{2}\ln R+\tfrac12\ln(J/T)+c_{\rm geom}+o(1)$,
where $c_{\rm geom}\coloneqq\tfrac12\ln(\pi\kappa_d/e)$, and the upper limit
in Eq.~\eqref{eq:invariant} is a limit:
\begin{equation}
 \boxed{\cC(T)=\frac12\ln\frac{\Upsilon(T)}T
 +\frac12\ln\frac{\pi\kappa_d}{e}}.
 \label{eq:value}
\end{equation}
Here $\Upsilon(T)=J$. The geometric term agrees with the Bose gas
because both estimate one common $2\pi$-periodic phase with the same
asymptotic block capacity [Fig.~\ref{fig:blocks}].

\subsubsection{The temperature distinction from the spins alone}
\label{sec:spin-reader}
Each observer reports the bin containing the direction of a weighted
spin average,
\begin{equation}
 U_R^{\mathrm{spin}}\coloneqq\operatorname{Arg}\sum_{x\in B_R}w_xs_x.
 \label{eq:spin-reader}
\end{equation}
The nonnegative weights sum to one. Spreading the optimal linear
weights over nearby sites preserves their leading variance and
suppresses nonlinear fluctuations. With the center spin as a fallback
if the sum vanishes, the principal angular error $\epsilon_R$ relative
to $\Theta$ has mean square $P_T(R)^{-1}(1+o(1))$
(SM Sec.~\ref{sm:classical-spin-error}).

At infinite separation the two errors depend on independent fields
$\eta$ and are independent of the common angle. Their variances are
bounded by the mean square above, and the squared bin width satisfies
$\Delta_R^2=o(P_T(R)^{-1})$. The finite-bin entropy bound
\eqref{eq:bin-information} therefore gives at least
$\tfrac12\ln(\pi P_T(R)/e)+o(1)$, matching the full-data upper bound.
Thus
\begin{equation}
 \lim_{R\to\infty}
 \bigl(M_T^{\mathrm{spin}}(R)-\tfrac{d-2}{2}\ln R\bigr)=\cC(T).
 \label{eq:spin-value}
\end{equation}

For $0<T_1<T_2$ the spin readers witness the same gap along the analytic path,
\begin{equation}
 \cC(T_1)-\cC(T_2)=\frac12\ln\frac{T_2}{T_1}>0.
 \label{eq:gap}
\end{equation}
Even a channel with access to all bond integers at $T_2$ cannot reproduce
the spin marginal at $T_1<T_2$. This excludes two-way conversion.

On finite tori a separate growing-block estimate excludes cooling
with vanishing error at causal radius $o(L)$ even for a spin-only
target. Nearest-neighbor stochastic gates with a fixed number of
unbounded registers per cell attain exact cooling in depth $O(L)$,
so the optimal fixed-range depth is $\Theta(L)$
(SM Sec.~\ref{sm:villain-cooling}).
For the reverse direction, an exact local bulk heating map exists for
$T\mapsto k^2T$ with integer $k\ge2$ (SM Sec.~\ref{sm:classical-heating}).

\subsection{Quantum model}
\label{sec:quantum}

A kinetic energy changes the short-distance fluctuations but leaves the
leading long-wavelength thermal covariance unchanged. The quantum model
has the same information constant and an analytic free energy for all
$T>0$. The bound below covers arbitrary quantum measurements, including
momenta and coherences.

\subsubsection{Quantization and thermodynamics}
\label{sec:q-thermo}

Set $\hbar=1$, $d\ge3$, and $J,\chi,T>0$. Each cell has Hilbert
space $L^2(S^1,\dd\theta_x/(2\pi))\otimes\ell^2(\Z^d)$:
the angle uses normalized angular measure, and its $d$ outgoing
bond integers use counting measure. Let $\mathcal H_L$ be the tensor
product of these spaces over all sites. Impose the flatness projector
of Eq.~\eqref{eq:flat-projector}, including all winding sectors:

\begin{equation}
 \mathcal H_{\mathrm{flat},L}\coloneqq P_{\mathrm{flat},L}\mathcal H_L.
 \label{eq:q-flat-space}
\end{equation}

The lifted coordinates $h_x$ are identified only under a common
$2\pi$ shift. Their momenta $P_x=-i\partial_{h_x}$ satisfy
$[h_x,P_y]=i\delta_{xy}$ on the real cover; globally,
$[P_x,s_y]=\delta_{xy}s_y$ for $s_y=e^{ih_y}$ and
$[P_x,a_{y,i}]=-i(\delta_{x,y+\hat e_i}-\delta_{xy})$.
A full turn of $h_x$ updates adjacent bond integers, so $P_x$ acts
in a one-bond neighborhood and preserves flatness and winding.
The integers record lifts, not independent momenta. With inertia
$\chi$ per site, define

\begin{equation}
 H_{{\rm q},L}\coloneqq \sum_x\frac{P_x^2}{2\chi}+\frac J2\sum_{x,i}a_{x,i}^2 .
 \label{eq:quantum-hamiltonian}
\end{equation}

The Gibbs state is $\rho_{{\rm q},L}=e^{-H_{{\rm q},L}/T}/Z_{{\rm q},L}$,
where $Z_{{\rm q},L}=\Tr_{\mathcal H_{\mathrm{flat},L}}e^{-H_{{\rm q},L}/T}$.
For local channels it is extended by zero to $\mathcal H_L$;
the channels need not preserve the flat subspace. Write $\rho_{\rm q}(T)$
for its symmetry-invariant thermodynamic limit.
With $\omega_k=\sqrt{J\lambda(k)/\chi}$, the free energy is

\begin{equation}
 f_{\rm q}(T)=T\int_k\ln\!\Bigl(2\sinh\frac{\omega_k}{2T}\Bigr).
 \label{eq:quantum-free-energy}
\end{equation}

It includes the zero-point energy and is real analytic for $T>0$,
with twist stiffness $\Upsilon(T)=J$. The selected-state covariance
of the real lifted heights has Fourier symbol, with the expansion taken as $k\to0$
at fixed $T,\chi>0$,

\begin{equation}
 \widehat C_{\chi,T}(k)\coloneqq \frac{\coth(\omega_k/2T)}{2\chi\omega_k}
              =\frac{T}{J\lambda(k)}+O(1).
 \label{eq:quantum-covariance}
\end{equation}

Thus $m_{\rm q}(T)=\exp[-\tfrac12\int_k\widehat C_{\chi,T}(k)]>0$,
and the rotation and reflection symmetry breaks as $O(2)\to O(1)$
at every $T>0$. The selected-state transverse and longitudinal
correlations follow the classical formulas in Table~\ref{tab:model-summary}
with $m$ replaced by $m_{\rm q}$.
The symmetric state's neutral connected correlation retains the
plateau $m_{\rm q}^2$.
Normal-mode, twist, and correlation
derivations are in SM Sec.~\ref{sm:quantum-details}.

\subsubsection{Information accessible to quantum measurements}
\label{sec:q-information}

Allow arbitrary block measurements on $\rho_{\rm q}(T)$, with at most
$N_R=\lceil R^\gamma\rceil$ outcomes and $\gamma>(d-2)/2$.
Using the same order of limits and logarithmic subtraction as in the
classical model gives

\begin{equation}
 \cC(T)=\frac12\ln\frac{\Upsilon(T)}{T}+\frac12\ln\frac{\pi\kappa_d}{e}.
 \label{eq:quantum-information}
\end{equation}

Spin-angle measurements attain this constant. For the upper bound,
write the thermal state as a positive mixture of Gaussian wave packets.
At infinite separation, the packets factorize conditional on their
centers, whose covariance retains the $T/(J\lambda(k))$ singularity.
Data processing then bounds every quantum reader by the classical
center observations (SM Sec.~\ref{sm:quantum}). The result is
independent of $\chi$. With trace distance
in the channel condition of Sec.~\ref{sec:local-readers},
it excludes two-way conversion by quasi-local channels between any distinct temperatures
along this analytic path.
Under the circuit and resource conventions of Sec.~\ref{sec:optimal-cooling},
the finite-size lower bound~\eqref{eq:cooling-depth-lower} and optimal
depth~\eqref{eq:cooling-depth-optimal} also hold. Local decoding
into rotor and oscillator coordinates, independent mode cooling, and
re-encoding give an exact $O(L)$ circuit
(SM Sec.~\ref{sm:villain-cooling}).
\subsection{Coupled multicomponent model}
\label{sec:multicomponent}

Several coupled angles provide an exact example with a multidimensional
order space. The information constant depends on the determinant of the
stiffness matrix, anticipating the general formulation in
Sec.~\ref{sec:general}.

\subsubsection{Model and thermodynamics}

On the same lattice with $d\ge3$, take $n$ angle components, labeled
by $a,b=1,\ldots,n$, and spatial directions $i=1,\ldots,d$. Put
$\boldsymbol\theta_x\in(\R/2\pi\Z)^n$ at each site and
$\boldsymbol n_{x,i}\in\Z^n$ on each oriented bond, imposing
$D_1\boldsymbol n=0$ componentwise. For a temperature-independent real
symmetric positive-definite coupling matrix $K$, define

\begin{align}
 \boldsymbol a_{x,i}&\coloneqq
 \boldsymbol\theta_{x+\hat e_i}-\boldsymbol\theta_x+2\pi\boldsymbol n_{x,i},
 \label{eq:multi-increment}\\
 H_L&\coloneqq\frac12\sum_{x,i}
          \boldsymbol a_{x,i}^{\mathsf T}K\boldsymbol a_{x,i}.
 \label{eq:multi-H}
\end{align}

The matrix $K$ has units of energy. The Gibbs law normalizes $e^{-H_L/T}$ on
the flat configurations, with angular measures
$\prod_{x,a}\dd\theta_x^a/(2\pi)$, integer counting measure, and all
windings included. Both interactions and constraints are local.
Uniform shifts give $U(1)^n$ symmetry, with order space
$\mathbb T^n=(\R/2\pi\Z)^n$.\footnote{Real diagonalization of $K$
generally changes the specified period lattice; see
SM Sec.~\ref{sm:multicomponent-details}.}

For every $T>0$, the free energy is real analytic and equals

\begin{equation}
 f(T)=\frac T2\biggl[
       \ln\det\!\Bigl(\frac{2\pi K}{T}\Bigr)+n\ell_d\biggr].
 \label{eq:multi-free}
\end{equation}

For total free energy $F_L(T,\boldsymbol\vartheta)$ with vector twist
$\boldsymbol\vartheta$ on one spatial cycle, the stiffness is

\begin{equation}
 \Upsilon_{ab}(T)\coloneqq
 \lim_{L\to\infty}L^{2-d}
 \left.\frac{\partial^2F_L(T,\boldsymbol\vartheta)}
 {\partial\vartheta^a\partial\vartheta^b}\right|_{\boldsymbol\vartheta=0}
 =K_{ab}.
 \label{eq:multi-stiffness}
\end{equation}

In the component with direction $\boldsymbol\Theta$,
$\langle e^{i\theta_x^a}\rangle=m_a e^{i\Theta^a}$ with
$m_a=\exp[-TG(0)(K^{-1})_{aa}/2]>0$.
Thus $U(1)^n\to\{1\}$ at every $T>0$. The same-component correlations
$C_{\parallel}^{aa},C_{\perp}^{aa}$ (no sum over $a$) have the
one-component decay powers. After division by $m_a^2$, the transverse
amplitude is set by the inverse thermal stiffness $T(K^{-1})_{aa}$
(Table~\ref{tab:model-summary}).
In the symmetric state the neutral same-component correlation instead
retains the plateau $m_a^2$. The full correlation formulas are derived
in SM Sec.~\ref{sm:multicomponent-details}.

\subsubsection{Exact information constant}

Allow readers of all cell variables with
$N_R=\lceil R^\gamma\rceil$, $\gamma>n(d-2)/2$.
With the same limits and channel conditions as in
Sec.~\ref{sec:local-readers}, subtracting $n(d-2)\ln R/2$
from $M_T(R)$ gives the ordinary limit

\begin{gather}
 \cC(T)
 =\frac12\ln\det\!\Bigl(\frac{\boldsymbol\Upsilon(T)}T\Bigr)
       +c_{\rm geom},
 \label{eq:multi-invariant}\\
 c_{\rm geom}\coloneqq\frac n2\ln\frac{\pi\kappa_d}{e}.
 \label{eq:multi-geometry}
\end{gather}

The determinant uses the specified $2\pi$-periodic coordinates.
For two independent Villain layers, $K=K_1\oplus K_2$, so the determinant
factorizes and the information constants add.
At fixed $K$, the difference between temperatures $T_1<T_2$ is
$\tfrac n2\ln(T_2/T_1)>0$, excluding two-way conversion by quasi-local channels along
this analytic path. Thermodynamic, correlation, and reader derivations are in
SM Sec.~\ref{sm:multicomponent-details}.
\section{Conjectures for general patterns of symmetry breaking}
\label{sec:general}

The exact results suggest that thermal stiffness controls information
about a shared order direction more generally.\footnote{For related
logarithmic terms in ground-state entanglement due to continuous
symmetry breaking, see Ref.~\cite{MetlitskiGrover2011}.}
We state separate conjectures about the information constant and the
free energy. SM Sec.~\ref{sm:general-details} gives the Gaussian
calculation and sufficient microscopic conditions for the information
formula.

\subsection{Stiffness as a metric on the ordered directions}
\label{sec:general-stiffness}

At $T>0$ in $d>2$, suppose the clustering ordered components of the equilibrium
ensemble under consideration form a single smooth compact connected orbit $\mathcal M=G/H$ of a compact
internal symmetry group $G$; $H$ fixes a chosen direction
$q_0$~\cite{WatanabeMurayama2014}.
The $n=\dim\mathcal M$ independent static tilts are described by
smooth dimensionless local coordinates
$\boldsymbol\vartheta=(\vartheta^1,\ldots,\vartheta^n)$,
with $q(\mathbf0)=q_0$.
Assume spatial isotropy. Pin one face of a cube of side $L$ to $q_0$
and the opposite face to $q(\boldsymbol\vartheta)$, with periodic
transverse boundaries. The corresponding microscopic total free energy
$F_{L,q_0}(T,\boldsymbol\vartheta)$ defines the stiffness components,
for $a,b=1,\ldots,n$, by~\cite{Fisher1973}

\begin{equation}
 \Upsilon_{ab}(T,q_0)\coloneqq\lim_{L\to\infty}L^{2-d}
 \left.\frac{\partial^2F_{L,q_0}(T,\boldsymbol\vartheta)}
 {\partial\vartheta^a\partial\vartheta^b}\right|_{\boldsymbol\vartheta=0}.
 \label{eq:coset-twist-metric}
\end{equation}

Dividing the stiffness by temperature defines the metric in thermal
units, $g_T(q)\coloneqq\boldsymbol\Upsilon(T,q)/T$, at each
ordered direction $q\in\mathcal M$.
Take the derivative before the limit, using the same twist-angle
normalization at all temperatures. Assume zero twist is stationary
and the limiting stiffness matrix is positive definite. Symmetry then
extends $g_T$ to a $G$-invariant metric on the ordered directions.

\subsection{Channel invariant}
\label{sec:general-information}

Suppose the symmetry-invariant thermodynamic Gibbs limit is the
uniform mixture of these ordered components, with respect to the
normalized $G$-invariant measure on $\mathcal M$. Choose a fixed
$G$-invariant reference metric $g_0$ and write
$\det_{g_0}g_T(q)\coloneqq\det\!\bigl(g_0(q)^{-1}g_T(q)\bigr)$.
Because $G$ relates any two order directions and preserves both metrics,
this relative determinant is independent of $q$.
For readers with $N_R=\lceil R^\gamma\rceil$,
$\gamma>n(d-2)/2$, define $\cC$ by subtracting
$n(d-2)\ln R/2$ from $M_T(R)$ and taking the upper limit as
$R\to\infty$.
We conjecture

\begin{gather}
 \cC(T)
 =\tfrac12\ln\det_{g_0}\!\Bigl(\frac{\boldsymbol{\Upsilon}(T,q)}{T}\Bigr)
   +c_{\rm geom},\label{eq:general-conjecture}\\
 c_{\rm geom}
 \coloneqq\tfrac n2\ln\frac{\kappa_d}{4\pi e}
       +\ln\operatorname{Vol}(\mathcal M,g_0).
 \label{eq:coset-geometry}
\end{gather}

Here $\operatorname{Vol}(\mathcal M,g_0)$ is the total volume of the
order space measured with $g_0$.
The reference metric cancels between the determinant and volume:
the constant measures the volume in the thermal stiffness metric.
It reproduces the exact circle and flat-torus results.
For two independent layers the dimensions add and the metric volumes
multiply, so the conjecture predicts
$\cC_{1\otimes2}(T)=\cC_1(T)+\cC_2(T)$.
For two independent condensed Bose gases with separately fixed densities,
the equality holds even if each observer measures both layers jointly.
The same proof allows either or both gases to be replaced by a
one-component classical Villain layer (SM Sec.~\ref{sm:mi-stacking-proof}).

A sufficient route compares each block with a noisy observation of
the shared direction: a measurement attains the proposed information,
and preparation from that observation bounds every competing reader.
SM Sec.~\ref{sm:general-details} gives the required conditional
independence and error bounds, the heat-kernel calculation, and
Heisenberg examples.\footnote{The $r^{2-d}$ decay of Gaussian static
tilts alone does not establish this comparison. Its microscopic
justification remains open for finite-spin magnets and the
unconstrained classical XY model.}

For the thermodynamic conjecture, fix a translation-invariant Hamiltonian
with finite-range or exponentially decaying interactions. Consider an
open interval of positive temperatures. At each temperature suppose
that the pure bulk equilibrium phases are all related by $G$, with the
same order space $G/H$ and positive-definite stiffness throughout. We further assume that the
order direction $q(x)$ accounts for all long-range static fluctuations.
In a long-wavelength description this means that, with the orientation
field held fixed, the remaining fields have exponentially decaying
connected correlations, with decay lengths bounded on each compact
subinterval. Products and gradients of $q(x)$ may themselves have
power-law correlations. Adding a decoupled Ising system at its own
critical temperature, for example, would violate this assumption. Under these physical hypotheses we conjecture that the
zero-field free-energy density is real analytic in temperature.
If both conjectures hold, unequal determinants of
$\boldsymbol\Upsilon(T,q)/T$ distinguish states along an analytic path;
distinguishing every pair further requires injectivity in $T$.
\section{Conclusions}
\label{sec:end}

Within a single phase with spontaneously broken continuous symmetry,
varying the temperature can take the system through a continuum of channel
classes while its bulk free-energy density remains analytic. We proved
this for the symmetry-invariant Gibbs states of
the canonical ideal Bose gas and vortex-free Villain model. In both
examples a quasi-local channel can only preserve or decrease the
information constant $\cC$. For the Bose gas at fixed density,
any $0<T_1<T_2<T_c$ obeys
\begin{equation}
 \cC(T_1)-\cC(T_2)
 =\frac12\ln\frac{\Upsilon(T_1)T_2}{\Upsilon(T_2)T_1}>0.
 \label{eq:comparison-gap}
\end{equation}
For the one-component classical and quantum Villain models, the
difference is $\tfrac12\ln(T_2/T_1)$ at any two positive temperatures.
Cooling would increase the invariant and is therefore impossible by
quasi-local channels.

The invariant measures a tangible property of the ordered state:
how precisely a finite region can determine the direction shared
with a distant region. Colder states contain more information about
this common direction. The leading logarithmic growth of the
information is unchanged within the phase, but its constant term
retains this difference. Optimizing over local measurements makes
that constant a property of the state and ties it to the thermal
stiffness $\Upsilon(T)/T$. In the classical Villain model, the spins
alone contain enough information to attain the invariant.

The distinction between thermodynamic and channel equivalence is
therefore structural. Requiring local reversibility cannot connect
states already separated by a quasi-local-channel invariant. Simply
allowing deeper circuits can also remove distinctions one intended
to retain: the linear-depth Bose preparation connects both sides of
the condensation transition. For cooling within the ordered regimes,
we prove that depth proportional to the system's linear size is
optimal, with the gate resources specified in
Sec.~\ref{sec:optimal-cooling}. The channel classification thus records
a spatial cost of changing the long-distance information that is
invisible to free-energy analyticity.

The quantum and multicomponent Villain extensions suggest that this
mechanism reaches beyond a single model. For a general breaking
pattern $G\to H$, we conjecture that the information constant is
controlled by the stiffness metric on the space of ordered directions.
If the determinant of the thermal stiffness matrix takes a
different value at each temperature, the conjecture distinguishes
every pair throughout that interval.
The resulting picture, illustrated in Fig.~\ref{fig:phase-qlc}, is
that a thermodynamically smooth ordered phase can contain a continuum
of operationally distinct states. Establishing this picture for
finite-spin XY and Heisenberg magnets is an important next step.
Our exact examples provide a starting point: phase classification
must account not only for thermodynamic singularities and broken
symmetries, but also for the information that local operations can
preserve or discard.
\begin{acknowledgments}
We thank Ryohei Kobayashi.
Generative AI tools assisted with derivation development, consistency checks,
and manuscript preparation.
\end{acknowledgments}
\section*{Data availability}
No data were created or analyzed in this study.
\bibliography{references}

\clearpage
\onecolumngrid
\setcounter{section}{0}
\setcounter{equation}{0}
\renewcommand{\theequation}{S\arabic{equation}}
\renewcommand{\theHequation}{supp.\arabic{equation}}
\renewcommand{\theHsection}{supp.\arabic{section}}
\setlength{\emergencystretch}{4em}
\pdfbookmark[0]{Supplemental Material}{supplement}
\begin{center}
{\Large\bfseries Supplemental Material\par}
\medskip
{\large Local-channel inequivalence within analytic symmetry-broken phases\par}
\end{center}

For the two information constants, the canonical Bose local limit is proved in
Sec.~\ref{sm:bose-local-details}, the common capacity and entropy estimates in
Sec.~\ref{sm:shared-information}, and the spin-only Villain lower bound in
Sec.~\ref{sm:classical-spin-error}. The last part treats finite-size cooling,
where the block radius grows with the lattice.

\medskip\noindent\textbf{Contents}\par
\medskip\noindent\textbf{The two models and their information constants}\par
\smcontentsentry{sm:bose-details}{Canonical Bose thermodynamics and field correlations}
\smcontentsentry{sm:bose-information-details}{Canonical Bose local limits}
\smcontentsentry{sm:shared-information}{Common estimates for the information constants}
\smcontentsentry{sm:classical-villain}{Spin readout and correlations in the classical Villain model}
\medskip\noindent\textbf{Extensions and the general conjecture}\par
\smcontentsentry{sm:quantum-details}{Quantum Villain thermodynamics and correlations}
\smcontentsentry{sm:quantum}{Quantum Villain information from Gaussian packets}
\smcontentsentry{sm:multicomponent-details}{Coupled multicomponent Villain derivations}
\smcontentsentry{sm:general-details}{General order spaces: the information constant}
\medskip\noindent\textbf{Finite-size cooling circuits}\par
\smcontentsentry{sm:bose}{Canonical Bose states on two growing blocks}
\smcontentsentry{sec:bose-circuits}{Optimal finite-size cooling of the canonical Bose gas}
\smcontentsentry{sm:villain-cooling}{Optimal cooling depth for the Villain models}

\clearpage
\section{Canonical Bose thermodynamics and field correlations}
\label{sm:bose-details}
An independent occupation law, conditioned on the particle number, gives both the fixed-density free energy and the canonical twist response.

\subsection{Fixed particle number and free energy}
\label{sm:bose-canonical-occupations}

Fix $0<T<T_c$. For each nonzero momentum take an independent geometric variable $M_k$ with probability $(1-q_k)q_k^m$ for $m=0,1,\ldots$, where $q_k\coloneqq e^{-\epsilon(k)/T}$. Its mean is $n_T(k)=q_k/(1-q_k)$. Write $\langle\cdot\rangle_{\rm ex}$ for this average and $N_{\rm ex}\coloneqq\sum_{k\ne0}M_k$. When $N_{\rm ex}\le N$, the remaining $N_0=N-N_{\rm ex}$ particles occupy the zero-energy mode. The canonical law is exactly this conditional distribution, with partition function
\begin{equation}
 Z_{L,N}(T)=\Pr(N_{\rm ex}\le N)\prod_{k\ne0}(1-q_k)^{-1}.
 \label{eq:bose-main-canonical-partition}
\end{equation}
The mean and variance of the total excited occupation satisfy
\begin{align}
 \overline N_{{\rm ex},L}&\coloneqq\langle N_{\rm ex}\rangle_{\rm ex}
 =\varrho_{\rm ex}(T)L^d+o(L^d),
 \label{eq:bose-main-occupation-mean}\\
 \sigma_L^2&\coloneqq\langle(N_{\rm ex}-\overline N_{{\rm ex},L})^2\rangle_{\rm ex}
 =\sum_{k\ne0}n_T(k)(1+n_T(k))\nonumber\\
 &=\begin{cases}
 O(L^4),&d=3,\\
 O(L^4\ln L),&d=4,\\
 O(L^d),&d\ge5.
 \end{cases}
 \label{eq:bose-main-occupation-variance}
\end{align}
Near zero $n_T(k)\le C|k|^{-2}$, and the shell $r\le L|k|/(2\pi)<r+1$ contains at most $Cr^{d-1}$ modes. The contribution from $|k|<\eta$ to $\overline N_{{\rm ex},L}/L^d$ is thus uniformly $O(\eta^{d-2})$; outside this ball the Riemann sums converge. Summing $|k|^{-4}$ on the same shells bounds the variance. Since $N-\overline N_{{\rm ex},L}=\varrho_0 L^d+o(L^d)$ and $\sigma_L^2=o(L^{2d})$, Chebyshev's inequality gives $\delta_L\coloneqq\Pr(N_{\rm ex}>N)\to0$. Hence $N_0/L^d\to\varrho_0$ in conditional probability. Taking the logarithm of Eq.~\eqref{eq:bose-main-canonical-partition} gives main Eq.~\eqref{eq:bose-free}, since the logarithmic singularity at $k=0$ is integrable.

\subsection{Canonical and grand-canonical stiffness}
\label{sm:bose-twist}
The twisted Hamiltonians are diagonal in the same occupation basis. Differentiating the canonical free energy twice gives
\begin{equation}
 \Upsilon_L(T)=\frac1{L^d}\biggl[\Bigl\langle\sum_k w_k\hat n_k\Bigr\rangle_L
 -\frac1T\Var_L\!\Bigl(\sum_k v_k\hat n_k\Bigr)\biggr].
 \label{eq:canonical-helicity-finite}
\end{equation}
Here the mean and variance are canonical, $\hat n_k=b_k^\dagger b_k$, $v_k=\partial_{k_1}\epsilon(k)$, and $w_k=\partial_{k_1}^2\epsilon(k)$. Reflection $k_1\mapsto-k_1$ makes the current mean zero.

The current square is unbounded, so we control the cost of canonical conditioning by an exponential tail. Under the independent occupation law, put $K_L(u)\coloneqq\ln\langle e^{u(N_{\rm ex}-\overline N_{{\rm ex},L})}\rangle_{\rm ex}$. Take $t=a/L^2$ with $a>0$ small enough that $n_u(k)\coloneqq(e^{\epsilon(k)/T-u}-1)^{-1}\le Cn_T(k)$ for $0\le u\le t$. Twice differentiating the geometric factors gives $K_L''(u)=\sum_{k\ne0}n_u(k)(1+n_u(k))\le C\sigma_L^2$. Since $K_L(0)=K_L'(0)=0$, the exponential Markov bound gives
\begin{equation}
 \delta_L\le e^{-t(N-\overline N_{{\rm ex},L})+Ct^2\sigma_L^2}
 \le e^{-cL^{d-2}}.
 \label{eq:bose-main-exponential-tail}
\end{equation}
Indeed $N-\overline N_{{\rm ex},L}\sim\varrho_0L^d$ and $t^2\sigma_L^2=o(L^{d-2})$ by Eq.~\eqref{eq:bose-main-occupation-variance}.

Substituting $N_0=N-N_{\rm ex}$, the two observables averaged in Eq.~\eqref{eq:canonical-helicity-finite} are $2JN+\sum_{k\ne0}(w_k-2J)M_k$ and $(\sum_{k\ne0}v_k M_k)^2$. For either observable $A$, $\langle A^2\rangle_{\rm ex}$ grows at most polynomially in $L$: each geometric moment up to order four is bounded by $C_p(1+n_T(k))^p\le C_p L^{2p}$, and there are at most $L^d$ modes. By Cauchy--Schwarz, conditioning on $N_{\rm ex}\le N$ changes its mean by at most $2\sqrt{\delta_L\langle A^2\rangle_{\rm ex}}/(1-\delta_L)=o(1)$. Averaging under the independent law and using the vanishing current mean therefore gives
\begin{equation}
 \Upsilon_L(T)=\frac{2JN}{L^d}+\frac1{L^d}\sum_{k\ne0}
 \biggl[(w_k-2J)n_T(k)-\frac{v_k^2}{T}n_T(k)(1+n_T(k))\biggr]+o(1).
 \label{eq:bose-canonical-response-asymptotic}
\end{equation}

The grand-canonical response in main Eq.~\eqref{eq:helicity-gc-finite} has the same sum with $N/L^d$ replaced by $\varrho$ and $n_T(k)$ by $n_{L,k}=(e^{(\epsilon(k)-\mu_L)/T}-1)^{-1}$, since $w_0=2J$, $v_0=0$, and the mean density is fixed. As $\mu_L\to0$, $n_{L,k}\le n_T(k)$. Both summands are bounded by $C|k|^{-2}$ near zero, so the contribution of $0<|k|<r$ is at most $Cr^{d-2}$, uniformly in $L$. Outside this ball the Riemann sums converge to the same integral. Taking $L\to\infty$ and then $r\downarrow0$, the integration by parts in main Eq.~\eqref{eq:helicity-cancellation} gives
\begin{equation}
 \lim_{L\to\infty}\Upsilon_L(T)
 =\lim_{L\to\infty}\Upsilon_L^{\rm gc}(T)=2J\varrho_0(T).
 \label{eq:bose-stiffness-ensemble-limit}
\end{equation}

\subsection{Bose thermodynamic regularity at the critical temperature}
\label{sm:bose-foundations}
\label{sm:bose-critical-thermo}

Fix the particle density $\varrho$, hopping strength $J$, and dimension $d$.
Write $n(T,\mu)\coloneqq-\partial_\mu g(T,\mu)$ for the grand-canonical
density and put $u\coloneqq-\mu>0$ on the normal side.
The dimension dependence comes from the small-momentum dispersion
$\epsilon(k)=J|k|^2+O(|k|^4)$~\cite{PathriaBeale2011}. At $T=T_c$,
\begin{equation}
 n(T_c,0)-n(T_c,-u)
 \sim\begin{cases}
 a_3\sqrt u,&d=3,\\
 a_4u\ln(J/u),&d=4,\\
 a_du,&d\ge5,
 \end{cases}
 \label{eq:bose-critical-density}
\end{equation}
with positive $a_d$. Since $\partial_T n(T_c,0)=\varrho_{\rm ex}'(T_c)>0$,
the fixed-density condition makes $u(T)$ vanish in proportion to
$(T-T_c)^2$, $(T-T_c)/|\ln((T-T_c)/J)|$, or $T-T_c$, respectively.

To check differentiability, take partial derivatives of the density integral
at $\mu<0$. Put $A\coloneqq\partial_T n$ and $D\coloneqq\partial_\mu n$.
Subscripts on $g,n,A,D$ below denote partial derivatives.
Along the normal branch, $\mu'=-A/D$ and $f'=g_T$. Two further derivatives give
\begin{align}
 f''&=g_{TT}+A^2/D,\label{eq:bose-critical-fsecond}\\
 f'''&=g_{TTT}+3A n_{TT}/D-3A^2D_T/D^2+A^3D_\mu/D^3.
 \label{eq:bose-critical-fthird}
\end{align}
To evaluate those derivatives near $(T_c,0)$, split the occupation at
$E=\epsilon(k)+u$ into $T/E$ and a remainder analytic in $E$ near zero,
uniformly for $T$ near $T_c$. Thus $D$ and $D_\mu$ have singular parts
$T\int_k E^{-2}$ and $2T\int_k E^{-3}$, respectively.
Radial integration near $k=0$ gives $D\sim c_3u^{-1/2}$ and
$D_\mu\sim(c_3/2)u^{-3/2}$ for $d=3$, while $D\sim c_4\ln(J/u)$
and $D_\mu\sim c_4/u$ for $d=4$, with $c_3,c_4>0$ along the normal branch.
For $d\ge5$, $D$ has a finite positive limit. The same split gives
bounded $n_{TT}$ and $D_T=O(1+D)$, and the pure temperature derivatives
of $g$ through third order are continuous up to $\mu=0$ by dominated convergence.
On the condensed side $f(T)=g(T,0)$. Comparing the one-sided limits in
Eqs.~\eqref{eq:bose-critical-fsecond} and~\eqref{eq:bose-critical-fthird}
therefore gives $f\in C^2$ at $T_c$ in $d=3,4$. Its third derivative has
a positive finite jump for $d=3$ and diverges to $+\infty$ from the normal
side for $d=4$. For $d\ge5$, $f\in C^1$ and $f''$ has a positive finite jump.

\subsection{Connected field correlations}
\label{sm:bose-correlations}
In Table~\ref{tab:model-summary}, $C_\parallel(x)$ and $C_\perp(x)$ are the connected correlations of the real and imaginary parts of $e^{-i\alpha}O_x$ in a state with selected direction $\alpha$, with $O_x=b_x$ for Bose particles and $O_x=s_x$ for Villain spins. In the gas $m^2=\varrho_0$, and main Eq.~\eqref{eq:bose-mixture} gives $\langle b_0^\dagger b_x\rangle=\varrho_0+C_T(x)$. Circular noise gives equal quadrature variances; including the commutator $[b_0,b_x^\dagger]=\delta_{0x}$ yields $C_\parallel(x)=C_\perp(x)=C_T(x)/2+\delta_{0x}/4$. For $r=|x|\to\infty$ the lattice Green function obeys~\cite{LawlerLimic2010}
\begin{equation}
 G(x)=A_d r^{2-d}+o(r^{2-d}),\qquad A_d\coloneqq\frac{\Gamma(d/2-1)}{4\pi^{d/2}}.
 \label{eq:lattice-G-tail}
\end{equation}
After removing its singularity at zero, $(e^{\epsilon/T}-1)^{-1}-T/\epsilon$ is analytic near the real band, so its Fourier coefficients decay exponentially. Thus $C_T(x)=(T/J)G(x)+O(e^{-r/\xi_T})$ for some $\xi_T<\infty$, and $\Upsilon=2Jm^2$ gives
\begin{equation}
 C_\perp(x)/m^2=(T/\Upsilon(T))G(x)+o(r^{2-d}).
 \label{eq:transverse-stiffness}
\end{equation}
\section{Canonical Bose local limits}
\label{sm:bose-information-details}

Trace-norm convergence of the reduced state controls every quantum measurement on a fixed block.

\subsection{The canonical local limit}
\label{sm:bose-local-details}

Fix $0<T<T_c$. Enclose any finite set $B$ in a cube of fixed
radius and place a second cube across the torus. The right side of
Eq.~\eqref{sm:eq:bose-regional-error} vanishes as $L\to\infty$ at
this fixed radius. Taking the partial trace over everything outside
$B$ can only decrease the distance. In the comparison state
\eqref{sm:eq:bose-target} it leaves the uniform phase and a Gaussian
noise restricted to $C_{T,B}$, proving main
Eq.~\eqref{eq:bose-mixture} in trace norm.
Lemma~\ref{sm:bose-block} starts directly from the finite-volume
canonical state via the exact mixture~\eqref{sm:eq:bose-hartree}.

\subsection{The noncondensed local limit}
\label{sm:bose-normal-details}

For $T>T_c$ fix $\mu=\mu(T)<0$ by main Eq.~\eqref{eq:bose-normal-density}. The local canonical limit is a centered Gaussian coherent-state mixture of covariance
\begin{equation}
 C_{T,\mu}(x-y)\coloneqq\int_k
 \frac{e^{ik\cdot(x-y)}}{e^{(\epsilon(k)-\mu(T))/T}-1}.
 \label{detail:eq:bose-normal-covariance}
\end{equation}
To verify this, test the state on the displacement $D_B(u)\coloneqq\exp[\sum_{x\in B}(u_x b_x^\dagger-\bar u_x b_x)]$. Put $\widetilde u_k\coloneqq L^{-d/2}\sum_{x\in B}u_x e^{-ik\cdot x}$ and $q_k\coloneqq e^{-(\epsilon(k)-\mu)/T}$. Averaging the normally ordered displacement against a geometric series, continuing its parameter to $q_ke^{it}$, and multiplying over all modes gives
\begin{equation}
 \frac{\Tr(e^{-(H_L-\mu\widehat N)/T}e^{it\widehat N}D_B(u))}
      {\Tr(e^{-(H_L-\mu\widehat N)/T}e^{it\widehat N})}
 =e^{-\|u\|^2/2-\sum_k\frac{q_ke^{it}}{1-q_ke^{it}}|\widetilde u_k|^2}.
 \label{detail:eq:bose-number-weyl}
\end{equation}
Here $\|u\|^2=\sum_{x\in B}|u_x|^2$ and the sum includes $k=0$. The canonical characteristic function is the ratio of the two $N$th Fourier coefficients. The $q_k$ are bounded away from zero and one, so by Parseval the insertion on the right differs from its value at $t=0$ by $O(|t|)$ uniformly in $L$. The normalized denominator $\prod_k(1-q_k)/(1-q_ke^{it})$ is bounded in modulus by $e^{-cL^dt^2}$ on $[-\pi,\pi]$. Replacing the insertion by its value at zero therefore costs $O(L^{-d})$ in the numerator coefficient after this common normalization.

The denominator coefficient is $(2\pi\sigma_L^2)^{-1/2}(1+o(1))$: analyticity and periodicity of the mean occupation give a total mean $m_L=N+O(1)$ by Fourier summation, and its variance $\sigma_L^2$ is proportional to $L^d$. Expand the logarithm of the product as $itm_L-t^2\sigma_L^2/2+O(L^d|t|^3)$ and rescale $t$ by $\sigma_L^{-1}$ in the Fourier integral, using the same Gaussian domination. The ratio thus converges to the $t=0$ limit, the Gaussian characteristic function of Eq.~\eqref{detail:eq:bose-normal-covariance}. Translation invariance bounds the probability of more than $K$ particles in $B$ by $\varrho|B|/(K+1)$. Compression to the finite-dimensional subspace with at most $K$ particles costs at most twice the square root of this bound in trace norm. This gives trace-norm precompactness, and the characteristic function identifies the unique limit.

The joint covariance of two fixed blocks approaches the direct sum
of its positive definite marginals as their separation tends to
infinity. The Gaussian mutual information, the logarithm of the
product of the marginal determinants divided by the joint
determinant, therefore vanishes. The Born-rule bound in main
Eq.~\eqref{eq:bose-born-reader} applies to every quantum reader,
giving $M_T(R)=0$ for $T>T_c$ at each fixed $R$.
\section{Common estimates for the information constants}
\label{sm:shared-information}

At fixed positive temperature the Bose and Villain calculations share
the cube capacity and the same phase-bin estimate. The Bose precision
and entropy formulas below assume $0<T<T_c$.

\subsection{Block precision and capacity}
\label{sm:capacity-details}

To obtain the two precisions in main Eq.~\eqref{eq:bose-precision-limit}, let $h_{\rm sp}$ have symbol $\epsilon(k)=J\lambda(k)$. The inverse of the full covariance $C_T+vI$ has symbol $q_0(\epsilon)\coloneqq e^{\epsilon/T}-1$ for $v=0$ and $q_1(\epsilon)\coloneqq 1-e^{-\epsilon/T}$ for $v=1$. Both are $\epsilon/T+O(\epsilon^2)$ on the lattice band. For smooth compactly supported $\varphi=1$ near the unit cube, set $\varphi_R(x)\coloneqq\varphi(x/R)$. Fourier-space Cauchy--Schwarz gives, for any real $c$ supported on $B_R$,
\begin{equation}
 |\one^{\mathsf T}c|^2\le
 [c^{\mathsf T}(C_{T,B_R}+vI)c]
 \langle\varphi_R,q_v(h_{\rm sp})\varphi_R\rangle.
 \label{eq:bose-capacity-cauchy-schwarz}
\end{equation}
Maximizing the left side divided by the covariance quadratic form gives $P_{T,v}(R)/(2\varrho_0)$, with the covariance restricted before inversion. The gradient energy is a Riemann sum and $\|h_{\rm sp}\varphi_R\|^2=O(R^{d-4})$. Using $\Upsilon=2J\varrho_0$ gives
\begin{equation}
 P_{T,v}(R)\le\frac{\Upsilon(T)}T R^{d-2}
 \int_{\mathbb R^d}|\nabla\varphi|^2\,\dd^dx+o(R^{d-2}).
 \label{eq:bose-precision-potential-bound}
\end{equation}
The infimum over $\varphi$ gives the cube capacity $\kappa_d$.

For the lower bound, choose a smooth unit-mass density $\nu\ge0$ supported inside the cube, with lattice weights $w_R(x)\propto\nu(x/R)$ normalized to sum to one. Since $(e^{\epsilon/T}-1)^{-1}+v-T/\epsilon$ is bounded and $\|w_R\|^2=O(R^{-d})$, the leading energy comes from the lattice Green function in Eq.~\eqref{eq:lattice-G-tail}, giving
\begin{equation}
 P_{T,v}(R)^{-1}\le\frac{w_R^{\mathsf T}(C_{T,B_R}+vI)w_R}{2\varrho_0(T)}
 =\frac T{\Upsilon(T)} R^{2-d}E(\nu)+o(R^{2-d}),
 \label{eq:bose-precision-charge-bound}
\end{equation}
where $E(\nu)\coloneqq\iint A_d |x-y|^{2-d}\nu(x)\nu(y)\,\dd^dx\,\dd^dy$. The bound $G(z)\le C(1+|z|)^{2-d}$ limits the contribution from $|x-y|<\eta R$ in the rescaled sum to $O(\eta^2+R^{-2})$. To see why the minimum unit-charge energy is $1/\kappa_d$, observe that integration by parts and Cauchy--Schwarz give $1\le E(\nu)\int|\nabla\varphi|^2$, and the equilibrium charge of the unit-potential cube has mass and energy $\kappa_d$. After normalization it attains equality, and shrinking and smoothing approximates it by the interior densities used above. This proves $P_{T,v}(R)/R^{d-2}\to\kappa_d\Upsilon(T)/T$. With covariance $G$ and inverse symbol $\lambda(k)$ the same two bounds give $\operatorname{cap}(B_R)/R^{d-2}\to\kappa_d$ (main Eq.~\eqref{eq:capacity-limit}).

\subsection{Scalar entropy, including the observed radius}
\label{sm:bose-entropy-details}

Main Sec.~\ref{sec:bose-upper-bound} gives the exact two-amplitude information $2S(Z_s)-S(Z_{\sqrt2s})-\ln(\pi e)$. The phase of $Z_s=se^{i\Theta}+\xi$ is uniform independently of its radius, distributed as $U_s=|s+\xi|$ for $\xi\sim\mathcal{CN}(0,1)$. With $f_s$ the density of $U_s$, the polar Jacobian gives
\begin{equation}
 S(Z_s)=\ln(2\pi)+S(U_s)+\int_0^\infty f_s(u)\ln u\,\dd u.
 \label{detail:eq:bose-polar-entropy}
\end{equation}
Angular integration yields
$f_s(u)=2u e^{-u^2-s^2}I_0(2su)$ for $u>0$, where
$I_0(t)\coloneqq\int_0^{2\pi}e^{t\cos\theta}\dd\theta/(2\pi)$.
Expanding at the maximum gives $I_0(t)\sim e^t/\sqrt{2\pi t}$ and
$f_s(s+x)\to\pi^{-1/2}e^{-x^2}$. The bound
$I_0(t)\le Ce^t/\sqrt{1+t}$ implies
$f_s(s+x)\le C\sqrt{1+|x|}e^{-x^2}$ for $s\ge1$ (zero for $x\le-s$).
It controls the entropy tails using $|f\ln f|\le C\sqrt f$ for small
$f$, so $S(U_s)=\tfrac12\ln(\pi e)+o(1)$.
For the logarithmic moment, use $|U_s-s|\le|\xi|$ and
$|\ln(U_s/s)|\le2|U_s-s|/s$ on $U_s\ge s/2$; on its complement,
$f_s(u)\le2u e^{-s^2/4}$. Thus $\int f_s\ln u=\ln s+o(1)$, proving
\begin{equation}
 I(Z_s\infosep Z'_s)=\ln s+\frac12\ln\frac{2\pi}{e}+o(1).
 \label{detail:eq:bose-scalar-information}
\end{equation}
This is an upper bound with the full radius information retained.
For the measured phase, the error $s\operatorname{Arg}(s+\xi)-\Im\xi$
is $O(|\xi|^2/s)$ on $|\xi|<s/2$. On its complement it is bounded
by $\pi s+|\xi|$, whose squared tail integral vanishes exponentially.
Thus the error tends to zero in mean square, giving angular variance
$1/(2s^2)+o(s^{-2})$.

\subsection{Wrapping and binning angular errors}
\label{sm:wrapped-entropy}\label{sm:bose-bin-details}

For $Y\sim\mathcal{N}(0,v)$ with $v\to0$, write
$Y=\bar Y+2\pi K$, with $\bar Y\in[-\pi,\pi)$ and integer $K$.
The entropy identity is
$S(Y)=S(\bar Y)+S(K\given \bar Y)$, with Shannon entropy for the
integer variable. Its tails satisfy
$\Pr(|K|\ge j)\le2e^{-(2j-1)^2\pi^2/(2v)}$, $j\ge1$, implying
$0\le S(K\given \bar Y)\le S(K)=o(1)$.
Thus wrapping changes $S(Y)=\tfrac12\ln(2\pi ev)$ by $o(1)$.
Two observations of a uniform phase with independent Gaussian errors
have mutual information $\ln(2\pi)-S(\bar Y)$, where $Y$ is their
error difference. This proves main Eq.~\eqref{eq:wrapped}.

Let the two angles be $\Theta+\epsilon,\Theta+\epsilon'$ modulo
$2\pi$, with independent identically distributed errors of variance
$\sigma^2$, independent of uniform $\Theta$. Partition the circle
into $N_R$ equal bins of width $\Delta_R\coloneqq2\pi/N_R$.
Denote the two bin labels by $U,V$.
The first bin label and fractional position $t\in[0,1)$ are independent
and uniform, even conditional on the errors. The second label is the
first plus $b\coloneqq\lfloor t+(\epsilon'-\epsilon)/\Delta_R\rfloor$
modulo $N_R$, where the floor means the greatest integer not exceeding
its argument. Conditional on the errors, $b$ has mean
$(\epsilon'-\epsilon)/\Delta_R$ and variance at most $1/4$; therefore
$\Var b\le2\sigma^2/\Delta_R^2+1/4$.

Adding independent uniform noise on $[0,1)$ to $b$ gives differential
entropy equal to its Shannon entropy and variance $\Var b+1/12$.
Gaussian maximal entropy at fixed variance, together with
$S(b\bmod N_R)\le S(b)$, proves
\begin{equation}
 I(U\infosep V)\ge\ln(2\pi)
 -\frac12\ln\!\bigl(2\pi e(2\sigma^2+\Delta_R^2/3)\bigr).
 \label{detail:eq:bin-information}
\end{equation}
Thus $\Delta_R=o(\sigma)$ preserves the constant in the small-noise
lower bound. This argument applies to the Bose heterodyne phase and
to both Villain readers; the errors need not be exactly Gaussian.
\section{Spin readout and correlations in the classical Villain model}
\label{sm:classical-villain}

\subsection{Nonlinear error of the spin reader}
\label{sm:classical-spin-error}
In the Gaussian representation of main Eq.~\eqref{eq:selected-state},
the weighted spin direction in main Eq.~\eqref{eq:spin-reader}
has the optimal linear angular precision to leading order. To see this,
take $b_R=\lfloor R/\ln R\rfloor$ and $D=B_{R-b_R}$. Extend the
unit equilibrium charges of $D$ by zero and average their translates
by all $z\in B_{b_R}$ to obtain weights $w_x$ supported in $B_R$.
The charges are nonnegative by the maximum principle for their potential,
so $\sum_x w_x=1$ and $\max_x w_x\le|B_{b_R}|^{-1}$.
The capacity variational bound on $B_R$ and convexity of the Green
energy under translation averaging give
\begin{equation}
 \frac{T}{J\operatorname{cap}(B_R)}
 \le\Var X_R\le\frac{T}{J\operatorname{cap}(D)},\qquad
 X_R\coloneqq\sum_x w_x\eta_x .
\end{equation}
The bounds agree to relative error $o(1)$ by Sec.~\ref{sm:capacity-details}.

Write $r_x=e^{i\eta_x}/m(T)-1-i\eta_x$, $Q_R=\sum_xw_xr_x$.
The weighted spin sum divided by $m(T)e^{i\Theta}$ is $1+iX_R+Q_R$.
With $\gamma_{T,B_R}$ the centered Gaussian law of covariance
$(T/J)G_{B_R}$ and $c_{xy}=(T/J)G(x-y)$, differentiation of its
characteristic function gives
\begin{equation}
 \int r_x\overline{r_y}\,\dd\gamma_{T,B_R}
 =e^{c_{xy}}-1-c_{xy}\le C G(x-y)^2.
 \label{sm:spin-residual-covariance}
\end{equation}
Since $G(z)\le C(1+|z|)^{2-d}$, the sum of $G(z)^2$ for
$|z|\le 2\sqrt d R$ is bounded by $C\sum_{j\le2\sqrt dR}(1+j)^{3-d}\le CR$,
where $j$ is a nonnegative integer. It follows that
$\int|Q_R|^2\,\dd\gamma_{T,B_R}\le CR/|B_{b_R}|=O(R^{1-d}(\ln R)^d)$.

Let $\epsilon_R$ be the principal angle of $1+iX_R+Q_R$, using the center
spin relative to $\Theta$ if the sum vanishes. For real $t$, the squared
difference between the principal angle of $1+it+z$ and $t$ is at most
$C(|z|^2+t^4)$. Indeed, when $|z|\le1/2$ the change due to $z$ is
at most $2|z|$ and $|\arctan t-t|\le t^2$; otherwise $(\pi+|t|)^2$
suffices, including the fallback. The fourth moment of $X_R$ is
$3(\Var X_R)^2$. Consequently
\begin{equation}
 \int|\epsilon_R-X_R|^2\,\dd\gamma_{T,B_R}
 =O\!\bigl(R^{1-d}(\ln R)^d+R^{4-2d}\bigr)=o(\Var X_R).
 \label{sm:spin-angle-error}
\end{equation}
By the triangle inequality in mean square, the spin error itself satisfies
\begin{equation}
 \int_{\R^{|B_R|}}\!\left(\prod_{x\in B_R}\dd\eta_x\right)
 \frac{\epsilon_R(\eta)^2\,e^{-J\eta^{\mathsf T}G_{B_R}^{-1}\eta/(2T)}}
 {(2\pi T/J)^{|B_R|/2}\sqrt{\det G_{B_R}}}
 =\frac{T}{J\operatorname{cap}(B_R)}(1+o(1)).
 \label{sm:spin-angle-variance}
\end{equation}
This is the variance-scale bound needed before binning.

\subsection{Longitudinal and transverse correlations}
\label{sm:classical-correlations}
At $\Theta=\alpha$, the transverse and longitudinal spin components are
$\sin\eta_x$ and $\cos\eta_x$, of means zero and $m(T)$. The variance of
$\eta_0\pm\eta_x$ is $2(T/J)[G(0)\pm G(x)]$. Taking the half-difference
and half-sum of the Gaussian characteristic functions and subtracting the
one-point products gives
\begin{align}
 C_\perp(x)&=m(T)^2\sinh[(T/J)G(x)],\label{sm:classical-transverse}\\
 C_\parallel(x)&=m(T)^2\bigl(\cosh[(T/J)G(x)]-1\bigr).
 \label{sm:classical-longitudinal}
\end{align}
Expansion at $G(x)\to0$ gives Table~\ref{tab:model-summary}.
Averaging over $\alpha$ leaves $\langle\bar s_0s_x\rangle$ unchanged
and makes $\langle s_x\rangle=0$, preserving the plateau in the
symmetric connected correlation.

\subsection{Finite-size test of stable clustering}
\label{sm:classical-bond-clustering}
The symmetric bond alignment $\cos a_e=\operatorname{Re}(\bar s_xs_y)$
is an allowed local interaction in family (ii) of Ref.~\cite{MKS2026}.
Take $e=(0,\hat e_1)$, $f=e+r\hat e_1$. In the zero-winding sector
their Gaussian increments have common variance $v\le T/J$ and covariance
$C_{ef}$. Thus $\Cov(\cos a_e,\cos a_f)=e^{-v}(\cosh C_{ef}-1)$.

The Fourier multiplier for $C_{ef}$, in units of $T/J$, is
$2(1-\cos k_1)/\lambda(k)$ with the zero mode omitted. For transverse
momentum $p=(k_2,\dots,k_d)\ne0$, define $u>0$ by
$\cosh u=1+\sum_{j=2}^d(1-\cos k_j)$. The multiplier becomes
$1-(\cosh u-1)/(\cosh u-\cos k_1)$. Summing the periodic images of
the inverse denominator's Fourier coefficients $e^{-|r|u}/\sinh u$
gives, including the $p=0$ contribution,
\begin{equation}
 -\frac JT C_{ef}=\frac1{L^d}+\frac1{L^{d-1}}\sum_{p\ne0}
 \tanh(u/2)\frac{e^{-ru}+e^{-(L-r)u}}{1-e^{-Lu}},\qquad 0<r<L .
 \label{eq:axial-torus-kernel}
\end{equation}
For $1\ll r\ll L$ the shell $1/r\le|p|\le2/r$ contains order
$(L/r)^{d-1}$ terms, each bounded below by $c/r$ since $u/|p|\to1$
near zero. Every summand is positive and bounded above by
$C|p|e^{-cr|p|}$ since $Lu$ is bounded away from zero. Shell summation
therefore bounds $|C_{ef}|$ above and below by multiples of $r^{-d}$.
At $r=\lfloor\sqrt L\rfloor$, the cosine identity makes the covariance
of the two alignments of order $L^{-d}$. Including nonzero windings
changes this by $O(e^{-2\pi^2JL^{d-2}/T})$ because the alignments
have norm at most one. A buffer of width $\lfloor r/2\rfloor$ then
violates the bound $P(L)e^{-\lfloor r/2\rfloor/\xi}$ for every fixed
polynomial $P$ and correlation length $\xi>0$.

\subsection{Lifted vortices and spin winding}
\label{sm:classical-vortices}
With successive spin angles $0,\pi/2,\pi,3\pi/2$ around a plaquette,
all four shortest turns are $+\pi/2$: the spins wind once.
Choosing these four turns as lifted increments gives circulation
$2\pi$ and violates flatness. Setting all bond integers to zero gives
the same spins but last increment $-3\pi/2$ and zero circulation.
Its cost on that edge is $J(3\pi/2)^2/2$. More generally, on a
square with $4R$ boundary steps of size $\pi/(2R)$, subtracting $2\pi$
from one lifted step gives zero total circulation as required by the
plaquettes inside.

\subsection{An exact local heating map}
\label{sm:classical-heating}
For an integer $k\ge2$, take $j_x=\lfloor k\theta_x/(2\pi)\rfloor$,
$\theta'_x=k\theta_x-2\pi j_x$ and
$n'_{x,i}=kn_{x,i}+j_{x+\hat e_i}-j_x$. This range-one map preserves
flatness and gives $a'_{x,i}=ka_{x,i}$. In main Eq.~\eqref{eq:selected-state},
$k\Theta\bmod2\pi$ is uniform, independent of $k\eta$ of covariance
$(k^2T/J)G$. The map therefore heats the bulk state exactly from $T$
to $k^2T$. At finite $L$ it only produces windings divisible by $k$,
whereas the warm Gibbs state includes all windings.
\section{Quantum Villain thermodynamics and correlations}
\label{sm:quantum-details}
The common angle, relative fluctuations and winding give three independent factors in the partition function. The same decomposition determines the equal-time spin correlations.

\subsection{Canonical variables and the partition function}
\label{sm:q-modes}
At fixed $d\ge3$ and $J,\chi,T>0$, use the trace over all flat sectors of main Sec.~\ref{sec:q-thermo}, with $\hbar=k_{\rm B}=1$. For winding $w\in\Z^d$ choose the integer field $n^w$ that equals $w_i$ on the bonds crossing the $i$th periodic seam and zero elsewhere. Every flat field in this sector is $n_{x,i}=n^w_{x,i}+m_{x+\hat e_i}-m_x$. Thus the lifted heights $h_x\coloneqq\theta_x+2\pi m_x$ are defined up to a common integer turn. A core for the translation generators $P_x=-i\partial_{h_x}$ consists of smooth functions periodic under $h\mapsto h+2\pi\one$ and rapidly decreasing in the relative heights; taking the closures gives their self-adjoint realizations. The commutator $[h_x,P_y]=i\delta_{xy}$ holds on the real cover. The globally defined spin and bond commutators in main Sec.~\ref{sec:q-thermo} follow by differentiation. A single turn changes incident bond integers by a gradient, preserving winding. Simultaneous turns act trivially, so the total momentum $Q\coloneqq\sum_xP_x$ has integer eigenvalues.

Subtract the slope $2\pi w\cdot x/L$ from $h_x$ in the coordinate cube $0\le x_i<L$. Denote the mean of the resulting real heights by $\bar h_w$ and their deviations from it by $\eta_x$. Then $\Theta\coloneqq\bar h_w\bmod 2\pi$ and $\pi_x\coloneqq P_x-Q/L^d$ obey
\begin{align}
 [Q,e^{i\Theta}]&=e^{i\Theta},\label{sm:q-rotor-ccr}\\
 [\eta_x,\pi_y]&=i(\delta_{xy}-L^{-d}).\label{sm:q-relative-ccr}
\end{align}
Both relative sums vanish and the relative operators commute with the rotor. In Fourier coordinates with the common convention $L^{-d/2}\sum_x e^{-ik\cdot x}$, $[\eta_k,\pi_{k'}]=i\delta_{k+k',0}$ for $k,k'\ne0$ modulo reciprocal vectors. This gives $L^d-1$ ordinary real canonical pairs in a real normal-mode basis. Since $a_{x,i}=\eta_{x+\hat e_i}-\eta_x+2\pi w_i/L$ and periodic gradients sum to zero,
\begin{equation}
 H_{{\rm q},L}\big|_w
 =\frac{Q^2}{2\chi L^d}+\sum_x\frac{\pi_x^2}{2\chi}
   +\frac J2\sum_{x,i}(\eta_{x+\hat e_i}-\eta_x)^2
   +2\pi^2JL^{d-2}|w|^2.
 \label{sm:q-canonical-H}
\end{equation}
The relative oscillators have frequencies $\omega_k\coloneqq\sqrt{J\lambda(k)/\chi}$. Tracing over rotor momentum, winding and oscillator occupations gives
\begin{equation}
 Z_{{\rm q},L}(T) =\sum_{Q\in\Z}e^{-Q^2/(2\chi L^dT)} \sum_{w\in\Z^d}e^{-2\pi^2JL^{d-2}|w|^2/T} \,\prod_{k\ne0}\bigl(2\sinh(\omega_k/2T)\bigr)^{-1}.
 \label{eq:quantum-partition}
\end{equation}
The rotor contributes $O(\ln L)$ to $\ln Z_{{\rm q},L}$ and the winding sum tends to one. Since $\ln\omega_k=\tfrac12\ln(J\lambda(k)/\chi)$ is integrable, the shell estimate of main Sec.~\ref{sec:free} gives the bulk free energy in main Eq.~\eqref{eq:quantum-free-energy}. To see analyticity, put $z=\omega_k/(2T)$ and write $\ln(2\sinh z)=\ln\omega_k-\ln T+\ln(\sinh z/z)$. The last logarithm has a bounded holomorphic branch on a common neighborhood of each $T>0$ for the entire compact frequency band, including $\omega_k=0$ by continuity.

\subsection{Equal-time correlations and stiffness}

The rotor supplies an independent uniform angle since its momentum eigenfunctions have constant modulus. At fixed direction $\alpha$, spins are $e^{i(\alpha+\eta_x)}$ with Gaussian covariance
\begin{equation}
 C_{\chi,T}(x)\coloneqq\int_k e^{ik\cdot x}\widehat C_{\chi,T}(k),\qquad
 \widehat C_{\chi,T}(k)\coloneqq\frac{\coth(\omega_k/2T)}{2\chi\omega_k}.
 \label{sm:q-C}
\end{equation}
Here the Fourier symbol is the position variance of one thermal oscillator. In the Gaussian identities of Sec.~\ref{sm:classical-correlations}, replace $(T/J)G(x)$ by $C_{\chi,T}(x)$ and $m(T)$ by $m_{\rm q}(T)=e^{-C_{\chi,T}(0)/2}$. This gives the neutral spin correlation $m_{\rm q}^2e^{C_{\chi,T}(x)}$ in both states and the selected transverse and longitudinal connected correlations $m_{\rm q}^2\sinh C_{\chi,T}(x)$ and $m_{\rm q}^2(\cosh C_{\chi,T}(x)-1)$. Only the symmetric connected neutral correlation retains the plateau. The difference $\widehat C_{\chi,T}-T/(J\lambda)$ is analytic near the lattice band, so its Fourier coefficients decay exponentially. Expansion at large distance gives the powers in Table~\ref{tab:model-summary}. At fixed $T$, expansion of $\coth$ gives a uniform $O(1/(\chi T))$ correction to the classical covariance as $\chi\to\infty$.

A twist shifts $2\pi w_1$ by $\vartheta$ in Eq.~\eqref{eq:quantum-partition}. The other factors cancel from the twisted-to-untwisted ratio, giving exactly the classical ratio at every $L$ and hence $\Upsilon(T)=J$. In particular the leading transverse coefficient is $m_{\rm q}^2T/\Upsilon$ times the lattice Green function.

On the torus the analytic correction and its bond differences still decay exponentially in torus distance, and differences cancel the zero mode. For axial bonds separated by $r=\lfloor\sqrt L\rfloor$, the classical estimate in Sec.~\ref{sm:classical-bond-clustering} thus gives a zero-winding increment covariance of order $r^{-d}$ in magnitude. The one-bond variance stays bounded. The Gaussian cosine identity in that section then bounds the alignment covariance above and below by positive multiples of $r^{-2d}$. The nonzero-winding correction for these bounded observables is $O(e^{-2\pi^2JL^{d-2}/T})$ by Eq.~\eqref{eq:quantum-partition}. The resulting order $L^{-d}$ violates the same finite-size exponential clustering bound.
\section{Quantum Villain information from Gaussian packets}
\label{sm:quantum}
The centers of thermal wave packets simulate arbitrary local measurements on separated blocks. The measured spins attain the resulting bound because their position covariance and the center covariance have the same long-wavelength term. For regional reductions we extend the flat Gibbs state by zero to the full tensor product of cells.

\subsection{Thermal wave packets}

Write $C_{\chi,T}$ for the measured position covariance in Eq.~\eqref{sm:q-C}, $A_{\chi,T}$ for its counterpart in a pure packet, and $X_{\chi,T}=C_{\chi,T}-A_{\chi,T}$ for the covariance of the random packet center on nonzero modes.

\begin{lemma}[Thermal wave packets]
\label{sm:q-packet-lemma}
In each winding sector the Gibbs state is a positive mixture of displaced
Gaussian packets, periodized under common $2\pi$ shifts of the heights.
\end{lemma}
\begin{proof}
The thermal oscillator has momentum variance $\widehat P_{\chi,T}(k)
=(\chi\omega_k/2)\coth(\omega_k/2T)$ and zero symmetrized
position--momentum covariance. Choose a real pure Gaussian packet with
the same momentum covariance and the corresponding minimum position
variance. Its position covariance $A_{\chi,T}$ and the covariance
$X_{\chi,T}$ of its random position shift have symbols
\begin{equation}
 \widehat A_{\chi,T}(k)=\frac1{4\widehat P_{\chi,T}(k)},\qquad
 \widehat X_{\chi,T}(k)=\widehat C_{\chi,T}(k)-\widehat A_{\chi,T}(k)
 =\frac1{\chi\omega_k\sinh(\omega_k/T)}\quad(k\ne0).
 \label{sm:q-seed}
\end{equation}
Independent Gaussian displacement adds $X_{\chi,T}$ to the position
covariance and leaves the momentum covariance unchanged, so the
Gaussian characteristic function reproduces the thermal state.
For the zero mode take $\widehat P_{\chi,T}(0)=\chi T$ and
$\widehat A_{\chi,T}(0)=1/(4\chi T)$. Periodizing its packet samples
the squared momentum wavefunction $e^{-p^2/(2\chi T)}$ at
$p=Q/L^{d/2}$. Averaging a uniform common angle kills the off-diagonal
terms in $Q$, giving precisely the rotor weights of
Eq.~\eqref{eq:quantum-partition}. The packet norm is independent
of its displacement. Restoring the slope and averaging the winding
with the weights in that equation completes the construction.
\end{proof}

\subsection{Reduction of separated blocks to classical centers}
\label{sm:q-packets}

\begin{lemma}[Local reduction and periodization]
\label{sm:q-folding-lemma}
Let $B_1,B_2$ be nonwrapping cubes with connected exterior and disjoint
one-layer enlargements $B_i^+$. Put $\mathcal B\coloneqq B_1^+\cup B_2^+$
and denote the set of sites outside $\mathcal B$ by $\mathcal B^c$.
In the zero-winding sector, for any displaced packet the two-cube
reduction is within $K_Te^{-c_T|\mathcal B^c|}$ in trace distance of
the image of its unperiodized $\mathcal B$ reduction under a product
of local channels $\mathcal E_1\otimes\mathcal E_2$.
Here and below $K_T,c_T>0$ are size-independent bounds at fixed
$d,J,\chi,T$ and may change between estimates. Constants in $O(\cdot)$
bounds may depend on these fixed parameters and on any subscripted quantity.
\end{lemma}
\begin{proof}
A cell contains its site's angle and the integers on its outgoing bonds.
Given the heights $h_{B^+}$, write $h_x=\theta_x+2\pi m_x$ with
$0\le\theta_x<2\pi$. In the zero-winding sector
$n_{x,i}=m_{x+\hat e_i}-m_x$. Denote the bonds with tail in $B$ by
$E_B$ and the bonds entering $B$ from outside by $E_{\rm in}$.
Let $h_{\rm collar}$ be the heights on the layer of sites $B^+\setminus B$.
With generalized position kets the coordinate isometry is
\begin{equation}
 W_B|h_{B^+}\rangle
  \coloneqq |\theta_B,n_{E_B}\rangle_{\rm phys}
     \otimes|h_{\rm collar},n_{E_{\rm in}}\rangle_{\rm env}.
\end{equation}
The collar coordinates fix a lift from which the interior integers can be recovered
along bonds. The continuous coordinate changes have unit Jacobian in
Lebesgue coordinates, with the constant normalization for Haar measure
absorbed in $W_B$. Tracing the environment defines
$\mathcal E_B(\rho)\coloneqq \Tr_{\rm env} W_B\rho W_B^\dagger$.
An observable on $B$ acts trivially on bonds entering $B$ from outside.
Keeping those integers in the environment enforces their equality
on the bra and ket sides, including for nondiagonal observables on $B$.

Under a common translation $\mathcal T_\ell$ by $2\pi\ell$, $W_B$ changes only the environment. Thus the lifted observable $\widetilde O=(\mathcal E_1\otimes\mathcal E_2)^*(O)\otimes I_{\mathcal B^c}$ commutes with $\mathcal T_\ell$. In the physical partial trace the exterior angles and all integers on bonds starting outside agree on bra and ket. Connectivity forces their exterior lifts to differ by a single common turn. Restricting one exterior coordinate in $\mathcal B^c$ to $[0,2\pi)$ selects a fundamental domain whose projector commutes with $\widetilde O$. Expand both periodized copies of the packet $g$ on this domain. Group the shifts $m,m'$ by $\ell=m'-m$ and sum over $m$: the translates of the domain tile $\mathbb R^{L^d}$. Normalizing by the same sum for $O=I$ yields
\begin{equation}
 \langle O\rangle_{\rm per}
  =\frac{\sum_{\ell\in\Z}\langle g,\widetilde O\mathcal T_\ell g\rangle}
          {\sum_{\ell\in\Z}\langle g,\mathcal T_\ell g\rangle}.
\end{equation}
For each unit vector $u$, write
$\rho_u^{\mathcal B^c}\coloneqq \Tr_{\mathcal B}|u\rangle\langle u|$ for its exterior marginal.
For unit vectors $a,b$, trace-norm duality
and the purification formula for root fidelity give
\begin{equation}
 \|\Tr_{\mathcal B^c}|a\rangle\langle b|\|_1
  =\|\sqrt{\rho_a^{\mathcal B^c}}\sqrt{\rho_b^{\mathcal B^c}}\|_1.
\end{equation}
Measuring the mean exterior position bounds the root fidelity of the
exterior marginals of $g$ and $\mathcal T_\ell g$ from above. The two Gaussian outcomes have equal
variance $\sigma^2\le\|A_{\chi,T}\|_{\rm op}/|\mathcal B^c|$ and means differing by
$2\pi\ell$. Here $\|A_{\chi,T}\|_{\rm op}=\max_k \widehat A_{\chi,T}(k)$ is the
operator norm of the finite-torus position covariance, including $k=0$.
Their classical root fidelity is
$\exp\bigl(-(2\pi\ell)^2/(8\sigma^2)\bigr)$. Thus every nonzero image in the
numerator is bounded by $\|O\|_{\rm op}$ times this exponential with the
variance upper bound. The same estimate at $O=I$ controls the
normalization. Summing the nonzero images and taking the supremum
over measurement effects $0\le O\le I$ proves the stated error.
\end{proof}

\begin{lemma}[Factorization of a displaced packet]
\label{sm:q-factor-lemma}
With the notation of Lemma~\ref{sm:q-folding-lemma},
let $s$ be the torus distance between the two enlarged regions. The
cover packet on their union differs from the product of its marginals
by at most $K_T|\mathcal B|^2 e^{-c_Ts}$ in trace distance, uniformly
in the displacement.
\end{lemma}
\begin{proof}
The Fourier symbols $\widehat A_{\chi,T}$ and $\widehat P_{\chi,T}$ are analytic functions of
the lattice Laplacian eigenvalue $\lambda(k)$ on a complex neighborhood
of its compact real band. Translating
the momentum contours gives exponential decay of their kernels,
including their periodic images. Theorem~3, Eq.~(4), of Ref.~\cite{Bittel2025}
uses unit vacuum covariance in dimensionless canonical coordinates.
In that convention $V_1$ is twice the two-block packet covariance
used here, and $V_2$ is its block-diagonal part. For the corresponding
Gaussian states $\rho_1,\rho_2$ with equal first moments, the bound gives
\begin{equation}
 \tfrac12\|\rho_1-\rho_2\|_1\le
 \frac{1+\sqrt 3}{8}\max_{i=1,2}\|V_i\|_{\rm op}\,\|V_1-V_2\|_1 .
 \label{sm:q-trace-distance-bound}
\end{equation}
The operator norms are uniformly bounded, and the matrix trace norm
of $V_1-V_2$ is at most the sum of its absolute entries. There are
at most $4|\mathcal B|^2$ entries, each bounded by $K_T e^{-c_Ts}$.
The displacements of the two states coincide.
\end{proof}

To specify the preparation available to a reader, let $g_i^0$ be
the centered packet density restricted to $B_i^+$, and denote a
position displacement there by $\mathcal D_i(\boldsymbol c_i)$. For a center vector $\boldsymbol c_i$
the physical output state is
\begin{equation}
 \Lambda_i(\boldsymbol c_i)\coloneqq \mathcal E_i\bigl(\mathcal D_i(\boldsymbol c_i)g_i^0\mathcal D_i(\boldsymbol c_i)^\dagger\bigr).
\end{equation}
For given centers the product of these maps simulates all the
readers up to the two lemma errors. Each $\Lambda_i$ is invariant under
a common $2\pi$ shift of its own center.

\begin{proposition}[Preparation of two separated bulk blocks]
\label{sm:q-comparison-prop}
Fix the block radius. Take $L\to\infty$ at fixed block positions,
then send the separation to infinity. In trace distance, the joint
state approaches the average of
$\Lambda_1(\boldsymbol c_1)\otimes\Lambda_2(\boldsymbol c_2)$ with
$\boldsymbol c_i=\Theta\one+\boldsymbol\zeta_i$. Here $\Theta$ is the common Haar angle and
$\boldsymbol\zeta_i$ are independent of $\Theta$ and of each other. Each has the law
of the centered bulk Gaussian field of covariance $X_{\chi,T}$ restricted to $B_i^+$.
The packet reductions defining $\Lambda_i$ are now those of the bulk.
\end{proposition}
\begin{proof}
At fixed block size $X_{\chi,T}$ has an integrable $T/(J\lambda(k))$ singularity plus a bounded analytic remainder by Eqs.~\eqref{sm:q-C} and~\eqref{sm:q-seed}. Dyadic shells with $0<|k|<\epsilon$ contribute $O(\epsilon^{d-2})$ to the normalized sum, so torus covariances converge entrywise. Their limit is strictly positive on each finite block because $\widehat X_{\chi,T}>0$ almost everywhere. Hence its Gaussian center law converges in total variation. The packet marginals converge in trace norm by exponential decay of the periodic images and Eq.~\eqref{sm:q-trace-distance-bound}. The folding error and nonzero-winding weight vanish by Lemma~\ref{sm:q-folding-lemma} and Eq.~\eqref{eq:quantum-partition}.

Now send the separation to infinity at fixed $\Theta$. Integrability gives vanishing cross covariances for the centers, hence convergence in total variation to independent block fluctuations. Lemma~\ref{sm:q-factor-lemma} factorizes the remaining packet fluctuations. Local preparation and averaging over $\Theta$ give the asserted trace-distance limit, uniformly over measurements. Before the angle average, the same argument gives clustering for all bounded local observables.
\end{proof}

\subsection{Block precision and mutual information}
\label{sm:q-readers}

For either bulk covariance $\Sigma=X_{\chi,T},C_{\chi,T}$ put $P_\Sigma(R)\coloneqq\one^T\Sigma_{B_R}^{-1}\one$. A block vector $\boldsymbol c$ of mean $\alpha\one$ and covariance $\Sigma_{B_R}$ has sufficient scalar $\one^T\Sigma_{B_R}^{-1}\boldsymbol c/P_\Sigma(R)$ with law $\mathcal N(\alpha,1/P_\Sigma(R))$. Subtracting that scalar leaves an independent Gaussian residual whose law is independent of $\alpha$. For the centers, $\Sigma=X_{\chi,T}$ on the one-layer enlargement. Invariance of $\Lambda_i$ under a common turn makes its output a randomized function of the scalar modulo $2\pi$.

\begin{lemma}[Precision of a cubic block]
\label{sm:q-precision-lemma}
Let $\Sigma$ be the bulk position or center covariance. For
$B_R$ the inverse variance of the optimal unbiased
linear estimate satisfies Eq.~\eqref{sm:q-cap} at fixed $J,\chi,T>0$.
The coefficient is the capacity of the continuum cube.
\end{lemma}
\begin{proof}
We prove the two sides of
\begin{equation}
 P_{\Sigma}(R)=\frac{J}{T}\operatorname{cap}(B_R)(1+o(1))
 =\frac{J}{T}\kappa_d R^{d-2}(1+o(1)).
 \label{sm:q-cap}
\end{equation}
The covariance symbols in Eqs.~\eqref{sm:q-C} and~\eqref{sm:q-seed}
obey $\widehat\Sigma=T/(J\lambda)+O(1)$ and
$\widehat\Sigma^{-1}=J\lambda/T+O(\lambda^2)$ on the lattice band.
The potential bound of Sec.~\ref{sm:capacity-details} uses the second
expansion and gives the upper bound on $P_\Sigma$; its smooth-charge
bound uses the first expansion and gives the lower bound.
Both apply to these positive covariance symbols and give
Eq.~\eqref{sm:q-cap}, with coefficient $J/T$ for the real angle.
For the spin reader we also need an attaining profile. Use the
smoothed equilibrium-charge weights of Sec.~\ref{sm:classical-spin-error},
with $b_R=\lfloor R/\ln R\rfloor$. Their Green energy is
$1/(\kappa_d R^{d-2})(1+o(1))$ and
$\sum_{x\in B_R}w_x^2\le1/|B_{b_R}|=o(R^{2-d})$. The bounded covariance
correction consequently gives
\begin{equation}
 P_{\Sigma}(R)^{-1}\le \boldsymbol w^T\Sigma_{B_R}\boldsymbol w
 =\frac{T}{J\kappa_d R^{d-2}}(1+o(1)).
 \label{sm:q-profile}
\end{equation}
\end{proof}

\begin{theorem}[Information from quantum and spin measurements]
\label{sm:q-information-theorem}
At fixed $d\ge3$ and $J,\chi,T>0$, let each reader have at most
$N_R=\lceil R^\gamma\rceil$ outputs with $\gamma>(d-2)/2$.
Spin-only measurements attain the constant term obtained by optimizing
the mutual information over all pairs of local quantum measurements.
\end{theorem}
\begin{proof}
After separation the two sufficient center angles are uniform, with a
wrapped Gaussian difference of variance parameter $2/P_{X_{\chi,T}}(R+1)$.
Local preparation, data processing, and the wrapped-Gaussian entropy
of Sec.~\ref{sm:wrapped-entropy} bound the information of every reader pair
by $\tfrac12\ln(\pi P_{X_{\chi,T}}(R+1)/e)+o(1)$.
Trace-distance convergence controls the optimum because the output
alphabet is finite at fixed $R$.

For the lower bound measure the direction of the weighted spin sum
in main Eq.~\eqref{eq:spin-reader}, with the same fallback and
equal bins as in the classical model. Given $\Theta$, let $\epsilon_R$
be its principal angular error and $X_R=\sum_xw_x\eta_x$ the linear
error of the measured positions with covariance $C_{\chi,T}$.
Since $|C_{\chi,T}(x)|\le K_T(1+|x|)^{2-d}$, the Gaussian
cancellation and square-sum estimate of Sec.~\ref{sm:classical-spin-error}
also give a mean-square difference between $\epsilon_R$ and $X_R$ of
$O(R^{1-d}(\ln R)^d+R^{4-2d})=o(\Var X_R)$.
The two blocks' errors are independent of one another and of $\Theta$
in the separated limit. As $2\pi/N_R=o(\sqrt{\Var X_R})$,
main Eq.~\eqref{eq:bin-information} and Eq.~\eqref{sm:q-profile}
give the matching lower bound. Together with Eq.~\eqref{sm:q-cap} this yields
\begin{align}
 M_T(R)&=\frac{d-2}{2}\ln R+\frac12\ln\frac{\pi\kappa_dJ}{eT}+o(1),
 \label{sm:q-information-growth}\\
 \cC(T)&=\frac12\ln\frac JT+\frac12\ln\frac{\pi\kappa_d}{e}.
 \label{sm:q-info}
\end{align}
Monotonicity follows from main Eq.~\eqref{eq:monotone} with
$s_R=o(R)$ and trace-distance error $\delta_R=o(1/\ln R)$.
\end{proof}
\section{Coupled multicomponent Villain derivations}
\label{sm:multicomponent-details}

\subsection{Thermodynamics and correlations}

Unfold the $n$ periodic angles of main Sec.~\ref{sec:multicomponent} to vector heights modulo a common shift in $2\pi\mathbb Z^n$. Each of the $L^d-1$ nonzero real modes has quadratic form $\lambda(k)K$ and Gaussian integral $[\lambda(k)^n\det(2\pi K/T)]^{-1/2}$. The normalized constant coordinates have period lattice $2\pi L^{d/2}\mathbb Z^n$, giving a factor $L^{nd/2}$. A winding $\mathbf w_i\in\mathbb Z^n$ around cycle $i$ is orthogonal to the periodic gradients. Multiplying the factors gives
\begin{equation}
 Z_L(T)=L^{nd/2}\,
 \frac{\displaystyle\sum_{\mathbf w\in\mathbb Z^{n\times d}}
 e^{-(2\pi^2 L^{d-2}/T)\sum_i\mathbf w_i^{\mathsf T}K\mathbf w_i}}
 {\displaystyle\det(2\pi K/T)^{(L^d-1)/2}
 \prod_{k\ne0}\lambda(k)^{n/2}} .
 \label{eq:multi-partition}
\end{equation}
For fixed $K>0$ and $T>0$, the winding sum is $1+O(e^{-cL^{d-2}})$ by the smallest eigenvalue of $K$. Thus $L^{-d}\sum_{k\ne0}\ln\lambda(k)\to\ell_d$ gives the analytic free energy in main Eq.~\eqref{eq:multi-free}. A twist adds $\boldsymbol\vartheta$ to $2\pi\mathbf w_1$. Differentiating the winding sum before taking the limit gives $\boldsymbol\Upsilon=K$, since polynomial moments of the nonzero windings are exponentially suppressed.

The thermodynamic limit of the symmetric periodic Gibbs laws has a
uniform direction $\boldsymbol\Theta$ on
$\mathbb T^n\coloneqq(\R/2\pi\Z)^n$ and an independent centered
Gaussian field $\boldsymbol\eta$:
\begin{align}
 \boldsymbol\theta_x&=(\boldsymbol\Theta+\boldsymbol\eta_x)
                            \bmod 2\pi\Z^n,
 \label{eq:multi-state}\\
 \langle\eta_x^a\eta_y^b\rangle_{\boldsymbol\Theta}
     &=T(K^{-1})_{ab}G(x-y).
 \label{eq:multi-covariance}
\end{align}
At fixed $\boldsymbol\Theta$, the component magnetizations are $m_a(T)e^{i\Theta^a}$ with $m_a(T)=\exp[-TG(0)(K^{-1})_{aa}/2]>0$. All directions are distinct and cluster because $G(x)\to0$. Applying the sine and cosine identities of Sec.~\ref{sm:classical-correlations} to components $a$ at the origin and $b$ at $x$ gives the transverse and longitudinal connected correlations
\begin{align}
 C_\perp^{ab}(x)
 &=m_a m_b\sinh\!\bigl[T(K^{-1})_{ab}G(x)\bigr],
 \label{eq:multi-transverse}\\
 C_\parallel^{ab}(x)
 &=m_a m_b\Bigl\{\cosh\!\bigl[T(K^{-1})_{ab}G(x)\bigr]-1\Bigr\}.
 \label{eq:multi-longitudinal}
\end{align}
At fixed $T>0$, for $a=b$, the Green-function tail and these formulas give
Table~\ref{tab:model-summary}; $(K^{-1})_{aa}>0$ ensures both leading
coefficients are nonzero.
The symmetric state averages over $\boldsymbol\Theta$; its neutral
same-component spin correlation retains the plateau $m_a^2$.

\subsection{Information in the vector of angles}
For a block $B=B_R$, apply the scalar equilibrium charges $q_x^B$ of
main Eq.~\eqref{eq:equilibrium-charge-components} to each component.
Their sum is one, so the weighted lifted height is well defined modulo the period lattice:
\begin{align}
 \mathbf U_B&\coloneqq\sum_{x\in B}q_x^B\mathbf h_x
     =\boldsymbol\Theta+\boldsymbol\xi_B\pmod{2\pi\mathbb Z^n},
 \label{eq:multi-statistic}\\
 \operatorname{Cov}(\boldsymbol\xi_B)&=TK^{-1}/\operatorname{cap}(B).
 \label{eq:multi-error}
\end{align}
For each $x\in B$, $\operatorname{Cov}(\boldsymbol\eta_x,\boldsymbol\xi_B)$
is the same matrix by Eq.~\eqref{eq:multi-covariance}. The Gaussian
residuals $\boldsymbol\eta_x-\boldsymbol\xi_B$ are therefore independent of
the weighted angle and the shared direction. Together with $\mathbf U_B$
they reconstruct the internal data. At infinite separation the residuals
are private, so data processing bounds every pair of readers by
$I(\mathbf U_B\infosep\mathbf U_{B'})$ for a distant translate $B'$.

The two angle errors become independent since $G(x)\to0$. Their unwrapped
difference $\mathbf Y$ has covariance $2TK^{-1}/\operatorname{cap}(B)$
and is independent of the uniform angle $\mathbf U_B$. Thus
\begin{align}
 I(\mathbf U_B\infosep\mathbf U_{B'})
 &= n\ln(2\pi)-S(\mathbf Y\bmod 2\pi\mathbb Z^n)
 \label{eq:multi-wrapped}\\
 &= \frac n2\ln\operatorname{cap}(B)+\frac12\ln\det(K/T)
     +\frac n2\ln(\pi/e)+o(1).
 \label{eq:multi-information-growth}
\end{align}
Here entropy uses angular Lebesgue measure. Write $Y_a=\bar Y_a+2\pi j_a$,
where $\bar Y_a\in[-\pi,\pi)$ is the wrapped angle and $j_a\in\mathbb Z$
counts the turns removed. By the Gaussian tail estimate of Sec.~\ref{sm:wrapped-entropy},
the entropy lost in wrapping is at most $\sum_a S(j_a)=o(1)$.

To attain the bound with at most $\lceil R^\gamma\rceil$ outputs, choose
$n(d-2)/2<\gamma'<\gamma$ and a grid of width $h_R$ proportional to
$R^{-\gamma'/n}$ in each angular coordinate. Conditional difference laws
from two points in a cell have relative entropy at most
$\operatorname{cap}(B)\,\boldsymbol\delta^{\mathsf T}K\boldsymbol\delta/(4T)$:
this is the unwrapped Gaussian bound for a displacement
$|\boldsymbol\delta|\le\sqrt n h_R$. Convexity and the chain rule bound
the information lost on binning both outputs by twice the supremum of this
bound, which tends to zero. The available cells of anchored spins and bonds
contain the internal data of $B_R$ and are contained in those of $B_{R+1}$.
Their capacity ratio thus gives the same constant in main
Eqs.~\eqref{eq:multi-invariant} and~\eqref{eq:multi-geometry}.

A real rotation of components generally changes the period lattice. To
preserve both the standard periods and the integer bonds, a basis change
$\boldsymbol\theta=A\boldsymbol\psi$ must have $A\in GL(n,\mathbb Z)$,
in which case the coupling becomes $A^{\mathsf T}KA$.
\section{General order spaces: the information constant}
\label{sm:general-details}

\subsection{Information from the stiffness metric}
On the compact orbit $\mathcal M=G/H$ of dimension $n$, a uniform tilt
$\vartheta^a/L$ costs $\tfrac12L^{d-2}\sum_{a,b}\Upsilon_{ab}\vartheta^a\vartheta^b$
in the quadratic gradient energy. This gives the normalization of main
Eq.~\eqref{eq:coset-twist-metric}. At a stationary twist the Hessian
transforms as a tensor, and invariance under $H$ makes its extension by
$G$ unique. Use $g_T=\boldsymbol\Upsilon/T$ and its volume measure below.

In the tangent Gaussian theory the covariance is $G(x-y)g_T(q)^{-1}$.
A block average with the equilibrium charges therefore has covariance
$g_T(q)^{-1}/\operatorname{cap}(B_R)$. To model this precision on the
orbit, choose $q$ uniformly in metric volume. Given $q$, draw $Y,Y'$
independently from the heat kernel $k_{s/2}(q,\cdot)$ of $\Delta_{g_T}$,
where $s=1/\operatorname{cap}(B_R)$. Each draw has leading tangent
covariance $s g_T(q)^{-1}$. Reversibility and the semigroup law make
the conditional density of $Y'$ given $Y=y$ exactly $k_s(y,\cdot)$.
Both marginals are uniform. Subtracting the short-time conditional
entropy $\tfrac n2\ln(4\pi e s)+o(1)$~\cite{Grigoryan2009,Menkovski2024}
from the uniform entropy gives
\begin{equation}
 I(Y\infosep Y')=\ln\operatorname{Vol}(\mathcal M,g_T)
              +\frac n2\ln\frac{\operatorname{cap}(B_R)}{4\pi e}+o(1).
 \label{eq:coset-heat-information}
\end{equation}
For a related asymptotic for repeated independent samples, see
Ref.~\cite{ClarkeBarron1990}. Finite outputs attain the same constant.
For $n(d-2)/2<\gamma'<\gamma$, partition the orbit into $O(R^{\gamma'})$
cells of diameter $h_R=O(R^{-\gamma'/n})$. Convexity of relative entropy
bounds the information lost by binning one output by the largest
relative entropy between conditional laws started in the same cell,
$O(h_R^2/s)$~\cite{Wang2010}. The chain rule bounds the loss for both
outputs by twice this quantity, which vanishes. The cell count also fits
the budget $\lceil R^\gamma\rceil$. Since the metric-volume ratio
$\sqrt{\det_{g_0}g_T(q)}$ is constant on the orbit, the capacity limit
in main Eq.~\eqref{eq:capacity-limit} gives the conjectured determinant formula.

\subsection{When a microscopic magnet has this constant}
First require the fixed-block thermodynamic limit at infinite separation
to be the invariant mixture of two independent block states conditional on
$q$, uniformly in the centers (including any known translation action on $q$).
Second, on each block require both a measurement that produces the ideal
heat-kernel draw from the selected state and a preparation from the draw
that reproduces the full selected state after averaging. Each map must
work for every unknown $q$, with error at most $\delta_R=o(1/\ln R)$
uniformly in direction and position, in total variation for classical
laws or trace distance for quantum states. The measurement and binning
give the lower bound; preparing and applying an arbitrary reader gives
the upper bound. The two-block error is at most $2\delta_R$, so finite-label
entropy continuity in main Eq.~\eqref{eq:reader-continuity} changes
the information by $o(1)$. This proves the formula under these conditions.

For a collinear Heisenberg antiferromagnet with labeled sublattices, the
order space of the unit N\'eel vector is the sphere $SO(3)/SO(2)$. Its
area $4\pi$ in the unit round metric gives
$\cC(T)=\ln(\Upsilon(T)/T)+\ln(\kappa_d/e)$ with the transverse stiffness.
The same prediction holds for a Heisenberg ferromagnet: its two static
tilts form a single dynamical type-B mode~\cite{WatanabeMurayama2014}.
The microscopic comparisons remain open for these magnets and the
unconstrained XY model.

\subsection{Additivity for independent layers}
\label{sm:mi-stacking-proof}

Take the product of two condensed canonical Bose gases at temperature $T$,
fixing the particle number in each gas separately. The two condensate
phases are independent. Use the same two blocks in both layers. After
the thermodynamic limit, keep their radius fixed and take their separation
to infinity. Let $Z_i,Z_i'$ be the lists of coherent
amplitudes preparing the two blocks of layer $i=1,2$. Given
$(Z_1,Z_2)$, the first observer can prepare both local coherent states
and perform any joint quantum measurement on them, and likewise for
the second observer. Thus for their outputs $U,V$, data processing and
independence of the two layers give
\begin{equation}
 I(U\infosep V)\le I((Z_1,Z_2)\infosep(Z_1',Z_2'))
 =I(Z_1\infosep Z_1')+I(Z_2\infosep Z_2').
 \label{sm:eq:stacked-reader-upper}
\end{equation}
By the single-layer calculation of main Sec.~\ref{sec:bose-upper-bound},
each term is $(d-2)\ln R/2+\cC_i(T)+o(1)$. A classical Villain layer
can replace either gas: its full local data supply the same upper
bound with its own constant (main Sec.~\ref{sec:full-information}).

For the lower bound let each observer measure the layers separately and
report the pair of binned angles. For a total budget $\lceil R^\gamma\rceil$
with $\gamma>d-2$, choose $\gamma_i>(d-2)/2$ with
$\gamma_1+\gamma_2<\gamma$. The product of the two bin counts
$\lceil R^{\gamma_i}\rceil$ fits the budget for large $R$.
Each measurement attains its single-layer constant, and the two
independent pairs of outputs have exactly additive mutual information.
The bounds therefore match through $o(1)$. Subtracting $(d-2)\ln R$
gives $\cC_{1\otimes2}(T)=\cC_1(T)+\cC_2(T)$, even with joint readers.
For independent general order spaces of dimensions $n_i$, the same
argument applies under the two comparison hypotheses above: take the
product of the preparation and measurement maps and choose
$\gamma_i>n_i(d-2)/2$ with $\gamma_1+\gamma_2<\gamma$. The simulation
errors add and remain $o(1/\ln R)$.
\section{Canonical Bose states on two growing blocks}
\label{sm:bose}

We prove a finite-volume approximation for every block radius $R=o(L)$,
including fixed radii. The resulting trace-distance bound controls arbitrary
measurements on the regional Fock spaces without an occupation cutoff.
Fix $J>0$, $\varrho>0$, and $0<T<T_c$, and set
$V\coloneqq L^d$, $N\coloneqq\lfloor\varrho V\rfloor$ and
$n_T(k)\coloneqq(e^{\epsilon(k)/T}-1)^{-1}$.
Constants may depend on these fixed parameters, not on $L,R$.
For a positive matrix $C$ on $r$ complex coordinates, write
$\nu_C(\dd z)\coloneqq e^{-z^\dagger C^{-1}z}\dd^{2r}z/(\pi^r\det C)$;
$\dd^{2r}z$ integrates their real and imaginary parts. The trace
distance is one half the trace norm. Define
\begin{equation}
 a_L\coloneqq\begin{cases}
 L^{-1},&d=3,\\
 L^{-2}\sqrt{\ln L},&d=4,\\
 L^{-d/2},&d\ge5,
 \end{cases}
 \qquad p_L\coloneqq a_L^2.
 \label{eq:bose-error-scales}
\end{equation}

\begin{lemma}[Regional canonical approximation]
\label{sm:bose-block}
Let $B_1,B_2$ be cubes of radius $O(R)$ separated by a fixed positive
fraction of $L$, and put $B\coloneqq B_1\cup B_2$. Define
\begin{equation}
 \sigma_{T;B_1,B_2}\coloneqq
 \int_0^{2\pi}\frac{\dd\Theta}{2\pi}
 \bigotimes_{j=1}^2\left(
 \int\nu_{C_{T,B_j}}(\dd\xi_j)\bigotimes_{x\in B_j}
 |\sqrt{\varrho_0(T)}e^{i\Theta}+\xi_{j,x}\rangle
 \langle\sqrt{\varrho_0(T)}e^{i\Theta}+\xi_{j,x}|\right).
 \label{sm:eq:bose-target}
\end{equation}
The two circular complex Gaussian noises are independent; their
covariances are $C_{T,B_j}$, and their nonconjugated second moments
vanish. The common uniform phase is retained, and $|z\rangle$ denotes
a normalized coherent state. For $1\le R\le cL$, with a sufficiently
small fixed $c>0$,
\begin{equation}
 \frac12\left\|(\rho_L(T))_B-\sigma_{T;B_1,B_2}\right\|_1
 \le C\left[(R/L)^{d/2}+a_LR^{(d-2)/2}+p_L+(R/L)^{d-2}\right].
 \label{sm:eq:bose-regional-error}
\end{equation}
Every term vanishes for every sequence $R=o(L)$.
\end{lemma}

\subsection{An exact sampler and the condensate radius}

Let $\Delta$ and $\Delta_L$ be the positive nearest-neighbor lattice
Laplacians on $\mathbb Z^d$ and the periodic torus, respectively, with
Fourier symbol $\epsilon(k)/J$. The finite-volume one-particle
Hamiltonian is $h_{{\rm sp},L}\coloneqq J\Delta_L$.
The finite thermal covariance, with its zero mode omitted, is
\begin{equation}
 (C_{T,L})_{xy}\coloneqq V^{-1}\sum_{k\ne0}n_T(k)e^{ik\cdot(x-y)}.
 \label{sm:eq:bose-finite-C}
\end{equation}
Shell counting near $k=0$ and periodic Green-kernel bounds give
\begin{equation}
 V^{-1}\sum_{k\ne0}n_T(k)=\varrho_{\rm ex}(T)+O(L^{2-d}),
 \qquad \sum_{k\ne0}n_T(k)^2=O(V^2p_L).
 \label{sm:eq:bose-spectral}
\end{equation}
Indeed $n_T(2\pi n/L)\le CL^2/|n|^2$ for nonzero momentum representatives,
and $C_{T,L}=(T/J)G_L+r_T(J\Delta_L)-r_T(0)P_0$,
where $G_L\coloneqq\Delta_L^+$ is the inverse on the zero-mean
subspace, extended by zero on constants,
$P_0\coloneqq\one\one^{\mathsf T}/V$ projects onto the constant mode,
and $r_T(E)\coloneqq(e^{E/T}-1)^{-1}-T/E$ is analytic across zero.
The Green comparison is proved below; the analytic remainder has
exponentially decaying matrix elements. In particular,
\begin{align}
 |(C_{T,L})_{xy}|&\le C[(1+d_L(x,y))^{2-d}+L^{2-d}],
 \label{sm:eq:bose-kernel}\\
 |(C_{T,L})_{xy}-C_T(x-y)|&\le CL^{2-d}\quad(|x-y|\le L/4),
 \label{sm:eq:bose-kernel-difference}
\end{align}
where $d_L$ is periodic distance. For later use, the independent
geometric occupations $\mathbf m=(m_k)_{k\ne0}$ have law
$p_{T,L}(\mathbf m)\coloneqq\prod_{k\ne0}(1-e^{-\epsilon(k)/T})e^{-m_k\epsilon(k)/T}$.
Their total is $m_{\rm ex}\coloneqq\sum_{k\ne0}m_k$; the corresponding
random variable $N_{\rm ex}$ has mean
$\overline N_{{\rm ex},L}=\varrho_{\rm ex}V+O(L^2)$ and variance $O(V^2p_L)$.

An exact positive mixture converts these estimates to quantum ones:
\begin{equation}
 \rho_L(T)=\int |N,u\rangle\langle N,u|\,\nu_{L,N}(\dd u),
 \qquad |N,u\rangle\coloneqq
 \frac{(\sum_xu_xb_x^\dagger)^N}{\sqrt{N!}}|0\rangle.
 \label{sm:eq:bose-hartree}
\end{equation}
Here $\nu_{L,N}$ is obtained by weighting a Gaussian of covariance
$e^{-h_{{\rm sp},L}/T}$ by its norm to the power $2N$ and normalizing
its direction. Gaussian pairings give Eq.~\eqref{sm:eq:bose-hartree}.
The following simpler sampling rule is exact. Independently draw
$z\sim\nu_{C_{T,L}}$ on the mean-zero complex subspace,
$W\ge0$ with gamma density $\gamma_N(\dd W)\coloneqq W^Ne^{-W}\dd W/N!$, and a
uniform $\Theta$. Condition on $\mathcal A\coloneqq\{W\ge s\}$,
where $s\coloneqq\|z\|^2$, and set
\begin{equation}
 \zeta\coloneqq z+\sqrt{(W-s)/V}\,e^{i\Theta}\one,
 \qquad u\coloneqq\zeta/\sqrt W.
 \label{sm:eq:bose-exact-sampler}
\end{equation}
To verify this, let $y$ be the zero-mode intensity in the norm-weighted
Gaussian. Its density is proportional to
$\nu_{C_{T,L}}(\dd z)(s+y)^Ne^{-(s+y)}\dd y\,\dd\Theta/(2\pi)$.
The substitution $W=s+y$ gives precisely the stated conditioned law.
Equation~\eqref{sm:eq:bose-spectral} and $\int(W-N)^2\gamma_N(\dd W)=N+2$
imply $\Pr(\mathcal A^c)=O(p_L)$ by the positive condensate-density gap.

Put $m\coloneqq|B|$. The regional Hartree state and the phase-averaged
coherent state with center $\sqrt N u_B$ occupy the same orbital, with
number laws $\operatorname{Bin}(N,p_u)$ and $\operatorname{Pois}(Np_u)$,
where $p_u\coloneqq\sum_{x\in B}|u_x|^2$. Here $\operatorname{Bin}(N,p)$
is the binomial law for $N$ trials of success probability $p$, and
$\operatorname{Pois}(Np)$ is the Poisson law of mean $Np$.
Writing $d_{\rm TV}$ for total variation distance, the sharp Poisson bound
\cite{BarbourHall1984} gives
\begin{equation}
 d_{\rm TV}(\operatorname{Bin}(N,p),\operatorname{Pois}(Np))
 \le p(1-e^{-Np})\le p.
 \label{sm:eq:bose-poisson-sharp}
\end{equation}
Translation invariance gives $\int p_u\nu_{L,N}(\dd u)=m/V$.
Thus the coherent-mixture replacement costs at most $m/V$; its global
phase is already uniform. Next replace $\sqrt{N/W}\,\zeta_B$ by
$\zeta_B$. The conditional average squared displacement is
$(m/V)$ times the conditional average of $(\sqrt W-\sqrt N)^2$,
which is $O(m/V)$. Since the trace distance of coherent states is
at most their center displacement, this costs $C\sqrt{m/V}$.
Extend the center using $\sqrt{(W-s)_+}$, where $(t)_+\coloneqq\max(t,0)$,
and remove the conditioning at cost $O(p_L)$.
On $s\le N-c_0V$, for fixed $0<c_0<\varrho_0(T)$,
\begin{equation}
 \left|\sqrt{(W-s)_+}-\sqrt{N-s}\right|^2
 \le\frac{(W-N)^2}{c_0V}.
 \label{sm:eq:bose-freeze-W}
\end{equation}
The complementary event has probability $O(p_L)$, so replacing $W$
by $N$ costs another $C\sqrt{m/V}+Cp_L$.

The remaining radius depends on the same field $z$ as the thermal
noise; it must not be replaced by an independent variable. Choose a
smooth mean-zero real vector $v$ equal to one on both blocks, with
$\|v\|^2\le CR^d$ and
$v^\dagger C_{T,L}^{-1}v\le CR^{d-2}$. A cutoff on doubled blocks,
minus its mean and then normalized, has these properties:
$C_{T,L}^{-1}=e^{h_{{\rm sp},L}/T}-1\le C h_{{\rm sp},L}$ on the
mean-zero subspace. Smoothly extend
$\delta(z)=\sqrt{(N-s)/V}-\sqrt{\varrho_0(T)}$ below
$(N-s)/V=\varrho_0/2$, making it constant below $\varrho_0/4$.
The change costs $O(p_L)$; the extension has derivative in $s$ bounded
by $C/V$, supported in $s\le CV$, and
$\int\delta^2\nu_{C_{T,L}}(\dd z)\le Ca_L^2$.
For fixed $\Theta$ define
$\mathcal T_\Theta(z)\coloneqq z+e^{i\Theta}v\delta(z)$ and
$j_\Theta\coloneqq\partial_{e^{i\Theta}v}\delta$, with the derivative
in the indicated real direction. Its Jacobian is $1+j_\Theta$ and
$\|j_\Theta\|_\infty\le C\sqrt{m/V}<1/2$ for small $R/L$.
Along each such line $\mathcal T_\Theta$ is an onto diffeomorphism.
Gaussian integration by parts cancels the linear displacement term,
giving the exact identity
\begin{equation}
 \begin{aligned}
 D((\mathcal T_\Theta)_*\nu_{C_{T,L}}\relentsep \nu_{C_{T,L}})
 &=\int\left[(v^\dagger C_{T,L}^{-1}v)\delta^2
       +j_\Theta-\ln(1+j_\Theta)\right]\nu_{C_{T,L}}(\dd z)\\
 &\le C\left(R^{d-2}a_L^2+m/V\right).
 \end{aligned}
 \label{sm:eq:bose-radius-transport}
\end{equation}
Here $D$ is classical relative entropy and the star denotes the
transported probability law. Pinsker's inequality and coherent-state
preparation turn this into a trace-distance cost. Integrating the same
$\Theta$ preserves the common phase. The total cost before changing
the thermal covariance is therefore
\begin{equation}
 C\left[(R/L)^{d/2}+a_LR^{(d-2)/2}+p_L\right].
 \label{sm:eq:bose-radius-error}
\end{equation}

\subsection{A common two-block Gaussian estimate}

The following estimate also applies to the Villain fields.
\begin{lemma}[Separated Gaussian restrictions]
\label{sm:two-block-gaussian}
Let $C^\infty=aG+b(\Delta)$ and
$C^L=aG_L+b(\Delta_L)-b(0)P_0$, where $a>0$, $b$ is analytic near
$[0,4d]$, and $C^\infty\ge cG$ for some $c>0$ as quadratic forms.
For the two blocks above, their joint real or circular complex
Gaussian law of covariance $(C^L)_B$ and the product law of covariances
$C^\infty_{B_1},C^\infty_{B_2}$ obey
\begin{equation}
 d_{\rm TV}\le C(R/L)^{d-2}
 \label{sm:eq:gaussian-two-block-tv}
\end{equation}
for sufficiently small $R/L$. This includes $C^\infty=G$ and the
Bose thermal covariance.
\end{lemma}

\begin{proof}
Put $G_0\coloneqq\operatorname{diag}(G_{B_1},G_{B_2})$,
$A\coloneqq\operatorname{diag}(C^\infty_{B_1},C^\infty_{B_2})$, and
$B_L\coloneqq(C^L)_B-A$. Its values, first differences and mixed
second differences are bounded by $M_j\coloneqq CL^{2-d-j}$,
$j=0,1,2$. For $G_L-G$, the heat-kernel representations of the two
Green functions give the value bound by splitting the time integral
at $L^2$~\cite{AruKorzhenkova2025}: below this time, periodic images
and the subtracted zero mode contribute $O(L^{2-d})$; above it,
the infinite-volume heat-kernel tail and the finite-volume
exponential relaxation contribute the same order. Moreover
$\Delta(G_L-G)=-V^{-1}$, so subtracting $|x|^2/(2dV)$ makes the
difference harmonic on a macroscopic neighborhood. Interior
harmonic difference estimates give the two derivative bounds
\cite[Theorem~6.3.8]{LawlerLimic2010}. The same reasoning applies to
$G_L$ between the distant blocks. The analytic remainder has
exponentially small image terms; its omitted zero mode is constant.

The inverse restricted Green matrix is a Schur complement of a
Dirichlet Laplacian in increasing boxes. It has nonpositive
off-diagonal entries and nonnegative row sums, and hence
\begin{equation}
 G_{B_j}^{-1}=\sum_x\beta_x e_xe_x^{\mathsf T}
 +\sum_{x<y}\omega_{xy}(e_x-e_y)(e_x-e_y)^{\mathsf T},
 \qquad \beta_x,\omega_{xy}\ge0.
 \label{sm:eq:green-inverse-decomposition}
\end{equation}
Here $e_x$ is a coordinate vector, and each sum is within its block.
The Dirichlet variational principle, applied to a cutoff equal to one
on the block and to cutoff coordinate functions centered there, yields
$\sum\beta_x\le CR^{d-2}$ and
$\sum\omega_{xy}|x-y|_1^2\le CR^d$ over both blocks.
The notation $|x-y|_1$ is the sum of the coordinate distances.
Expanding the squared matrix norm using this nonnegative decomposition
and telescoping the first and mixed differences gives
\begin{equation}
 \begin{aligned}
 \|G_0^{-1/2}B_LG_0^{-1/2}\|_{\rm F}^2
 &\le C\left(M_0^2R^{2d-4}+M_1^2R^{2d-2}+M_2^2R^{2d}\right)\\
 &\le C(R/L)^{2(d-2)}.
 \end{aligned}
 \label{sm:eq:green-relative-error}
\end{equation}
The Frobenius norm is the square root of the sum of squared entries.
Since $A\ge cG_0$, multiplication on both sides by
$A^{-1/2}G_0^{1/2}$ transfers this bound to
$E_L\coloneqq A^{-1/2}B_LA^{-1/2}$ without a block-volume factor.
For small $R/L$, its operator norm is below $1/2$.
The complex Gaussian relative entropy is
$\Tr[E_L-\ln(I+E_L)]\le C\|E_L\|_{\rm F}^2$; the real case has an
additional factor $1/2$. Pinsker's inequality proves the lemma.
\end{proof}

For Bose fields $a=T/J$, $b(\lambda)=r_T(J\lambda)$, and
$C_T\ge[4d/(e^{4dJ/T}-1)]G$. Adding the common phase and preparing
coherent states contracts total variation to quantum trace distance.
Combining the lemma with Eq.~\eqref{sm:eq:bose-radius-error} proves
Eq.~\eqref{sm:eq:bose-regional-error}. No uniformity at $T=0$ or $T_c$
is required.

\subsection{Bounded symmetric probes violate exponential clustering}
\label{sm:bose-bounded-clustering}

On a finite condensed torus, take the norm-one empty-site projector $\Pi_x\coloneqq|0\rangle_x\langle0|$. It commutes with the on-site number and belongs to the symmetric observable algebra of case (ii) in Ref.~\cite{MKS2026}, as can be checked with finite matrices. The total-$N$ space embeds exactly into the tensor product of on-site spaces spanned by $|0\rangle,\ldots,|N\rangle$. The algebra $\mathcal A$ generated by local $U(1)$-invariant operators contains the compressed $H_L$ and $\widehat N$, hence also $P_N$ and the canonical state by finite-dimensional spectral calculus. The algebra on $X$ is generated by partial traces of $\mathcal A$ over $X^c$. With $I_Y$ the identity on $Y$, for $x\in X$,
\begin{equation}
 \Tr_{X^c}(\Pi_x\otimes I_{\{x\}^c})
 =(N+1)^{L^d-|X|}(\Pi_x\otimes I_{X\setminus\{x\}}),
 \label{eq:bose-projector-local-algebra}
\end{equation}
which puts the probe in this algebra. The Gibbs-path theorem of that reference does not apply to the embedding: its local dimension and interaction norms grow with $L$, and the total-number projection remains global.

In main Eq.~\eqref{eq:bose-mixture}, the probability of an empty site given its amplitude $z_x$ is $e^{-|z_x|^2}$. At any fixed common phase put $a_T\coloneqq1+C_T(0)$ and $c\coloneqq C_T(x)$ for $x\ne0$. The independent Gaussian variables $(z_0\pm z_x)/\sqrt2$ have variances $C_T(0)\pm c$ and mean amplitudes of magnitudes $\sqrt{2\varrho_0(T)}$ and zero, respectively. Completing the square gives
\begin{align}
 \langle\Pi_0\rangle
 &=\frac{e^{-\varrho_0(T)/a_T}}{a_T},
 \label{eq:bose-vacuum-one}\\
 \langle\Pi_0\Pi_x\rangle
 &=\frac{e^{-2\varrho_0(T)/(a_T+c)}}{a_T^2-c^2}.
 \label{eq:bose-vacuum-two}
\end{align}
These expressions are independent of the phase and thus hold in the symmetric state. Expanding at $c=0$ gives
\begin{equation}
 \Cov(\Pi_0,\Pi_x)
 =\frac{2\varrho_0(T)e^{-2\varrho_0(T)/a_T}}{a_T^4}C_T(x)
   +O(C_T(x)^2).
 \label{eq:bose-vacuum-tail}
\end{equation}
The coefficient is positive and $C_T(x)\sim(T/J)A_d|x|^{2-d}$ by Sec.~\ref{sm:bose-correlations}. On the finite torus choose $r_L\coloneqq\lfloor(\ln L)^2\rfloor$ (rounding down) and $x_L\coloneqq r_L\hat e_1$. Apply Eq.~\eqref{sm:eq:bose-regional-error} to a cube of radius $r_L$ containing $0,x_L$ and trace out the distant cube. Each error term is $o(r_L^{2-d})$ by Eq.~\eqref{eq:bose-error-scales}, so the trace distance $\eta_L$ is of this order. Both projectors and their product lie between $0$ and $I$, giving a covariance error at most $3\eta_L$. The finite-torus covariance is therefore at least a positive constant times $r_L^{2-d}$. A fixed cube containing the origin and a buffer of width $\ell_L=\lfloor r_L/2\rfloor$ leave $x_L$ in the exterior. For every fixed polynomial $P$ and $\xi>0$, $P(L)e^{-\ell_L/\xi}=o(r_L^{2-d})$, excluding the exponential clustering bound even with its size-dependent prefactor.

\section{Optimal finite-size cooling of the canonical Bose gas}
\label{sec:bose-circuits}

Fix $0<T_-<T_+<T_c$. Each circuit layer consists of disjoint local
gates. The initial ancillas are independent of each other and of the
input. The backward radius $r_L$ counts all quantum and classical
wires, including feedforward; gates may depend on $L$ and need not
respect any symmetry. For opposite output cubes take
\begin{equation}
 r_L=o(L),\qquad R\coloneqq\lceil\sqrt{L(r_L+1)}\rceil,
 \qquad r_L=o(R),\quad R=o(L).
 \label{eq:bose-radius-regime}
\end{equation}
The independent readers pull back to disjoint warm cubes of radius
$R+r_L$. On both the cold and the enlarged warm cubes the trace error
in Lemma~\ref{sm:bose-block} is $o(1)$. We use a bounded event so that
this error need not decay faster than $1/\ln R$.

Put $c_\pm\coloneqq\sqrt{J\varrho_0(T_\pm)\kappa_d/T_\pm}$ and choose
$M=\lfloor\alpha R^{(d-2)/2}\rfloor$ labels for fixed $\alpha>0$,
within the allowed budget for large $R$. By main
Sec.~\ref{sec:bose-upper-bound}, any warm measurement can be
simulated from $Z_s=se^{i\Theta}+\xi$, $\xi\sim\mathcal{CN}(0,1)$,
with $s/R^{(d-2)/2}\to c_+$. Rotating by $-\Theta$, put
$x=|Z_s|-s$ and $y=s\operatorname{Arg}(e^{-i\Theta}Z_s)$.
Their polar density on $x>-s$, $|y|<\pi s$ is
$(1+x/s)e^{-x^2-2s(s+x)(1-\cos(y/s))}/\pi$.
It converges pointwise to $e^{-x^2-y^2}/\pi$; both integrate to one,
so the convergence is in total variation. Thus for every event we
may replace the radius by private noise and use the angle with
independent Gaussian error of variance $1/(2s^2)$ modulo $2\pi$,
at cost $o(1)$ for both blocks together.

Relative to two independent uniform angles, the ideal warm pair has
density ratio $2\pi g_{1/s^2}(t)$ at angle difference $t$, where $g_v$
is the wrapped normal density of variance parameter $v$. Its maximum
is $\sqrt{2\pi}s(1+o(1))$, by summing the Gaussian images at zero.
Independent readers preserve this ratio bound on every output event.
For balanced readers (all labels equally probable), independent
uniform angles give agreement $1/M$. Thus ideal warm agreement is
at most $\sqrt{2\pi}c_+/\alpha+o(1)$.

On the cold state the heterodyne angle has the same Gaussian
approximation with $s/R^{(d-2)/2}\to c_-$. The first angle is uniform
and independent of the difference $t$. For two equal-width bin
readers, the conditional chance of agreement is
$\max(1-M|t|/(2\pi),0)$ for the principal difference. Averaging gives
the limiting cold agreement
\begin{equation}
 p(\alpha,c_-)\coloneqq\frac{\sqrt{2\pi}c_-}{\alpha}
 \int_{-1}^1(1-|u|)e^{-2\pi^2c_-^2u^2/\alpha^2}\,\dd u.
 \label{eq:bose-agreement-target}
\end{equation}
Let $\varepsilon_L$ be the full trace distance to the cold target
after cooling. All approximations above place the warm simulation
within $\varepsilon_L+o(1)$ of the cold joint law. The cold marginals
are uniform by rotation symmetry. Each warm reader can restore its
own uniform marginal by relabeling with probability at most
$\varepsilon_L+o(1)$ (move only excess mass to deficient labels).
Together with the original comparison, this costs at most
$3\varepsilon_L+o(1)$ in agreement. Since $c_->c_+$ and
$\alpha p(\alpha,c_-)\to\sqrt{2\pi}c_-$, some fixed $\alpha$ gives
\begin{equation}
 \liminf_{L\to\infty}\varepsilon_L\ge
 \tfrac13\bigl[p(\alpha,c_-)-\sqrt{2\pi}c_+/\alpha\bigr]>0 .
 \label{eq:bose-error-floor}
\end{equation}
Applying this on any subsequence with $r_L/L\to0$ proves the linear
radius lower bound for arbitrary vanishing error. At fixed gate
diameter the depth bound is attained by the following construction.

\subsection{Exact number-conserving preparation in linear depth}
\label{sm:bose-linear-preparation}

For any $T_->0$ and any input $\sigma$ on the physical total-number-$N$
sector, three number-conserving operations give
\begin{equation}
 \mathcal E_{L,T_-}(\sigma)=\rho_L(T_-),\qquad D_L\le4d(L-1)\quad(L\ge2).
 \label{sm:eq:bose-preparation-depth}
\end{equation}
First gather all particles at the origin. On each child--parent bond
use Kraus operators $|a+b,0\rangle\langle a,b|$ (parent first,
$a,b\ge0$). Their adjoint products sum to the identity. Sweep each
line from $L-1$ down to zero in the first coordinate, then repeat on
the remaining hyperplanes for the other coordinates. Parallel lines
are disjoint, so $d(L-1)$ layers leave the root in $|N\rangle$ and
every other site empty, including for coherent inputs.

Next reverse the sweep. Give site $x$ the temporary momentum label
$k(x)\coloneqq2\pi x/L$. For each nonempty set $A$ of final sites let
\begin{equation}
 Z_A(M)\coloneqq\sum_{\substack{n_x\ge0,\ x\in A\\\sum_{x\in A}n_x=M}}
 e^{-\sum_{x\in A}\epsilon(k(x))n_x/T_-}.
 \label{sm:eq:bose-subtree-partition}
\end{equation}
When a parent representing $A$ splits into the disjoint descendant
sets $A_1,A_2$, assign $m$ of its $M$ particles to the child with probability
\begin{equation}
 p(m\given M)\coloneqq Z_{A_2}(m)Z_{A_1}(M-m)/Z_A(M),\qquad 0\le m\le M.
 \label{sm:eq:bose-split-probability}
\end{equation}
Convolution normalizes this law. On any input $|q,M-q\rangle$, the
Kraus operators $\sqrt{p(m\given M)}\,|M-m,m\rangle\langle q,M-q|$
define a trace-preserving channel and preserve the pair total.
After $d(L-1)$ reverse layers the probabilities telescope to the
canonical weights $e^{-\sum_x\epsilon(k(x))n_x/T_-}/Z_{L,N}(T_-)$.

Finally apply the mode Fourier transform
$b_x^\dagger\mapsto L^{-d/2}\sum_y e^{ik(x)\cdot y} b_y^\dagger$,
turning the sampled site-Fock vectors into energy eigenvectors. In each
of $d$ directions perform length-$L$ transforms on parallel lines
with adjacent two-mode rotations~\cite{Clements2016}. To see the
depth, eliminate the entry in row $r$, column $c<r$ ($0\le c<r<L$)
using rows $r-1,r$ at layer $2c+L-1-r$. Simultaneous pairs are
disjoint and required zeros are already in place. Reversing the
$2L-3$ elimination layers and including a diagonal phase layer
completes the claimed depth.

The root must hold $N=O(L^d)$ particles. Splits use precomputed
$L$-dependent probabilities and arbitrary occupation-dependent gates
with independent local environments; only spatial depth is counted. With $O_d(L^{d+1})$ gates, per-gate
trace-distance error $\eta_L$ uniform on local number at most $N$,
including arbitrary references, gives total error $O_d(L^{d+1}\eta_L)$
by telescoping. Thus $\eta_L=o(L^{-d-1})$ suffices.
\section{Optimal cooling depth for the Villain models}
\label{sm:villain-cooling}

We prove the finite-torus lower bound and matching constructions for
the scalar classical and quantum models. Fix $d\ge3$, $J>0$,
$\chi>0$ in the quantum case, and $0<T_-<T_+$ independently of $L$.
Use the circuit and error conventions of Sec.~\ref{sec:bose-circuits}:
local gates have disjoint supports within each layer, and the initial
auxiliaries are independent of the input and of one another; classical
communication is included in the circuit. All winding sectors
are included. The upper bounds allow exact local channels on
unbounded registers; they count spatial depth, not arithmetic cost
or bounded-strength evolution time.

\subsection{A spin-only test excludes sublinear causal radius}
\label{sm:villain-linear-lower}

The growing-block Gaussian estimate of
Lemma~\ref{sm:two-block-gaussian} also applies to the Villain fields.
Classically the zero-winding height covariance is $(T/J)G_L$.
Quantum mechanically both $X_{\chi,T}$ and $C_{\chi,T}$ of
Eqs.~\eqref{sm:q-C} and~\eqref{sm:q-seed} have symbol
$T/(J\lambda)+h(\lambda)$ with $h$ analytic near the lattice band.
Their symbols are positive and bounded below by a positive multiple
of $1/\lambda$. Lemmas~\ref{sm:q-folding-lemma} and
\ref{sm:q-factor-lemma} are already uniform in block size: their
additional trace-distance errors on opposite radius-$R$ blocks are
$O(e^{-cL^d})$ and $O(R^{2d}e^{-cL})$, respectively. Thus the quantum
two-block state can be prepared by a product of local channels from
two independent bulk center fields and one common uniform angle,
with error at most
\begin{equation}
 C_T\bigl[(R/L)^{d-2}+R^{2d}e^{-c_TL}
                      +e^{-c_TL^{d-2}}\bigr].
 \label{sm:villain-growing-error}
\end{equation}
The last term bounds the nonzero-winding probability. The classical
statement holds for the full cell data with the same bound. A
one-layer enlargement supplies all outgoing links. Hence the error
vanishes for every $R=o(L)$, uniformly over all local readers. For
quantum spin measurements, apply the same Gaussian comparison to
$C_{\chi,T}$; periodization changes their law by an exponentially
small error, by the same folding estimate.

Let the circuit's backward causal radius satisfy $r_L=o(L)$, and
choose $R=\lceil\sqrt{L(r_L+1)}\rceil$, where the ceiling means the
smallest integer not less than its argument. Then $r_L=o(R)$ and
$R=o(L)$. The two output cubes at separation $\lfloor L/2\rfloor$
have disjoint backward neighborhoods, contained in cubes of radius
$b=R+r_L+1$. Here the floor denotes the greatest integer not exceeding
its argument. The warm sufficient angle is a common uniform angle
plus wrapped Gaussian noise; its independent residuals can be
absorbed into private reader randomness. Before wrapping, the angle difference has variance
\begin{equation}
 v_L=\frac{2T_+}{J\operatorname{cap}(B_b)}(1+o(1)).
 \label{sm:villain-warm-variance}
\end{equation}
The expression is exact without $o(1)$ classically. In the quantum
case, Gaussian sufficiency and Eq.~\eqref{sm:q-cap} give the formula
for the center fields, so it bounds arbitrary quantum processing.

For the cold target use the weighted spin reader of
Sec.~\ref{sm:classical-spin-error}, and its quantum version in
Sec.~\ref{sm:q-readers}. Its angular error differs in mean square by
$o(R^{2-d})$ from a real Gaussian. The two linear errors are
independent conditional on the common angle and have difference
variance $u_L=2T_-/[J\kappa_dR^{d-2}](1+o(1))$. Thus
$u_L/v_L\to q\coloneqq T_-/T_+<1$.

For $t>0$ use the bounded score
\begin{equation}
 K_t(z)\coloneqq Z(t)^{-1}\sum_{k\in\Z}e^{-tk^2/2}e^{ikz},
 \qquad Z(t)\coloneqq\sum_{k\in\Z}e^{-tk^2/2}.
 \label{sm:villain-score}
\end{equation}
This normalized wrapped Gaussian takes values in $[0,1]$. If the first
angle is uniform and the second differs from it by an independent centered
Gaussian of variance $v$ modulo $2\pi$, the mean score is $Z(t+v)/Z(t)$.
Independent readers retaining uniform marginals cannot increase this
mean. Indeed, their conditional averages define $L^2$ contractions
$A,B$ on the uniform circle by Jensen's inequality. Write $H_v$ for
convolution by the wrapped Gaussian of variance $v$; its Fourier
eigenvalues are $e^{-vk^2/2}$. The mean after processing, multiplied
by $Z(t)$, is $\Tr(H_t A^*H_vB)$. The singular values of $A^*H_vB$
are bounded termwise by those of $H_v$. The trace inequality for singular
values therefore bounds this trace by $\sum_k e^{-(t+v)k^2/2}=Z(t+v)$.

Let $\varepsilon_{L,\mathrm{spin}}$ be the total variation distance
from the cold two-block spin-angle law. The cold marginals of the
reader angle are uniform by symmetry. As in Sec.~\ref{sec:bose-circuits},
the two warm marginals can be made uniform by private relabeling,
so the score comparison costs at most $3\varepsilon_{L,\mathrm{spin}}+o(1)$,
including Eq.~\eqref{sm:villain-growing-error}. Set $t=\alpha v_L$.
The cold Gaussian approximation costs $o(1)$ since
$\|K'_t\|_\infty=O(t^{-1/2})$ and its root-mean-square angular error
is $o(\sqrt{v_L})$. Even binning the angle into $N_R=\lceil R^\gamma\rceil$
arcs and redrawing inside each changes the score by at most
$4\pi\|K'_t\|_\infty/N_R=o(1)$ for $\gamma>(d-2)/2$.
Using $Z(t)\sim\sqrt{2\pi/t}$ and $q=T_-/T_+$ gives
\begin{equation}
 \liminf_{L\to\infty}\varepsilon_{L,\mathrm{spin}}
 \ge\frac13\sup_{\alpha>0}
 \left(\sqrt{\frac{\alpha}{\alpha+q}}
              -\sqrt{\frac{\alpha}{\alpha+1}}\right)
 =\frac{1-q^{1/3}}{3\sqrt{1+q^{1/3}+q^{2/3}}}>0 .
 \label{sm:villain-linear-floor}
\end{equation}
Full-state error dominates this finite-output test. The subsequence
argument of Sec.~\ref{sec:bose-circuits} yields $r_L=\Omega(L)$ for
vanishing-error cooling even when only the spins are tested.

\subsection{An exact classical circuit}
\label{sm:villain-classical-upper}

Use a nearest-neighbor spanning tree rooted at $o=0$. In the cube
$\{0,\ldots,L-1\}^d$, choose each nonroot site's parent
$p(y)\coloneqq y-\hat e_{i(y)}$ by decreasing a positive coordinate $i(y)$.
The tree avoids periodic seams and has height at most $d(L-1)$.
It computes the following coordinates and distributes the results. Read the winding $w_i$ by summing the integer links on
one coordinate loop, and broadcast $w$ and the root angle $\theta_o$.
Subtract the uniform slope $2\pi w_i/L$ from each lifted increment
and integrate along the tree to obtain a real field $h_o=0$. Its law
is centered Gaussian, with covariance
\begin{equation}
 \Cov_T(h_x,h_y)=\frac TJ
 [G_L(x-y)-G_L(x)-G_L(y)+G_L(0)].
 \label{sm:villain-root-covariance}
\end{equation}
The root angle is independent and uniform; the law of $h$ is
independent of $w$. These are coordinates of the original unpinned
Gibbs state. Draw $w'$ privately at the root from
\begin{equation}
 p_{T,L}(w)\coloneqq
 \frac{e^{-2\pi^2JL^{d-2}|w|^2/T}}
 {\sum_{v\in\Z^d}e^{-2\pi^2JL^{d-2}|v|^2/T}},
 \qquad w'\sim p_{T_-,L},
 \label{sm:villain-winding-law}
\end{equation}
and broadcast it. Put $s\coloneqq\sqrt{T_-/T_+}$ and
$\ell_w(x)\coloneqq2\pi w\cdot x/L$ in the coordinate cube
$\{0,\ldots,L-1\}^d$. Define
\begin{equation}
 \begin{aligned}
 \phi'_x&\coloneqq\theta_o+sh_x+\ell_{w'}(x),
 &j'_x&\coloneqq\lfloor\phi'_x/(2\pi)\rfloor,\\
 \theta'_x&\coloneqq\phi'_x-2\pi j'_x,
 &n'_{x,i}&\coloneqq j'_{x+\hat e_i}-j'_x
                  +w'_i\one_{\{x_i=L-1\}} .
 \end{aligned}
 \label{sm:villain-classical-circuit}
\end{equation}
The rescaled field has exactly the cold covariance, and the last
formula restores all links, including seams. Thus the full output
law is exactly cold. Loop sums, inward tree additions, and outward
broadcasts use $O_d(L)$ layers; a bounded edge coloring schedules
the final neighboring exchanges. A fixed number of unbounded real
and integer registers per cell suffices. The random winding is
transmitted by gates, not supplied as free shared randomness.

\subsection{Quantum encoding and cooling in linear depth}
\label{sm:villain-quantum-upper}

At fixed winding, Eq.~\eqref{sm:q-canonical-H} separates a rotor
of inertia $\chi L^d$ and $L^d-1$ real oscillators of frequencies
$\omega_k$. Let $\Gamma_T$ be their product thermal state and
$E_w$ the coordinate unitary encoding it in the physical sector.
The required transformation is
\begin{equation}
 \rho_{{\rm q},L,w}(T_+)
 \xrightarrow{E_w^\dagger}\Gamma_{T_+}
 \xrightarrow{\text{local channels}}\Gamma_{T_-}
 \xrightarrow{E_{w'}}\rho_{{\rm q},L,w'}(T_-).
 \label{sm:villain-quantum-circuit}
\end{equation}
We specify the locality of $E_w$, since Fourier diagonalization
alone would not establish a local circuit on the physical registers.

Add one dummy real oscillator at the zero-mode label and retain the
compact rotor at the root. Tensor products of one-dimensional real
Laplacian eigenbases transform these oscillators into real site
coordinates $h_x$. Rotations of two neighboring coordinates implement each length-$L$
orthogonal transform in $2L-3$ layers plus local signs, by the same
adjacent-row elimination schedule used for the Bose Fourier circuit.
All parallel lines are processed together. The mean
$\bar h=L^{-d}\sum_xh_x$ is the dummy coordinate divided by
$L^{d/2}$. Tree shears replace each nonroot $h_y$ by
$\delta_y=h_y-h_{p(y)}$, processing leaves first. If $\zeta_y$ is
the number of sites in the subtree rooted at $y$, including $y$,
divided by $L^d$, reversible subtree sums
implement
\begin{equation}
 \bar h-h_o=\sum_{y\ne o}\zeta_y\delta_y,
 \qquad \theta_o=[\Theta+h_o-\bar h]_{2\pi}.
 \label{sm:villain-quantum-root}
\end{equation}
Here $[a]_{2\pi}$ is the representative of $a$ in $[0,2\pi)$.
Add the sum to the real root register, subtract it from the compact
rotor, and undo the sums. This separates the independent dummy
register while retaining the differences and $\theta_o$.
A one-site dilation $\bar h\mapsto L^{d/2}\bar h$, with its unitary
Jacobian factor, restores the original dummy coordinate and normalized
wave packet before removal; the inverse inserts that same packet.

Set $j_o=0$ and traverse the tree outward. With
$z_y=\delta_y+\ell_w(y)-\ell_w(p(y))$, the controlled change
\begin{equation}
 \theta_y=[\theta_{p(y)}+z_y]_{2\pi},\qquad
 j_y=j_{p(y)}+\left\lfloor
                  \frac{\theta_{p(y)}+z_y}{2\pi}\right\rfloor
 \label{sm:villain-quantum-fold}
\end{equation}
is a unitary identification
$L^2(\R)\simeq L^2(S^1)\otimes\ell^2(\Z)$ at the child,
controlled by retained parent coordinates. It measures or copies
neither quantum angle. Initialize links in the seam field $n^w$ of
Sec.~\ref{sm:q-modes}. At site $x$, add $j_x$ to incoming links
and subtract it from outgoing links. The result is
$n_{x,i}=n^w_{x,i}+j_{x+\hat e_i}-j_x$.
Clear the integer scratch registers leaves first, using
$j_y=j_{p(y)}+n_{p(y),i(y)}$ on seam-free tree edges. This encodes
precisely $\Theta+h_x-\bar h+\ell_w(x)$ modulo a common turn.
Real shears include their unitary Jacobian factors. Each tree sweep
has linear depth. Each site updates only its incident links, so its
update overlaps only a bounded number of others at fixed $d$. All link
updates therefore fit in a number of layers independent of $L$. Reversing these operations
gives the decoder, with the same normalized dummy wave packet.

Read the commuting winding label by reversible loop additions and
measure only their total; this does not change the thermal input.
After decoding, replace the rotor by its cold law proportional to
$e^{-Q^2/(2\chi L^dT_-)}$. On each nonzero oscillator use a vacuum
ancilla of the same frequency and a beam splitter with transmissivity
\begin{equation}
 t_k\coloneqq\frac{\bar n_k(T_-)}{\bar n_k(T_+)},\qquad
 \bar n_k(T)\coloneqq\frac1{e^{\omega_k/T}-1}.
 \label{sm:villain-mode-cooling}
\end{equation}
Tracing its second output multiplies the covariance above vacuum by
$t_k$, producing the exact cold thermal oscillator. These channels
act in parallel. Resample and broadcast $w'$ with the cold law
\eqref{sm:villain-winding-law}, replace the seam labels, and encode
with $E_{w'}$. The full output is exactly cold. With arbitrary
ambient-cell gates a star can be split into nearest-neighbor link
additions, so the entire circuit has fixed range and depth $O_d(L)$.
Its fixed number of work registers per cell may be infinite
dimensional; exact coordinate unitaries and local state preparations
are allowed.

\end{document}